\documentclass[
    ,final            
  ]
  {aipproc}

\layoutstyle{6x9}

\usepackage[T1]{fontenc}
\usepackage{lmodern}
\usepackage{graphicx}
\usepackage{amsmath}
\usepackage{color}
\usepackage{amssymb}

\newcommand{\be}{\begin{equation}}
\newcommand{\ee}{\end{equation}}
\newcommand{\bea}{\begin{eqnarray}}
\newcommand{\eea}{\end{eqnarray}}
\newcommand{\beq}{\begin{equation}}
\newcommand{\eeq}{\end{equation}}

\newcommand{\e}{{\varepsilon}}

\begin{document}

\title{Superconductivity from repulsive interaction}



\author{Saurabh Maiti}{
address={Department of Physics, University of Wisconsin, Madison,
Wisconsin 53706, USA} }

\author{Andrey V. Chubukov}{
address={Department of Physics, University of Wisconsin, Madison,
Wisconsin 53706, USA} }

\begin{abstract}
The BCS theory of superconductivity named electron-phonon
interaction as a glue that overcomes Coulomb repulsion and binds
fermions into pairs which then condense and super-conduct.  We
review recent and not so recent works  aiming to understand
whether a nominally repulsive Coulomb interaction can by itself
give rise to a superconductivity.  We first discuss  a generic
scenario of the pairing by electron-electron interaction, put
forward by Kohn and Luttinger back in 1965, and then turn to
modern studies of  the electronic mechanism of superconductivity
in the lattice models for the cuprates,  the Fe-pnictides, and the
doped graphene.  We show that the pairing in all three classes of
materials can be viewed as lattice version of Kohn-Luttinger
physics, despite that the pairing symmetries are different.  We
discuss under what conditions the pairing occurs and rationalize
the need to do parquet renormalization-group analysis. We also
discuss the interplay between superconductivity and density-wave
instabilities.
\end{abstract}

\maketitle


\section{Introduction}

Superconductivity (SC) is one of most remarkable aspects of
quantum physics of interacting electrons. Discovered in 1911 by
Kamerlingh Onnes and his team of technicians~\cite{Onnes} it
preoccupied the minds of the most prominent physicists of the 20th
century and remains at the forefront of condensed-matter physics
in the 21st century.  Many ideas developed first in the studies of
superconductivity like the mass generation of the gauge field (the
Meissner effect)~\cite{london,LandauLifshitz}, and the mass
generation of superconducting phase fluctuations~\cite{anderson},
were later extended to other fields of physics and served as
paradigms for the works by Higgs~\cite{higgs} and others to
explain mass generation of the electro-weak gauge W and Z bosons
in particle physics~\cite{wilczek}.

In simple words, superconductivity is the ability of fermions to
carry electric current without dissipation. In quantum physics
such phenomenon is generally associated with the appearance of a
macroscopic condensate, i.e., a quantum state in which $10^{23}$
particles ``hold together" at the lowest quantum level and do not
allow individual particles to get swiped away by impurities,
interactions with boundaries, etc.   Bosons are capable to do this
because any number of them can occupy a single quantum level, and
the appearance of a macroscopic condensate of bosons is a
well-known phenomenon of Bose-Einstein condensation
~\cite{LandauLifshitz}.  Fermions, however, are "lone wolves"-- by
Pauli principle, only two of them (with opposite spins) can occupy
a single quantum level, others are expelled. As a result,
$10^{23}$ fermions occupy a comparable number of energy levels. In
this situation, any coherent motion of fermions (e.g., a current)
will only survive for a limited time, before fermions will be
individually affected by impurities, walls, etc.

There is a way to change this unwanted situation.  If, somehow,
fermions form bound pairs, quantum mechanics tells us that each
pair has spin $S=0$ or $1$, i.e., it becomes a boson. Bosons can
condense and behave as one monolithic object, i.e., if they are
forced to move in one direction by an applied electric field, they
will continue moving even after the field is turned off because
$10^{23}$ bound fermionic pairs  will not allow an individual
fermion  to change its direction due to, e.g., impurity
scattering.

This simple reasoning tells us that the phenomenon of
superconductivity can be straightforwardly explained if there will
be an explanation why fermions form bound states.  This is where
the real difficulty is.  An electron-electron interaction is
repulsive and generally should not allow fermions to pair.   That
remained the mystery for almost  50 years after the discovery of
superconductivity.

The breakthrough came in 1957 in a paper by Bardeen, Cooper, and
Schrieffer (BCS) \cite{BCS}.  They found that the interaction
between fermions and lattice vibrations effectively creates an
attraction between fermions. An electron creates a disturbance of
a lattice structure around it, another electron ``feels" this
disturbance and through it ``feels" the other electron.  BCS have
demonstrated that the effective electron-electron interaction,
mediated by quanta of lattice vibrations -- phonons, is attractive
at energies smaller than characteristic phonon frequency
$\omega_D$. On a first glance this may not be enough as electrons
still repel each other by Coulomb interaction. However, Coulomb
repulsion is known to become progressively smaller at smaller
energies due to screening in the particle-particle channel
\cite{BardeenPines,McMillan_68,Tolmachev}, and it drops between
fermionic bandwidth, which is typically of order few
electron-volts, and Debye frequency, which is typically a few tens
of millivolts. If the drop down to $\omega_D$ is strong enough,
electron-phonon interaction overshoots electron-electron repulsion
and the total interaction becomes attractive.  The BCS theory was
preceded by the observation by Cooper\cite{Cooper} that there is
no threshold for the pairing, i.e., an arbitrary weak attractive
interaction already gives rise to fermionic pairing. As a result,
all what is required for pairing is that  at energies of order
$\omega_D$ electron-phonon interaction must exceed screened
Coulomb interaction.

Electron-phonon mechanism of SC has been successfully applied to
explain the pairing in a large variety of materials, from $Hg$ and
$Al$ to recently discovered and extensively studied MgB$_2$ with
$T_c$ as high as  $39K$~\cite{mgb2}. The phonon density of states,
obtained by inelastic neutron scattering experiments, and the
spectrum of the bosons which mediate pairing, as deduced from
tunneling experiments, agree very well in systems like, e.g.,
Pb~\cite{McMillan69,Scalapino69}. This comparison of two
independent experiments, together with the isotope
effect\cite{isotope1,isotope2}, are generally considered to be a
very reliable proof of a phonon--mediated pairing state.

BCS theory also stimulated efforts to develop theoretical
frameworks to describe the phenomenon of SC, and the outcomes were
the fundamental Gorkov's theory of the SC state involving normal
and anomalous Green's functions\cite{AGD}, and the controlled
Eliashberg theory\cite{GEliashberg,Scalapino69,Carbotte11} of
superconductivity, which goes beyond the BCS theory and includes
fermionic self-energy and the dynamical part of the pairing
interaction.

Non-phononic mechanisms of the pairing has been also discussed,
most notably in connection with the superfluidity of
$^3He$~\cite{3he,3he2}, but didn't become the mainstream before
the breakthrough discovery of SC in LaBaCuO in 1986
\cite{bednorz}. That discovery, and subsequent discoveries of
superconductivity at higher $T_c$ in other cuprates, signaled the
beginning of the new era of ``high-temperature
superconductivity''~\cite{M2S}. The discovery, in 2008, of
superconductivity in $Fe$-based pnictides\cite{bib:Hosono} (binary
compounds of the elements from the 5th group: N, P, As, Sb, Bi)
with maximum $T_c$ near $60K$ quickly established another
direction of research in this field.

The high superconducting transition temperature is important but
not the central feature of the phenomenon of "high-temperature
superconductivity".  After all, $T_c$ in MgB$_2$ is higher than in
many Fe-pnictides.  What truly created an enormous interest to new
superconductors is the observation, shared by most
scientists(although not all of them)  that electron-phonon
interaction is too weak to account for observed $T_c$ in these
materials. The same belief holds for organic and heavy-fermion
superconductors, for which $T_c$ is smaller,  but electron-phonon
interaction is not the ``glue" for superconductivity, by one
reason or another.

If electron-phonon interaction is not the pairing glue, then what
binds electrons together? The only other option is Coulomb
interaction.  But it is repulsive, how can it give rise to the
pairing?  It turns out, it can. This set of lecture notes is an
attempt to present a comprehensive story of electron-electron
pairing by the nominally repulsive Coulomb interaction.

The study of the pairing due to electron-electron interaction
(often termed as electronic mechanism of superconductivity) has a
long history. It has been known from early 1950th  that screened
Coulomb potential has a long-range oscillatory tail $\cos(2k_F r +
\phi_0)/r^3$ at large distances $r$ ($k_F$ is Fermi momentum).
These oscillations are often called Friedel
oscillations~\cite{friedel}. Due to these oscillations, the
screened Coulomb interaction gets over-screened at some distances
and becomes attractive. Landau and Pitaevskii analyzed the pairing
at non-zero orbital momentum $l$ of the pair and found that the
pairing problem decouples between different $l$ (see Ref.
\cite{LandauLifshitz}). Because of this decoupling, even if only
one partial component of the interaction is attractive and all
other repulsive, the system still undergoes a pairing instability
into a state with $l$ for which the interaction is attractive.
Because the components of the interaction with large $l$ come from
large distances, it is conceivable that occasional over-screening
of the Coulomb interaction at large distances may make some of
partial interaction components with large $l$ attractive.

A bold next step in this direction has been made by Kohn and
Luttinger (KL) in 1965 \cite{kl,L}. They analyzed the form of the
fully screened irreducible pairing interaction at large $l$ in
three-dimensional, rotationally isotropic systems with $k^2/(2m)$
dispersion and separated the non-analytic $2k_F$ screening and the
regular screening from other momenta. They incorporated the latter
into the effective interaction $U({\bf q}) = U(q)$ ($q =|{\bf
q}|$) and made no assumptions about the form $U(q)$ except that it
is an analytic function of $q^2$.  The full irreducible pairing
interaction is $U(q)$ plus extra terms of $O\left(U^2(q)\right)$
coming from $2k_F$ screening. KL  argued that contributions to
partial components of the irreducible interaction from $2k_F$
scattering scale as $1/l^4$ due to the non-analyticity of the
$2k_F$ screening, while the partial components of analytic $U(q)$
behave at large $l$ as $e^{-l}$, i.e., are much smaller. This
smallness overshoots the fact that KL interaction is of second
order in $U$ and makes KL contribution larger than the direct
first-order interaction term.

KL found that, for large $l$, partial components of the full
irreducible interaction with even $l$ are attractive when
$U(0)/U(q=2k_F) > \sqrt{3} -1$, and components  with odd $l$ are
attractive  no matter what is the form of $U(q)$. As a result, any
rotationally-invariant system with repulsive Coulomb interaction
is unstable against pairing, at least at large enough odd $l$.
When $U(0)= U(2k_F) =U$, both odd and even components are
attractive.

The situation at smaller $l$ is less definite as one no longer can
separate the non-analytic  $2k_F$ contribution to the irreducible
pairing vertex and regular contributions from other momenta. In
this situation, one can only do perturbation theory to second
order in some bare $U(q)$. For momentum-independent $U(q) =U$, KL
attraction survives down to $l=1$ which is, by far, the largest of
attractive components~\cite{Fay,kagan}. For momentum-dependent
interaction, a bare $U(q)$ has components for all $l$ and whether
second-order KL contribution can overshoot bare interaction is not
obvious and depends on the details. One case when KL term
definitely wins and again leads to $l=1$ pairing instability, is
when the Born parameter is of order one, i.e., the radius of the
interaction in real space is about the same as $s-$wave scattering
length $a$, and $ak_F/\hbar$ is small.  In this situation, partial
components of $U(q)$ scale as $(a k_F/\hbar)^{2l+1}$, while KL
terms are of order $(ak_F/\hbar)^2$ for all $l$, i.e., the KL
components are parametrically larger for all $l >0$.

KL applied their results to $^3He$. Back in 1965, it was widely
believed that the pairing in $^3He$ should be with $l=2$, so they
approximated $U(q)$ by a constant $U$, expressed $U$ in terms of
s-wave scattering length $a$, used $ak_F/\hbar \approx 2$, known
for $^3He$, and obtained a ridiculously small $T_c \sim
10^{-17}K$.  A few years later, in 1968, Fay and Layzer~\cite{Fay}
extended KL calculations to $l=1$, which a few years later (in
1972) was found experimentally~\cite{osherov} to be the actual
pairing state in $3He$. For $p$-wave, the KL result for $T_c$ is
$\sim 10^{-3}K$, which by order of magnitude is the same as
experimental $T_c \sim 2.5 \times 10^{-3}K$ (Ref. \cite{osherov}).

The KL analysis in 2D  is more involved. If the regular
interaction $U(q)$ is momentum-independent, the $2k_F$ part is
also momentum independent for all $q \leq 2k_F$, which are
relevant for pairing (the pairing interaction connects momenta on
the Fermi surface).  However, a picture similar to that in 3D gets
restored once we apply perturbation theory and go to third order
in $U$ \cite{chubukov93}. Now the $2k_F$ part becomes momentum
dependent and non-analytic at $q \to 2k_F$ from below. Its partial
components at large $l$ scale as $1/l^2$ and are attractive. Like
in 3D, the largest KL attraction is for $l=1$, but now it scales
as $U^3$ rather than $U^2$. As an additional complication, the
relation between $U(q)$ and the scattering amplitude in 2D is also
logarithmically singular, but this does not affect the statement
about KL attraction down to $l=1$, only $U^3$ is replaced by
$(1/\log{ak_F})^3$.


We are now in a position to list the number of topics covered in
these lecture notes. In the presence of a lattice, rotational
symmetry is broken and one cannot simply expand in angular
harmonics, but has to consider discrete irreducible
representations for a particular lattice. We discuss how to
analyze pairing in lattice systems  and show that modified KL
mechanism works here as well, particularly when the bare
interaction is momentum-independent.  We consider three examples
of lattice physics in two dimensions:  the 2D model with two Fermi
pockets in different parts of the Brillouin zone, and two
different models with a single Fermi surface (FS), but with highly
anisotropic density of states, which is peaked at particular
points on the FS -- the "patch models". We consider models with
two and three non-equivalent patches. The model with Fermi pockets
is applicable to Fe-pnictides~\cite{2p_pnict}, the two-patch model
is applicable to overdoped cuprates~\cite{Furukawa}, and
three-patch model is applicable to doped graphene near $3/8$ and
$5/8$ filling, and to fermions on triangular lattice near $3/4$
filling~\cite{andreyd+id}.

We first discuss what is the condition on superconductivity in
lattice systems, assuming that we deal with short-range repulsive
interaction.  We show that, in many aspects, the situation is
similar to isotropic systems. Namely, for Hubbard $U$ model, the
bare pairing interaction is repulsive in a conventional $s-$-wave
channel, and zero in other channels, which in the cases we
consider are either d-wave like (in terms of how many times the
pair wave function changes sign along the FS), or, for the case of
Fe-pnictides, another $s-$wave, often called $s^{+-}$, in which
the pair wave function changes sign between different FS pockets
(the conventional $s-$wave is called $s^{++}$). We show that
second-order KL contributions to the pairing interaction are
attractive in these "other" channels, much like they are
attractive in all channels with $l>0$ in isotropic systems.

We then consider a more realistic case of a bare $U(q)$, which is
still short-range, but does have momentum dependence and is larger
at small $q$ than at large $q$, as it is expected for a screened
Coulomb interaction. We show that, in this situation, the bare
interaction is repulsive in all non-$s^{++}$ channels, and KL
contributions alone cannot cure the situation. We discuss three
approaches which give rise to pairing even in this case. All three
explore the idea that there is another order which the system
wants to develop in either spin or charge density-wave channel,
and fluctuations of this order increase the strength of the
attractive KL contribution and make it larger than the repulsive
contribution from the bare interaction.

The first one a phenomenological ``collective mode"
approach~\cite{acs}.  It abandons the controllable weak coupling
limit (i.e. expansion in $U$) and assumes that the pairing
interaction between fermions can be thought of as being mediated
by soft collective fluctuations of some density-wave order whose
fluctuations develop at energies much larger than those at which
superconductivity sets in. It is assumed that collective
excitations are soft enough such that this effective interaction
is large and exceeds the bare repulsive interaction. The form of
the static pairing interaction mediated by soft bosons is obtained
phenomenologically, based on physics intuition and experimental
results.

The second approach assumes that superconductivity and
density-wave instabilities are competing orders, which grow
together and develop at about the same energies/temperatures.
There is no pairing "mediated by collective bosons" in this case
because collective bosons by themselves develop at the same scale
as superconductivity.  Still, the idea is that, as these
fluctuations develop, they give progressively larger contribution
to the pairing channel via KL mechanism, and below some energy,
which is internally set by the system,  the attractive KL
interaction gets larger than the bare repulsion. This progressive
increase of the KL interaction with decreasing energy can be
analyzed within the parquet renormalization-group (RG)  approach
(either conventional~\cite{2p_pnict,andreyd+id,rg2} or
functional~\cite{frg1,frg2}), which is still a weak-coupling
approach, but it goes beyond second order in $U(q)$  ana allows
one to sum up series of logarithmically singular KL contributions
to the pairing interaction. One thing one should analyze in the RG
approach is whether superconductivity is the leading instability,
or density-wave order develops first.

And the third approach is to obtain the effective pairing
interaction in an approximate computation scheme, called a random
phase approximation (RPA), which amounts to a summation of a
particular class of ladder diagrams~\cite{Scal}.  In this
effective interaction  the bare repulsion and the KL contributions
are the first and the second terms in the expansion in $U(q)$,
however higher-order terms are not assumed to be small and in many
cases the RPA pairing interaction is attractive in one of
non-$s^{++}$ channels.  This approach is non-controlled, but its
advantage is that it can be equally applied to the case when
density-wave fluctuations develop before SC fluctuations and to
the case when density-wave and SC fluctuations develop at the same
energies.

Crudely speaking, in all these models, the KL effect is enhanced
by enhanced density-wave fluctuations in either spin or charge
channel.  For repulsive interaction, the enhancement in the  spin
channel is a natural choice.

\section{Background information: Fermi Liquid Theory}

To understand and describe  superconductivity, one needs to know
the concept and the mathematical apparatus on how to treat
interacting fermions at low temperatures. We first remind our
readers the basic facts about a non-interacting (free) Fermi gas
and then present a brief summary of properties of a
weakly/moderately interacting system of fermions, which is often
called a Fermi-liquid. We will  not attempt to derive the well
known Fermi-Liquid theory here (we direct the reader to
Ref.\cite{LandauLifshitz}) and only state some of the important
properties that will help us advance towards our goal of
understanding the phenomenon of superconductivity from
electron-electron interaction.

\subsection{A Free Fermi gas}

A classical gas is well described by the Boltzmann
statistics~\cite{LandauLifshitz1}, according to which an average
number of particles at an energy $E_k$ (the distribution function)
is given by \beq n (k) =e^{(\mu -E_k)/T}, \eeq where $\mu$ is the
chemical potential, and $T$ is the temperature. Throughout the
text we set the Boltzmann constant $k_B =1$ and measure
temperature in units of energy.

The classical statistics is valid when thermal de Broglie
wavelength of particles $\lambda_T \sim {\hbar}/(m T)^{1/2}$ is
much smaller than their separation. At lower temperatures,
$\lambda_T$ becomes comparable to a separation between particles
and quantum effects become important. Depending on whether
particles have half-integer spin or integer spin, they obey either
Fermi statistics
\beq\label{eq:FD}n_F(k)=\frac{1}{e^{(E_k-\mu(T))/T}+1} \eeq or
Bose statistics \beq
\label{eq:FD_1}n_B(k)=\frac{1}{e^{(E_k-\mu(T))/T}-1} \eeq We will
focus on electrons, whose spin is $S=1/2$, and only consider Fermi
statistics.

The total number $N$ of electrons in a volume $V$ is, at any
temperature is given by \beq\label{eq:mu}
\frac{N}{V}=2\int\frac{d^dk}{(2\pi\hbar)^d}n_F(k),\eeq where the
factor $2$ is due to a summation over the two spin projections. If
the ratio $N/V$ is fixed, this relation determines $\mu$ as a
function of $T$. At low $T$, $\mu$ is positive and is of order
\beq T_F = \frac{{\hbar}^2}{m} \left(\frac{N}{V}\right)^{2/3} \eeq
At temperatures $T \geq T_F$, $\mu$ changes sign and becomes
negative. At somewhat higher $T$,   $e^{-\mu(T)/T}$ gets large and
Fermi distribution function transforms into Boltzmann distribution
function. We are interested in fermionic properties at  low
temperatures, so we will deal with Fermi statistics, Eq.
(\ref{eq:FD}).

Eq. (\ref{eq:FD}) has the important property that at $T=0$, $n_F
(k) =0$ when $E_k>\mu$ and $n_F (k) =1$ for $E_k <\mu$. Because
$E_k=k^2/2m$ for a free gas, and $k$ is a good quantum number, the
condition $E_k =\mu$ determines a sharp boundary in $k-$space (a
FS) beyond which all states are empty and inside all states are
filled. Its radius is called $k_F$,  and the corresponding energy
$E_F = k^2_F/(2m)$ is called the Fermi energy.  By order of
magnitude, $E_F \sim T_F$. The Fermi energy at $T=0$ obviously
coincides with the chemical potential $\mu$.

In 3D,  the FS is a sphere. The value of $k_F$ is obtained from
(\ref{eq:mu}) and is given by $k_F=\left(\frac{3\pi^2 \hbar^3
N}{V}\right)^{1/3}$. In 2D, the FS is a circle, and
$k_F=\left(\frac{2\pi^2 \hbar^2 N}{V}\right)^{1/2}$.

The region near $k_F$ is the most relevant one for the low-energy
physics. The fermionic dispersion at $k \approx k_F$ can be
approximated by \beq\label{eq:lin}E_k -\mu  \approx
\frac{k_F}{m}(k-k_F) = v_F (k-k_F) \eeq where $v_F$ is called
Fermi velocity.

The total  ground state energy $E_0$ is the sum of $k^2/2m$ over
all occupied states. In 3D \beq\label{eq:energy_1}
E_0=\frac{3}{5}N\e_F \eeq

At a finite $T$, a Fermi-Dirac distribution function deviates from
the step function, and this gives rise to temperature dependencies
of observables. In particular, specific heat in linear in $T$. In
3D \beq C(T) = N T \frac{m \pi^2 \hbar^2}{k^2_F} \eeq

\subsection{Systems of interacting fermions}

\subsubsection{Fermi liquid theory}

For interacting fermions, quantum states of the full system,
$\epsilon_k$ do not reduce to a sum of quantum states of
individual particles.  Each $\epsilon_k$ should be understood as a
collective excitation of the whole system of $10^{23}$ particles.
In principle, there is no guarantee that an interacting system
still has well defined excitations with a given momentum $k$.
Nevertheless, the presence of such excitations is the key
postulate of a Fermi-liquid theory developed by Landau in
mid-1950s ~\cite{LandauLifshitz}. In Landau Fermi liquid (LFL)
theory, excitations behave much like free fermions, despite the
fact that each excitation is collectively produced by all fermions
in the system. These new excitations, called quasiparticles, obey
Fermi-Dirac statistics, and their number is the same as the number
of the actual particles, which implies that $k_F$  does not change
with the interaction.  The dispersion of quasiparticles
$\epsilon_k$ is not known for a generic $k$ but near $k_F$ is
assumed to have the same form as $E_k$ for free particles, but
with renormalized Fermi velocity: \beq\label{eq:lin_1}
\epsilon_k-\mu=v_F^*(k-k_F) \eeq Because $k_F$ is not
renormalized,  $v_F^* = k_F/m^*$, where $m^*$ is the quasiparticle
mass.  The ratio $m^*/m$ can be extracted from the measurements of
the specific heat $C(T)$ and is one of the indicators of the
strength of the interaction between fermions.

The change of the dispersion from $E_k$ to $\epsilon_k$ can be
viewed as if other fermions combine and create an effective field
which acts on a given quasiparticle. But this is not the only
effect. Interaction between fermions additionally forces them to
jump from one level to the other, i.e., they only spend a  finite
amount of time at a given (renormalized) level $\epsilon_k$. This
implies  that a lifetime of a quasiparticle with a given momentum
is finite. A finite lifetime means that the quasiparticle energy
has both real and imaginary parts.  The imaginary part of
$\epsilon_k$ can be obtained only perturbatively for $k$ far away
from $k_F$, but near $k_F$ the result is \beq Im \epsilon_k
\propto (k-k_F)^{\alpha} \eeq The condition that quasiparticles
are well defined near the FS implies that $Re \epsilon_k$ must be
larger that $Im \epsilon_k$, i.e., that $\alpha$ must be larger
than one. In the original Landau treatment of a 3D FL, $\alpha
=2$, and the extension of his arguments to arbitrary dimension,
$D$, shows that it is the case for all $D >2$. FLs with $\alpha
=2$  are sometimes called canonical FLs. In dimensions $D$ between
$1$ and $2$, $\alpha = D$, except for special cases, and in $D=2$,
$Im \epsilon_k \propto (k-k_F)^{2} \log{|k-k_F|}$. Such systems
are called non-canonical FLs. Still, by Landau criterion, all
these systems are Fermi liquids. In $D=1$ real and imaginary parts
of $\epsilon_k$ are of the same order, and the assumption of
well-defined quasiparticles near the FS becomes invalid.

\subsubsection{Microscopic treatment}

Microscopic treatment of interacting fermions is under control
when there is small parameter to justify perturbative expansion.
There are two situations when perturbative expansion is possible.
First is the case of Coulomb interaction $V(q)$ at small $r_s$,
where in 3D $r_s = (3/4\pi)^{1/3} (V/N)^{1/3} (e^2 m/\hbar^2)$
($r_s$ is small when the density $N/V$ is large enough. A
perturbative expansion in powers of $r_s$ is a bit tricky as
interaction between fermions accounts for the renormalization of
fermionic dispersion from $E_k = k^2/(2m)$ to a quasiparticle
$\epsilon_k$ and, at the same time, screens static long-range
Coulomb interaction and transforms  $V_q = \frac{4\pi e^2}{q^2}$
into \beq\label{eq:screened pot} V_{\text{screened}}=
\frac{V_q}{\epsilon(q)} =  \frac{4\pi e^2}{q^2 + \kappa^2}, \eeq
where $\kappa^2 \propto k^2_F r_s$.  In real space,
$V_{\text{screened}}(r)$ becomes a Yukawa potential
$V_{\text{screened}}(r)=\frac{4\pi e^2}{r}e^{-\kappa r}$.

Another situation for which perturbative expansion is possible is
the case when $r_s \geq 1$, but the magnitude of the screened
interaction is small. For $r_s \geq 1$, screened interaction does
not actually extend beyond a few nearest neighbors (a limiting
case when only on-site interaction is essential is known as the
Hubbard model). Once $U(r)$ is short-ranged and $U(q) = \int U(r)
e^{i{\bf qr}} d^3 r$ remains finite for all $q$, including $q=0$,
one can assume semi-phenomenologically that the overall magnitude
of the screened $U(q)$ is smaller than the Fermi energy, and use
$(N/V) U(0) /E_F$ as a small dimensionless parameter. In Born
approximation, this small parameter can be written as
$ak_F/\hbar$, where, as before, $a$ is s-wave scattering length
($a = m U(0)/4\pi \hbar^2$). Because $k^3_F$ is proportional to
fermionic density, one can also view the smallness of $ak_F/\hbar$
as the consequence of  small density rather than small
interaction.
Beyond Born approximation, $a$ is not linearly proportional to
$U(0)$ and $ak_F/\hbar$ can still be a small parameter at small
enough density even when $U$ is not small.

Fermi liquid postulates have been verified and confirmed in both
high-density and low-density expansions.  In particular, in small
$ak_F$ the effective  mass $m^*$ is, in 3D
\beq\label{eq:m*}\frac{m^*}{m}=1+\frac{8}{15\pi^2}(7ln2-1)\left(\frac{ak_F}{\hbar}\right)^2,
\eeq and expansion of the total  energy in powers of $ak_F$ is
\beq\label{eq:energy}
E_0=\frac{3}{5}N\e_F\left[1+\frac{10}{9\pi}\left(\frac{ak_F}{\hbar}\right)+\frac{4(11-2ln2)}{21\pi^2}\left(\frac{ak_F}{\hbar}\right)^2\right]
\eeq

\subsubsection{Green's Function Approach}

The mathematical apparatus to treat interacting fermions was
developed at about the same time as FL
theory~\cite{LandauLifshitz}. The two key elements in the theory
are single-particle and two-particle fermionic Green's functions.
A single-particle Green's function is defined in terms of the
time-ordered product of the two coordinate- and time-dependent
operators $\Psi =\Psi (x, t)$ in Heisenberg representation:
\bea\label{eq:GF}
&&G_{\alpha\beta}(X_1,X_2) = -i \langle T \Psi_{\alpha}(X_1)\Psi^{\dag}_{\beta}(X_2)\rangle \nonumber \\
&=&
\begin{array}{c}
-i\langle \Psi_{\alpha}(X_1)\Psi^{\dag}_{\beta}(X_2)\rangle~~~~~
\text{for}~~~~
t_1>t_2\\
i\langle \Psi^{\dag}_{\beta}(X_2)\Psi_{\alpha}(X_1)\rangle~~~
\text{for}~~~~
t_1<t_2\\
\end{array}
\eea where $X=(\vec{x},t)$, $T$ stands for time-ordering, the
brackets imply averaging over the ground state of an interacting
system, and $\alpha,\beta$ are the spin indices. Translational
invariance makes $G$ a function of only one variable $X\equiv
X_1-X_2$ and one can conveniently introduce its Fourier transform
$G(\vec{k},\omega)$.

\begin{figure}[htp]
\includegraphics[width=2.5in]{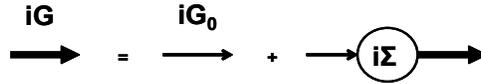}
\caption{\label{fig:G}Diagrammatic representation of the full
Green's Function in terms of the self energy $\Sigma$. The thin
line is the bare Green's function $G_0$ and the thick lines are
the fully renormalized Green's functions $G$. }
\end{figure}

The few things we need to know about single-particle Green's
function are
\begin{itemize}
\item
Green's function of free fermions $G(\vec{k},\omega)$ near the FS
is \beq\label{eq:G_FL}G_0(\vec{k},\omega)=\frac{1}{\omega-v_F
(k-k_F)/\hbar +i\delta sgn(\omega)}\eeq
\item
The Green's function of interacting fermions has poles and branch
cuts. Poles describe quasiparticle excitations, and the
quasiparticle energy spectrum
 $\omega = \epsilon_k$ is the solution of
\beq\label{eq:disp}G^{-1}(\omega,\vec{k})=0\eeq The branch cuts
describe fully incoherent excitations which cannot be attributed
to a single quasi-particle with a given momentum.
\item
In a generic FL,  Green's function near the FS has the form
\beq\label{eq:G_FL_1}G(\vec{k},\omega)=\frac{Z}{\omega-v^*_F
(k-k_F)/\hbar+i\delta sgn(\omega)}+G_{\text{inc}}\eeq where $Z$ is
the residue of the pole, $v^*_F$ is the renormalized Fermi
velocity, and $G_{\text{inc}}$ describes the incoherent part. The
damping term $(k-k_F)^{\alpha}$ is incorporated into the
incoherent part. Alternatively, near the mass shell ($\omega =
v^*_F (k-k_F)/\hbar$) it can be converted into $|\omega|^\alpha
sgn(\omega)$ term and added to the pole part as a replacement to
$i\delta sgn(\omega)$ term. In general, FL description is valid as
long as $0< Z <1$. For free fermions, incoherent part is absent
and $Z=1$. At small $ak_F/\hbar$  in 3D, \beq Z = 1 - \frac{8 \log
2}{\pi^2} (\frac{ak_F}{\hbar})^2 \eeq
\item
The momentum distribution function is obtained from the Green's
function by integrating over frequencies
\beq\label{eq:momentum_dist1}n_k=-i\lim_{t\rightarrow
-0}\int_{-\infty}^{\infty}\frac{d\omega}{2\pi}G(\omega,\vec{k})e^{-i\omega
t}\eeq The particular limit in this expression implies that the
integral can be extended only to the upper frequency half-plane.
For free fermions, frequency integration gives $n_k =
\theta(k_F-k)$, as it should be ($\theta (x) =0$ for $x <0$ and
$\theta(x) =1$ for $x >0$).
\item
The full Green's function $G$ is related to the Green's function
of free fermions $G_0$ by \bea\label{eq:dyson} G^{-1}
(k,\omega)&=&G_0^{-1}(k,\omega)+\Sigma(k,\omega) \eea where
$\Sigma$ is called self-energy (see Fig. \ref{fig:G}). The
quasiparticle residue and the effective mass are expressed via
partial derivatives of $\Sigma$ as \bea\frac{1}{Z}&=&1+
\frac{\partial \Sigma}{\partial
\omega} \label{eq:Z2}  \\
v^*_F &=&Z \left(v_F^*-\hbar\frac{\partial \Sigma}{\partial
k}\right) \label{eq:Z3} \eea
\end{itemize}

\begin{figure*}[htp]
$\begin{array}{ccc}
\includegraphics[width=2.7in]{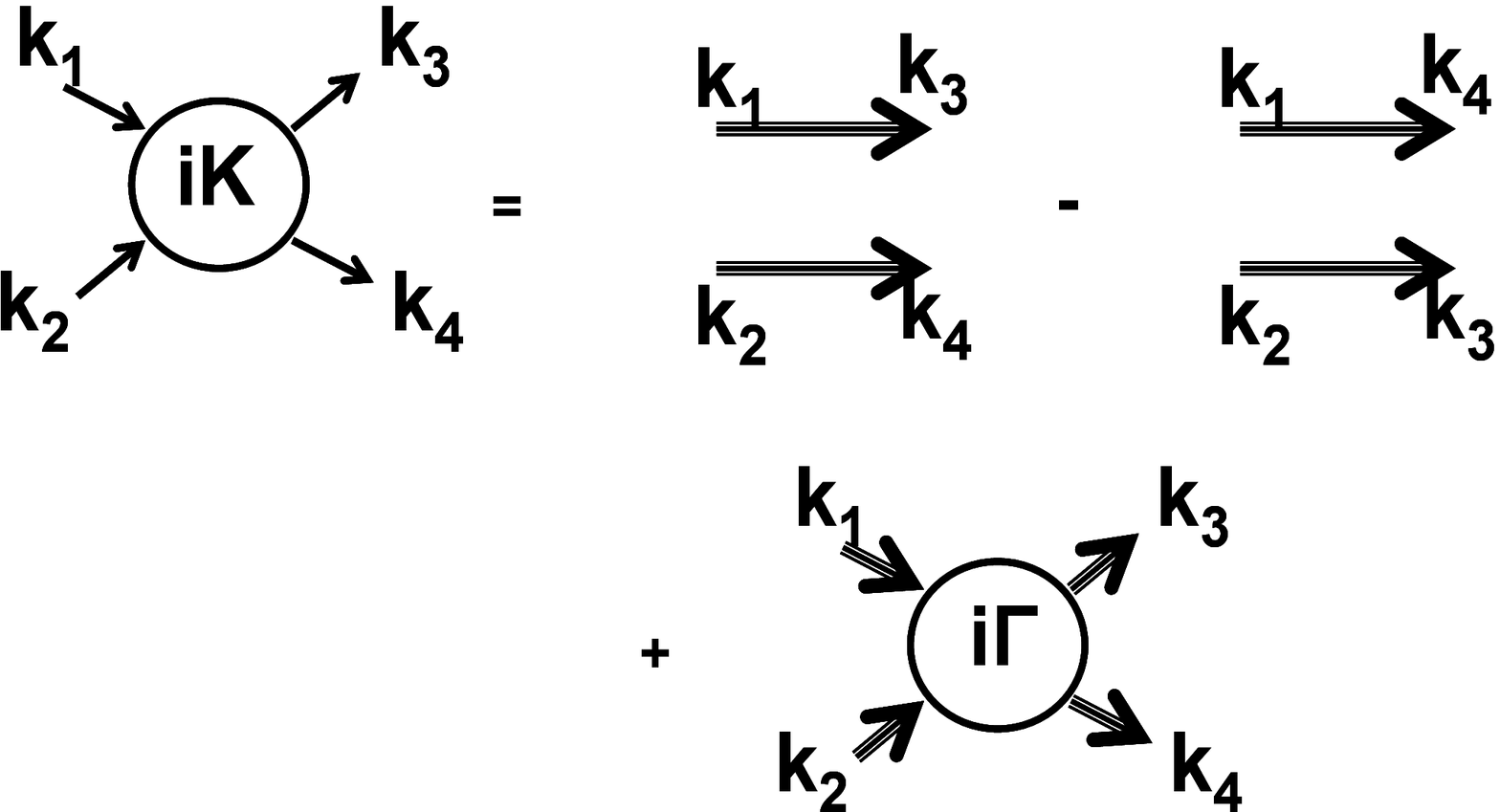}&~~~~~~~~&
\includegraphics[width=2.7in]{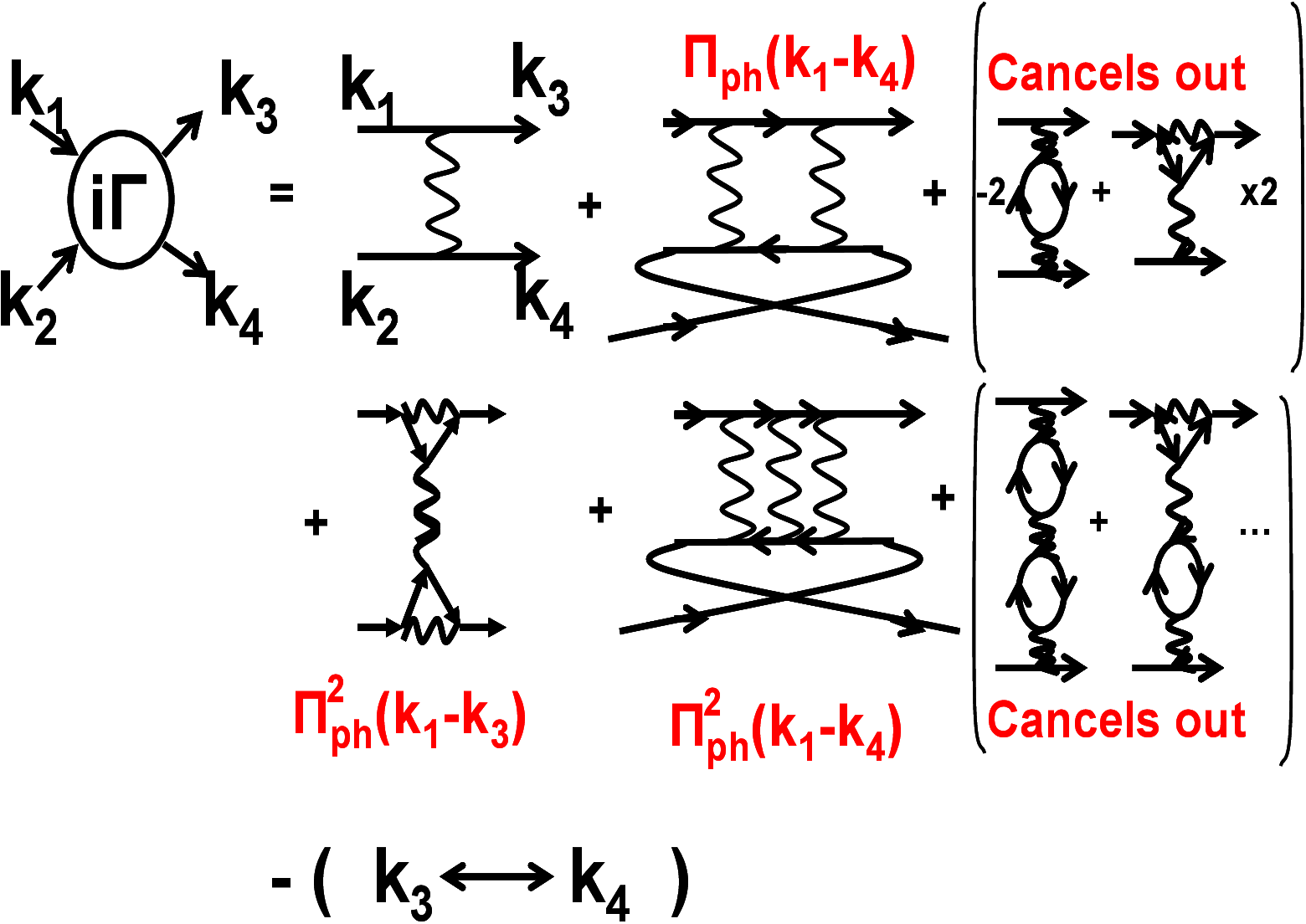}
\end{array}$
\caption{\label{fig:2part_green}(Left) Diagrammatic representation
of the full two particle Green's function as the anti-symmetrized
product of two Green's functions plus the vertex function
$\Gamma$. (Right) Diagrams that are included in the vertex
function $\Gamma$. (All labels correspond to a case of a constant
interaction.) }
\end{figure*}

\begin{figure}[htp]
\includegraphics[width=4in]{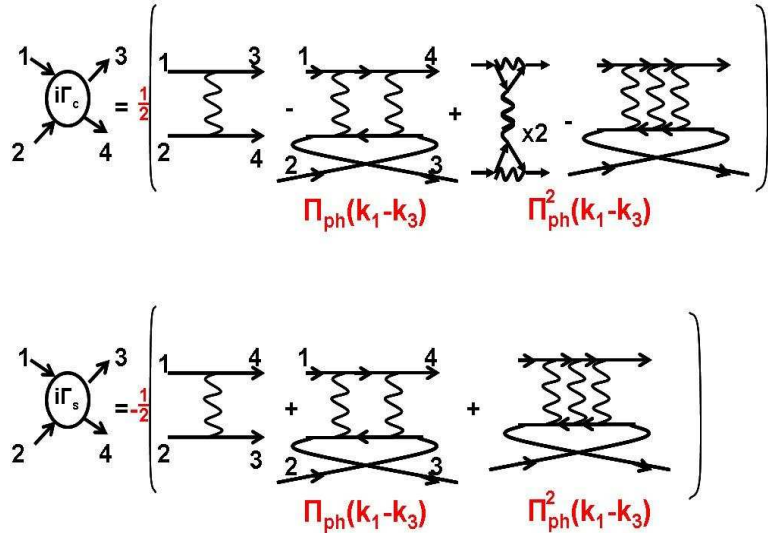}
\caption{\label{fig:S_C}Diagrams up to third order in the
interaction $U(q) =U$, that contribute to the charge and spin
components of the full vertex function, $\Gamma_c$ and $\Gamma_s$,
respectively. Only diagrams with small $q = {\bf k}_1-{\bf k}_3$
are shown.}
\end{figure}

The other element of the mathematical apparatus is a two-particle
Green's function \beq\label{eq:2gf}
K_{\gamma\delta,\alpha\beta}=\langle T \Psi_{\gamma} \Psi_{\delta}
\Psi^{\dag}_{\alpha} \Psi^{\dag}_{\beta} \rangle \eeq The
properties of the two-particle Green's functions are the
following:
\begin{itemize}
\item
For free fermions, $K_{\gamma\delta,\alpha\beta}$ reduces to an
anti-symmetrized product of two Green's functions and does not
provide additional information compared to a single-particle
Green's function.
\item
For interacting  fermions, $K$ contains an additional term $\Gamma
G^4$ (with a proper summation over spin indices), which cannot be
factorized into the product of two Green's functions (see
Fig.\ref{fig:2part_green}). The quantity $\Gamma$ is called the
vertex function. It contains information about interactions and
generally depends on frequencies and momenta of two incoming and
two outgoing fermions subject to momentum and frequency
conservation.
\item
To leading  order in the interaction $U(q)$, the  vertex $\Gamma$
coincides with the antisymmetrized interaction
\bea\label{eq:scat_vertex}
&& \Gamma^0_{\alpha\beta,\gamma\delta} (k_1,k_2;k_3,k_4) \nonumber \\
&&= -
U(\vec{k}_1-\vec{k}_3)\delta_{\alpha\gamma}\delta_{\beta\delta}+
U(\vec{k}_1-\vec{k}_4)\delta_{\alpha\delta}\delta_{\beta\gamma}\eea
Using the identity
\beq\label{eq:sigmaidentity}\vec{\sigma}_{\alpha,\gamma}\cdot\vec{\sigma}_{\beta\delta}=
-\delta_{\alpha,\gamma}\delta_{\beta\delta}+2\delta_{\alpha,\delta}\delta_{\beta\gamma},
\eeq we can decompose $\Gamma^{0}$ into   spin and charge
components: \bea\label{eq:scat_vertex2}
&&\Gamma^0_{\alpha\beta,\gamma\delta} (k_1,k_2;k_3,k_4) =
\Gamma^{0}_c \delta_{\alpha\gamma}\delta_{\beta\delta}+
\Gamma^{0}_s
\vec{\sigma}_{\alpha,\gamma}\cdot\vec{\sigma}_{\beta\delta}\eea
where \bea
&&\Gamma^{0}_c = \frac{-2U(\vec{k}_1-\vec{k}_3)+U(\vec{k}_1-\vec{k}_4)}{2} \nonumber \\
&&\Gamma^{0}_s = \frac{U(\vec{k}_1-\vec{k}_4)}{2} \eea The
decomposition into spin and charge parts is the consequence of
$SU(2)$ spin invariance and survives beyond the leading order,
i.e., the full $\Gamma_{\alpha\beta,\gamma\delta}
(k_1,k_2;k_3,k_4)$ has the same form as Eq.
(\ref{eq:scat_vertex2}), but $\Gamma^0_c$ and $\Gamma^0_s$ are
replaced by fully renormalized $\Gamma_c$ and $\Gamma_s$, which
generally depend on momenta and frequency, i.e., ${\bf k}_i \to
({\bf k}_i,\omega_i)$.
\item
In a generic FL, $\Gamma$, viewed as a function of either a total
frequency of two incoming fermions or a transferred frequency, may
have poles and branch cuts. The poles in $\Gamma_c$ (or
$\Gamma_s$) viewed as functions of a transferred frequency
describe weakly damped collective excitations in the charge (or
the spin) channel.  Examples of such excitations at $T=0$ are zero
sound waves in the charge channel and spin waves in the spin
channel.  To illustrate this, we present the expressions for spin
and charge components of $\Gamma_{\alpha\beta,\gamma\delta}
(k,p;k+q,p-q)$ for small transferred momentum $q$ and frequency
$\Omega$. For a constant interaction $U$,  one obtains, keeping
the terms which depend only on $q$, (see Fig\ref{fig:S_C})
\beq\label{eq:rpa} \Gamma_c=-\frac{U}{2} \frac{1}{1+U\Pi_{ph}(q,
\Omega)}~~~\Gamma_s = \frac{U}{2} \frac{1}{1-U\Pi_{ph}(q,\Omega)}
\eeq where $\Pi_{ph} (q,\Omega)$ is a particle-hole polarization
operator (the product of two Green's functions with relative
momentum $q$ and relative frequency $\Omega$). At zero frequency
and in the limit of zero momentum, $\Pi_{ph}(0,0)$ coincides with
the density of states at a Fermi level $N_0 = m
k_F/2\pi^2\hbar^3$. At zero momentum and any finite frequency,
$\Pi_{ph} (0,\Omega) =0$. This last condition is the consequence
of the conservation of the total number of particles. When both
$q$ and $\Omega$ are small, but the ratio $x\equiv \Omega/v_Fq$ is
arbitrary, \beq \Pi_{ph}(q,\Omega)=\Pi (x) =
\frac{mk_F}{2\pi^2\hbar^3}\left(1-\frac{x}{2} \log
\left|\frac{1+x}{1-x}\right|\right). \eeq

In particular, $\Pi_{ph} (q,\Omega)$ can be made arbitrary large
(and negative) when $\Omega$ approaches $v_F q$ from above. This
property of $\Pi_{ph} (q, \Omega)$ ensures the existence of the
pole in $\Gamma_c$ at $\Omega \approx v_F q$ (a zero sound mode).
\item  The poles in $\Gamma$ viewed as a function of a  total frequency of two fermions describe two-particle (or two-hole) collective  excitations
of $S=1/2$ fermions with  the total spin of a pair either $S=0$ or
$S=1$.  These poles are most relevant for superconductivity.
\item
A fermionic system is stable as long as the poles of collective
bosonic excitations are located in the lower half-plane of a
complex  frequency (either total or transferred).  Under this
condition, bosonic excitations decay with time. If the poles are
located in the upper frequency half-plane, excitations increase
with time and fermionic system becomes unstable.
\end{itemize}

\section{Superconductivity}

Like we said in the Introduction, superconductivity is the ability
of fermions to carry a persistent non-dissipative current in
thermodynamic equilibrium, without an applied field.  This does
not hold in conventional metals in which the current $\vec{j}$
appears as a response of a system to an external electric field
$\vec{E}$,  and  is related to $\vec{E}$ by the Ohm formula
\beq\label{eq:diss_curr} \vec{j}=\sigma \vec{E}\eeq where $\sigma$
is a conductivity. Resistivity, $\rho$, is inverse of
conductivity. In the Drude theory, $\sigma=\frac{ne^2\tau}{m}$,
where $\tau$ is the scattering time -- the time between collisions
imposed by either impurities or electron-electron interaction in
non-isotropic systems. The current is dissipative in the sense
that to carry a current $j$ one needs to grab the energy from the
field at the rate proportional to $\sigma E^2$.

How can one get a non-dissipative current?  Quantum mechanics
teaches us that such a possibility exists, at least in  principle.
Namely, if a quantum-mechanical system is described by a
macroscopic complex wave function $\Psi = |\Psi| e^{i\phi}$, it
contains a density current associated with the phase of the
complex wave function: \bea\label{eq:current}
\vec{j}&=&\frac{\hbar}{2m
i}\left(\Psi^*\vec{\nabla}\Psi-\Psi\vec{\nabla}\Psi^*\right)=\frac{|\Psi|^2}{m}\hbar\vec{\nabla}\phi\eea

In a conventional metal, a macroscopic wave function is, to first
approximation, an antisymmetrized product of single-fermion wave
functions (the Slater determinant). The phases of individual wave
functions are not correlated, and $\vec{\nabla}\phi$ vanishes
after averaging over particles. The situation changes, however, if
there a macroscopic number of particles in the same
quantum-mechanical state.  In this case,  a wave function of the
{\it whole} system has the form $\Psi = |\Psi| e^{i\phi}$, and
$\vec{\nabla}\phi$ determines a macroscopic, non-dissipative
current in the thermodynamic equilibrium.  If particles are
charged, a density current is proportional to a charge current,
hence a persistent electric current also exists in an equilibrium.

The existence of a macroscopic condensate is a well-known property
of bosonic systems, where a condensate emerges as a result of
Bose-Einstein condensation.  But electrons are fermions, and Pauli
exclusion principle prevents fermions to accumulate on a single
quantum level. But what if fermions are bound into pairs?  A pair
of fermions has spin $S=0$ or $S=1$ and  therefore behaves as a
boson.  Bosons do condense, as we just said, and the condensation
of bosons generates a persistent current (which is often called a
supercurrent).

So, once fermions form bound pairs, they will eventually generate
a supercurrent. The issue then is to identify the mechanism by
which fermions are bound into pairs.  This is a challenging task
as electron-electron interaction is repulsive, and repulsive
interaction is not expected to produce a bound state.

The solution\cite{BCS}, proposed by Bardeen, Cooper, and
Schrieffer in 1957, is to use electron-phonon interaction as a
pairing glue. A passing electron creates a perturbation of the
lattice, another electron passing through the same area "feels"
the perturbation, and through it "feels" another electron. In a
more scientific language, lattice vibrations (phonons) mediate the
retarded interaction between fermions. One can show quite
generally that such effective interaction is attractive
 at energies smaller or comparable to a Debye
frequency $\omega_D$. Still, to give rise to bound pairs,
electron-phonon attraction has to overcome a supposedly much
larger Coulomb repulsion. What helps is  that Coulomb repulsion
 progressively gets smaller at smaller frequencies.  If $\omega_D$ is small
enough compared to the fermionic bandwidth, $W$, Coulomb
interaction already gets reduced a lot  between $W$ and
$\omega_D$, and detailed calculations have found that
electron-phonon attraction, which emerges below $\omega_D$, well
may overshoot this reduced Coulomb repulsion.

The weak coupling theory of superconductivity based on
electron-phonon interaction has been developed by Bardeen, Cooper,
and Schrieffer and is known as BCS theory. Gorkov and
Melik-Barkhudarov~\cite{gmb} extended weak coupling theory one
step ahead and found exact expression for $T_c$.  The BCS theory
has been further extended by Eliashberg\cite{GEliashberg}, who
demonstrated that  the dynamical part of electron-phonon
interaction can be incorporated into the theory in a controllable
way.  Both BCS and Eliashberg theories are actually not specific
to electron-phonon interaction and, with some modifications, can
be applied to any pairing mechanism.

Electron-phonon mechanism of superconductivity works well for some
superconductors, but, as most of researchers believe, it does not
account  for the pairing symmetry and high $T_c$ obtained in Cu
and Fe-based superconducting  materials and in several other
families of superconductors.  The only other alternative is
superconductivity originating directly from repulsive
electron-electron interaction.

To set the stage for the analysis of the pairing by
electron-electron interaction, in the remainder of this section we
will briefly review two generic issues about the pairing in
isotropic (rotationally-invariant) systems:  (i) that it emerges
already for arbitrary weak attraction and (ii) that the pairing
problem decouples between different angular momenta $l$ and it is
sufficient to have an attraction for just one angular component of
the interaction (i.e., for just one value of $l$).

\subsection{Pairing instability at arbitrary weak interaction}

\subsubsection{A constant interaction}

\begin{figure}[htp]
\includegraphics[width=4in]{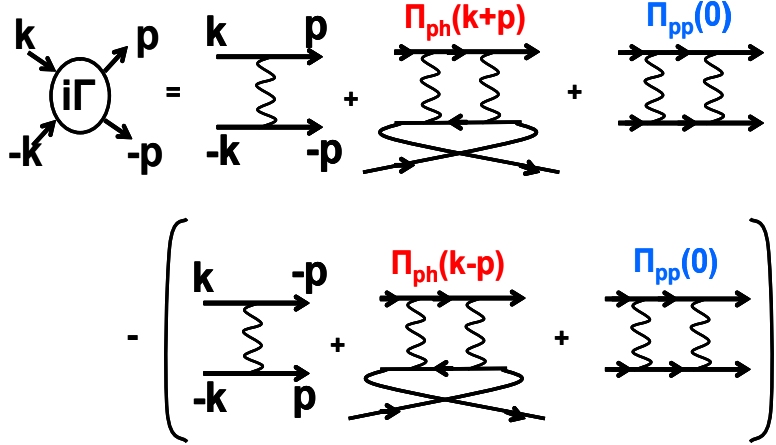}
\caption{\label{fig:Gamma_pairing}Diagrams to second order in the
interaction $U(q) =U$, which contribute to the vertex function
with zero total incoming momentum. There are contributions from
both particle-hole and particle-particle channels. Other diagrams
to order $U^2$ cancel each other and we do not show them.}
\end{figure}

In the mathematical apparatus developed to study interacting
fermions, the information about the potential bound pairs is
encoded in the 2-particle vertex  function $\Gamma$ which is a
fully renormalized and antisymmetric interaction between
quasi-particles. Consider first a system of fermions with a small
constant attractive interaction $U$ and compute $\Gamma$ in
perturbation theory. To first order in $U$, $\Gamma$ is given by
Eq. (\ref{eq:scat_vertex}). Earlier, we had decoupled $\Gamma$
into spin and charge parts.  To study pairing, i.e., possible pair
states with $S=0$ or $S=1$, it is more convenient to decompose
$\Gamma$ into singlet and triplet components. A singlet component
has spin structure \beq \delta_{\alpha\gamma}\delta_{\beta\delta}
- \delta_{\alpha\delta}\delta_{\beta\gamma} \eeq and a triplet
component is \beq \delta_{\alpha\gamma}\delta_{\beta\delta} +
\delta_{\alpha\delta}\delta_{\beta\gamma} \eeq For a constant $U$,
only singlet component is present at the leading order, i.e., \beq
\Gamma^{0}_{\alpha\beta,\gamma\delta}=-U
\left(\delta_{\alpha\gamma}\delta_{\beta\delta} -
\delta_{\alpha\delta}\delta_{\beta\gamma}\right) \eeq Let's go to
next order.  The four diagrams which give rise to the
renormalization of $\Gamma$ to order $U^2$ are shown in Fig.
\ref{fig:Gamma_pairing}.

Two of these diagrams contain a pair of fermionic Green's
functions with opposite directions of the arrows. This pair is
called a particle-hole bubble because one can immediately check
that the momentum and frequency integral over the two Green's
functions of intermediate fermions is non-zero at $T=0$ only when
one fermion is above FS, i.e., is a particle, and the other one is
below FS, i.e., is a hole. The other two diagrams contain the
product of two fermionic G's with the same direction of arrows.
This combination is called particle-particle bubble because  the
momentum and frequency integral over these two Green's functions
is non-zero at $T=0$  when both fermions are above or below FS,
i.e. both are particles, or both are holes.

The two diagrams with particle-hole bubbles contain particle-hole
bubbles with different momentum combination ($\Pi_{ph} (k-p)$ and
$\Pi_{ph} (k+p)$ in Fig\ref{fig:Gamma_pairing}) and does affect
the spin structure of the interaction, generating the triple
component. These diagrams are singular  when transferred frequency
and transferred $v_F q$ are nearly equal, and this singularity
gives rise to zero-sound waves.  But we are interested in
potential poles as a function of the total frequency of two
fermions, while transverse frequencies and transferred momenta can
be arbitrary. It is easy to verify that the particle-hole bubble
is not singular for a generic transferred momentum and  frequency
and therefore is incapable to substantially modify $\Gamma$ at
small $U$.

The particle-particle bubble is a different story. Suppose we set
the total momentum $q$ of two fermions to be zero (by momentum
conservation total incoming and outgoing momenta are both zero). A
straightforward computation shows that in this situation the
renormalization in the particle-particle channel does not affect
the spin structure of the interaction, and that at $T=0$, each
particle-particle bubble is logarithmically singular in the limit
of small total frequency and behaves as

\bea
\Pi_{pp} (\Omega, q=0)&=&-i \int\frac{d^3k d \omega}{(2\pi\hbar)^4}G_{k,\omega}G_{-k, -\omega + \Omega}\nonumber\\
&=&\frac{m
k_F}{2\pi^2\hbar^3}\left[ln\left(\frac{\omega_c}{\Omega}\right) +
i\frac{\pi}{2}\right] \eea

where $\omega_c$ is the upper limit of the integration over $v_F
(k-k_F)$, which, physically, is the upper end of the energy range
in which $U$ can be approximated by a constant.

The logarithmical divergence of $\Pi_{pp} (q=0,\Omega)$ at $\Omega
\to 0$ implies that the product $U \Pi_{pp}$ cannot be neglected
even when the interaction is weak. Keeping this renormalization,
we find that, to order $U^2$, \beq
\Gamma^{0}_{\alpha\beta,\gamma\delta}=-U \left(1- U \Pi_{pp}
(0,\Omega)\right) \left(\delta_{\alpha\gamma}\delta_{\beta\delta}
- \delta_{\alpha\delta}\delta_{\beta\gamma}\right) \eeq We assume
and then verify that the most relevant $\Omega$ are the ones for
which $U \Pi_{pp} (q=0,\Omega)$  are of order one. Let's go now to
next order of $U$. The number of diagrams increase, but for $U
\Pi_{pp} = O(1)$ all of them are small in $U$,  except for the two
diagrams with two particle-particle bubbles which give $(U
\Pi_{pp})^2$ with prefactor equal to one  The same holds for
fourth and higher orders in $U$. One can easily make sure that
perturbative series form geometric progression, hence the full
$\Gamma$ in this approximation is \beq
\Gamma^{full}_{\alpha\beta,\gamma\delta}=- U \frac{1}{1 + U
\Pi_{pp} (0,\Omega)}
\left(\delta_{\alpha\gamma}\delta_{\beta\delta} -
\delta_{\alpha\delta}\delta_{\beta\gamma}\right) \eeq For positive
(repulsive) $U$,  $\Gamma$ has no poles, but for negative
(attractive) $U$,  $\Gamma$ has a pole at $\Omega=i\Omega_p$ where
\beq \Omega_p=\omega_c e^{-2\pi^2 \hbar^3/|U|mk_F} =\omega_c
e^{-\frac{\pi \hbar}{2 |a|k_F}}. \eeq where, $a =
mU/(4\pi\hbar^2)$ is s-wave scattering length in Born
approximation. The pole exists at arbitrary small $U$ and, as we
see, is located in the {\it upper} half-plane of complex frequency
$\Omega$.  A pole in the upper half-plane implies that, if we
create an excitation with $\Omega_p$, its amplitude will
exponentially grow with time and destroy a Fermi liquid state that
we departed from. What does it mean physically?  The excitations,
which grow with time, describe fluctuations in which a pair of
fermions behaves as a single boson with total spin $S=0$ and zero
momentum.  A natural suggestion would be that the new state, which
replaces a Fermi liquid, contains a macroscopic number of such
bosons in the same $q=0$ state, i.e., the ground state has a
macroscopic condensate. This is precisely what is needed for
super-current.

The analysis of the pole in $\Gamma$ can be extended to a to a
non-zero total momentum $q$ and to a finite temperature.
Calculations show~\cite{AGD} that at a finite $q$ the pole is
located at
$\Omega=i\Omega_p\left(1-\frac{v_F^2q^2}{6\hbar^2\Omega^2_p}\right)$.
Once $q$ exceeds the critical value $\sqrt{6} \hbar\Omega_p/v_F$,
the pole moves to the lower half-plane in which case a collective
excitation decays with time and does not destroy a Fermi liquid.
The consequence is that, for moving fermions,  the pairing
instability exists only when their velocity is below the critical
value.   A finite $T$ leads to the same effect:  the pole is
located in the upper frequency half-plane only at $T < T_c$, where
$T_c$ is comparable to $\Omega_p$. At larger $T$, the pole is in
the lower frequency half-plane, and a Fermi-liquid state is
stable. Note by passing that in weak coupling theory bound pairs
appear and condense at the same $T$.   Beyond weak coupling, pairs
condense at a lower $T$ than the one at which they appear. This
difference between the two temperatures may be large at strong
coupling in lattice systems.  This phenomenon is often termed as
BCS-BEC crossover (BEC stands for Bose-Einstein condensation)~\cite{mohit}
The meaning is that at strong coupling pairs of fermions appear at
high $T = T_{pair}$, and condense at low $T = T_{BE}$, and between
$T_{BE}$ and $T_{pair}$ the system can be described as a
weakly/moderately interacting gas of uncondensed bosons.

The two main messages here are (i) the pairing instability can be
detected from Fermi-liquid analysis as the appearance of the pole
in $\Gamma$ in the upper half-plane of the total frequency of two
fermions, and (ii)  there is no threshold for such phenomenon --
Fermi liquid state gets destroyed already at infinitesimally small
attraction between fermions.

\subsubsection{Momentum-dependent interaction}

How this helps our consideration of a possible pairing due to
repulsive electron-electron interaction? If fully screened
electron-electron interaction was a positive constant, we surely
would not get any superconductivity as attraction is still a must
condition for the pairing.  But the  screened electron-electron
interaction $U(q)$ is generally a function of $q$. Let's see what
we obtain for the pairing when the interaction $U(q)$ is still
weak, but momentum-dependent.

\begin{figure}[htp]
\includegraphics[width=4in]{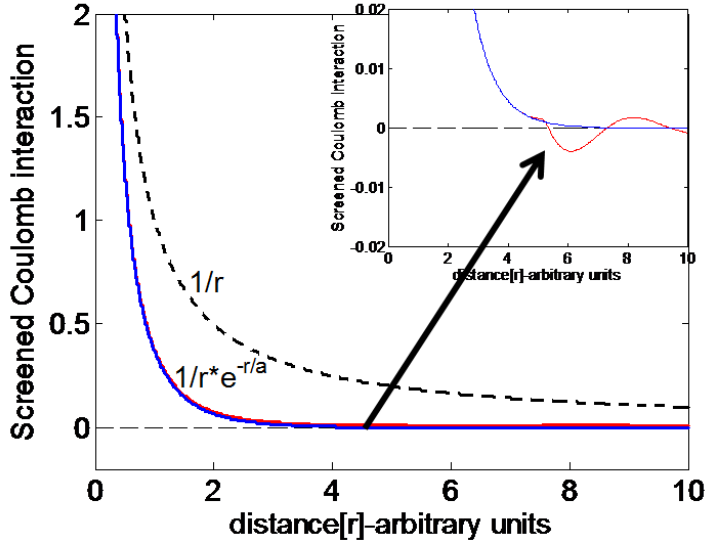}
\caption{\label{fig:friedel}The screened coulomb potential as a
function of $r$. $\frac{1}{r}$ (dashed line) is the bare coulomb
potential. $\frac{1}{r}e^{-r/a}$ (blue line) is the Yukawa
potential which includes regular screening and dies of
exponentially ($a$ is some characteristic screening length). The
fully screened potential (red line) includes the contribution from
the $2k_F$ scattering which gives rise to Friedel oscillations at
large $r$. These oscillations are responsible for the attraction
in large angular momentum channels. The inset is a zoomed in
version, which shows the oscillations.}
\end{figure}

The input for the analysis is the observation that the
logarithmical singularity in $\Pi_{pp}$ comes from fermions in the
immediate vicinity of the Fermi surface. To logarithmic accuracy,
the interaction between fermions with incoming momenta ${\bf k},
-{\bf k}$ and outgoing momenta ${\bf p}$ and ${-\bf p}$ can then
be constrained to particles on the FS, such that $U(q = |{\bf k} -
{\bf p}|)$ depends only the angle $\theta$ between incoming ${\bf
k}_F$ and outgoing ${\bf p}_F$. The decomposition of the vertex
function $\Gamma$ into spin-singlet and spin-triplet channels now
gives \bea
&&\Gamma^{0}_{\alpha\beta,\gamma\delta}=-U(\theta) \delta_{\alpha\gamma}\delta_{\beta\delta}  + U(\pi-\theta) \delta_{\alpha\delta}\delta_{\beta\gamma} \nonumber \\
&& = - \frac{U(\theta)+ U(\pi -\theta)}{2}  \left(\delta_{\alpha\gamma}\delta_{\beta\delta} -  \delta_{\alpha\delta}\delta_{\beta\gamma}\right) \nonumber \\
&& - \frac{U(\theta)- U(\pi -\theta)}{2}
\left(\delta_{\alpha\gamma}\delta_{\beta\delta} +
\delta_{\alpha\delta}\delta_{\beta\gamma}\right) \label{fr_1} \eea
The way to proceed is to expand the interaction $U(\theta)$ into
angular momentum harmonics. In 3D, we have \beq\label{eq:spherical
harmonis} U(\theta)=\sum_l (2l+1)P_l(\theta)U_l,\eeq where
$P_l(\theta)$ are the Legendre polynomials: $P_0 (\theta) =1$,
$P_1 (\theta) = \cos \theta$, $P_2 (\theta) = (3 \cos^2 \theta
-1)/2$, etc. Even components $\Pi_{l=2m} (\theta)$ satisfy
$\Pi_{2m} (\theta) = \Pi_{2m} (\pi -\theta)$, odd components
satisfy $\Pi_{2m+1} (\theta) = -\Pi_{2m+1} (\pi -\theta)$.
Substituting into  (\ref{fr_1}), we obtain that spin-singlet
contribution is the sum of even components, and spin-triplet
contribution is the sum of odd components. \bea
&&\Gamma^0_{\alpha\beta,\gamma\delta} (\theta)= \nonumber \\
&&- \sum_{m=0}^\infty \left[(4m+1) P_{2m} (\theta) U_{2m} \left(\delta_{\alpha\gamma}\delta_{\beta\delta} -  \delta_{\alpha\delta}\delta_{\beta\gamma}\right)\right. \nonumber \\
&& \left. +  (4m+3) P_{2m+1} (\theta) U_{2m+1}
\left(\delta_{\alpha\gamma}\delta_{\beta\delta} +
\delta_{\alpha\delta}\delta_{\beta\gamma}\right)\right]
\label{fr_2} \eea One can easily make sure that the spin structure
of $\Gamma$ is reproduced at every order, if we restrict with
renormalizations in the particle-particle channel, i.e., even and
odd angular momentum components do not mix. As a result, the full
$\Gamma$ in this approximation is given by \bea
&&\Gamma^{full}_{\alpha\beta,\gamma\delta} (\theta)= \nonumber \\
&& - \sum_{m=0}^\infty \left[(4m+1) P_{2m} (\theta)
{U}^{full}_{2m}
\left(\delta_{\alpha\gamma}\delta_{\beta\delta} -  \delta_{\alpha\delta}\delta_{\beta\gamma}\right)\right. \nonumber \\
&& \left. +  (4m+3) P_{2m+1} (\theta)
{U}^{full}_{2m+1}\left(\delta_{\alpha\gamma}\delta_{\beta\delta} +
\delta_{\alpha\delta}\delta_{\beta\gamma}\right) \right]
\label{fr_2_1} \eea Even more, using the property of the Legendre
polynomials \beq \int \frac{d\Omega_{q}}{4\pi} P_{m} (\cos
\theta_{k,q}) P_{n} (\cos \theta_{q,p}) =
\frac{1}{2m+1}\delta_{m,n} P_m (\cos\theta_{k,p}) \eeq where
$\theta_{k,q}$ is the angle between fermions with momenta ${\bf
k}_F$ and ${\bf q}_F$ and $d\Omega_q$ is the element of the solid
angle for ${\bf q}_F$, one can show that  components with
different $m$ also do not mix up, i.e., each partial component
${U}^{full}_l$ of the full interaction is expressed only via
$U_l$.  The relations are the same as at $l=0$, i.e., \beq
{U}^{full}_l  (q=0,\Omega) = \frac{U_l}{1 + U_l
\Pi_{pp}(q=0,\Omega)} \label{sa_1} \eeq This result is very
important for our story. It states that, even if the
angular-independent component $U_{l=0}$ is repulsive, the pairing
instability may still occur at some finite angular momentum $l$.
All what is needed is that just one partial channel is attractive,
either for even or for odd $l$. This may, in principle, occur even
if overall the interaction is repulsive. A good hint comes from
the analysis of screened Coulomb interaction. As a reader surely
knows, a  screened potential far away from a charge contains
Friedel oscillations -- ripples of positive and negative regions
of charge density (see Fig.\ref{fig:friedel}) Overall, screened
interaction is indeed repulsive, but the negative regions can
provide attraction at some angular momenta, particularly at large
$l$, because dominant contributions to components $U_l$ with $l
>>1$ come from $U(r)$  at large distances. The magnitude $U_l$ is
not an issue because, as we know, $\Pi_{pp}$ is logarithmically
singular at small frequencies. Furthermore, an attraction in just
one channel is a sufficient condition for a superconducting
instability, because if one of ${U}^{full}_l$ has a pole in the
upper half-plane of $\Omega$, the full vertex
$\Gamma^{full}_{\alpha\beta,\gamma\delta} (\theta)$ also has such
pole. The only difference with the case of a constant attractive
$U$ is that  when the instability occurs at some $l>0$ a
two-fermion bound pair has a non-zero angular momentum $l$.

\section{Kohn-Luttinger Mechanism}

Friedel oscillations at large distances occur by the special
reason -- the static particle-hole polarization bubble $\Pi_{ph}
(q)$ is non-analytic at $q =2k_F$.  For free fermions with
spherical FS, \bea\label{eq:ph bubble}
\Pi_{ph}(q,\Omega =0)&=&i\int\frac{d^3k d \omega}{(2\pi\hbar)^4} G(k,\omega)G(k+q,\omega)\nonumber\\
&=&\frac{mk_F}{2\pi^2\hbar^3}\left(
\frac{1}{2}+\frac{1-x^2}{4x}ln{\left|\frac{1+x}{1-x}\right|}
\right) \eea where $x\equiv \frac{q}{2k_F}$.  Near $q =2k_F$ (or
$x=1$),  $\Pi (x) \propto (1-x) \log{|1- x|}$, and its derivatives
over $x$ are singular at $x=1$. This $2k_F$ non-analyticity is a
universal property of a FL and it  survives even if one adds
self-energy and vertex corrections to the bubble. One can also
show quite generally that  the screening due to $2k_F$ scattering
acts on top of "conventional" screening which transforms Coulomb
potential into Yukawa-type short-range potential.  In this
respect, Friedel oscillations can be considered starting from
either bare Coulomb, or Yukawa, or even Hubbard interaction
potential.

Note in passing that the $2k_F$ non-analyticity is an example of
the special role played by  "hidden" 1D processes in a
multi-dimensional FL~\cite{masl_1d}.  Indeed, when $q = |{\bf k}_F
-{\bf p}_F|$ is  near $2k_F$,  ${\bf p}_F$ is antiparallel to
${\bf k}_F$.  One can make sure that the two internal fermions,
which contribute to $(1-x) \log{|1-x|}$ term in $\Pi_{ph} (x)$,
are also located near ${\bf k}_F$ and  $-{\bf k}_F$, i.e.,
everything comes from fermions moving in direction along or
opposite to ${\bf k}_F$

\begin{figure}[htp]
\includegraphics[width=4in]{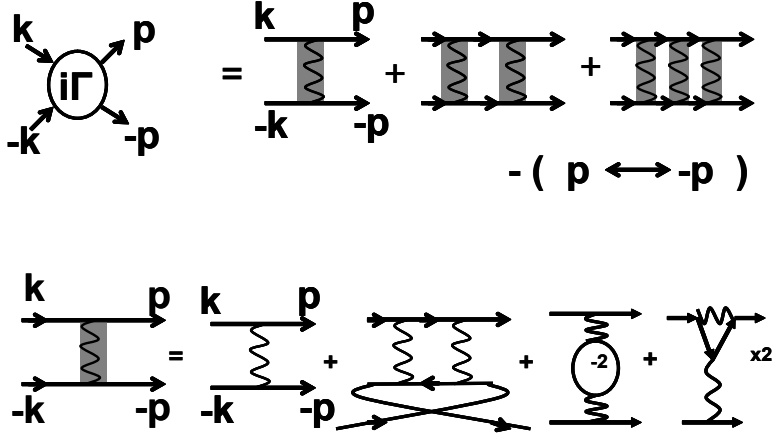}
\caption{\label{fig:KL}(Top) The fully renormalized vertex
function in the Cooper channel is the sum of the particle-particle
diagrams. The boxed wavy line is the irreducible pairing vertex
which is the sum of all diagrams with the structure different from
the Cooper channel. (Bottom) The expansion of the irreducible
vertex to second order in the interactions. The second order terms
are particle-hole channel contributions (Kohn-Luttinger diagrams)}
\end{figure}

The effect of $2k_F$ oscillations on superconductivity was first
considered by Kohn and Luttinger~\cite{kl,L}, and the result is
known as Kohn-Luttinger (KL) mechanism of superconductivity. The
idea of KL was the following: let's incorporate all non-singular
corrections to the interaction into new $U(\theta)$ and  treat it
as unknown, but  regular function of $\theta$.  A simple exercise
with Legendre polynomials shows that for any regular function of
$\theta$, partial components with angular momentum $l$ ($U_l$ in
our case) scale as $e^{-l}$, i.e., are exponentially small at
large $l$.  It is natural to assume that this bare interaction is
entirely repulsive, i.e., all $U_l >0$. If we substituted these
$U_l$ into (\ref{sa_1}), we would obviously not obtain any pairing
instability.  However, the input for the pairing problem is the
full irreducible anti-symmetrized vertex function ${\bar
\Gamma}^{0}$ in which incoming fermions have momenta $({\bf
k}_F,-{\bf k}_F)$ and outgoing fermions have momenta $({\bf
p}_F,-{\bf p}_F)$ (the word "irreducible" means that this vertex
function does not contain contributions with the particle-particle
bubble at zero total momentum). Such irreducible ${\bar
\Gamma}^{0}$ contains additional contributions from non-analytic
$2k_F$ scattering. KL computed $2k_F$ contribution to irreducible
${\bar \Gamma}^0 (\theta)$ to second order in the renormalized
$U(\theta)$. The corresponding diagrams are shown in Fig
\ref{fig:KL} The result is

\bea
&&{\bar \Gamma}^{0}_{\alpha\beta,\gamma\delta} (\theta) = \nonumber \\
&&- A (\theta) \left(\delta_{\alpha\gamma}\delta_{\beta\delta}  -
\delta_{\alpha\delta}\delta_{\beta\gamma}\right) -
B(\theta)  \left(\delta_{\alpha\gamma}\delta_{\beta\delta} +  \delta_{\alpha\delta}\delta_{\beta\gamma}\right) \nonumber \\
&& \label{fr_3_1} \eea where

\bea
&&A(\theta)  =  \frac{U(\theta)+ U(\pi -\theta)}{2}  - \left(2 U^2(\pi) - 2 U(0)U(\pi) -U^2(0)\right) \frac{\Pi_{ph} (\theta) + \Pi_{ph} (\pi-\theta)}{2}\nonumber \\
&& B(\theta) =  \frac{U(\theta)- U(\pi -\theta)}{2}  - \left(2
U^2(\pi) - 2 U(0)U(\pi) +U^2(0)\right) \frac{\Pi_{ph} (\theta)
-\Pi_{ph} (\pi-\theta)}{2} \label{sa_2} \nonumber\\\eea

\beq \Pi_{ph} (\theta) \approx -(mk_F/16\pi^2 \hbar^3) (1+ \cos
\theta) \log{(1+ \cos \theta)}, \eeq and in the factors $(\Pi_{ph}
(\theta) \pm \Pi_{ph} (\pi-\theta)$ in Eq. (\ref{sa_2}) one should
keep only the term $\Pi_{ph} (\theta)$ for $\theta$ close to $\pi$
and only the term $\Pi_{ph} (\pi-\theta)$ for $\theta$ close to
$0$.

Note that  $U(0)$ terms in the prefactors for $\Pi_{ph} (\theta)$
in $A(\theta)$ and $B(\theta)$  have different signs. This is the
consequence of the fact that these terms are of exchange type (two
crossed interaction lines), and in the internal parts of the
corresponding diagrams $\theta$ and $\pi-\theta$ are interchanged
compared to other terms.  Note also that for a constant $U$ the
exchange terms are the only ones which do not cancel out.

Now expand $A(\theta)$ and $B(\theta)$ into harmonics and consider
large $l$.   Like we said, regular terms coming from $U(\theta)$
are exponentially small. However, the terms of order $U^2$ are
non-analytic, and integrating them with Legendre polynomials one
finds that partial harmonics decay as $1/l^4$ rather than
exponentially.  Specifically, at large $l$, \beq
\label{eq:attractive1} S_l = -\int \frac{d\Omega_q}{4\pi}
\Pi_{ph}(\theta) P_l(\cos\theta) \approx  \frac{mk_F}{8\pi^2
\hbar^3} \frac{(-1)^{l}}{l^4} \eeq One can easily verify that
$A_l$ are again non-zero when $l=2m$ and $B_l$ are non-zero when
$l =2m+1$.  In explicit form we obtain \bea
&&A_{l=2m} =  U_{2m}  + \nonumber \\
&&\left(2 U^2(\pi) - 2 U(0)U(\pi) -U^2(0)\right) S_{2m} \nonumber \\
&&B_{l=2m+1} = U_{2m+1} + \nonumber \\
&&\left(2 U^2(\pi) - 2 U(0)U(\pi) + U^2(0)\right) S_{2m+1} \eea Al
large $l$, $U_l$ is exponentially small and can be neglected
compared to $S_l \propto 1/l^4$. Because $S_{2m+1}$ is negative
and $2 U^2(\pi) - 2 U(0)U(\pi) + U^2(0) = (U(2k_F) - U(0))^2 +
U^2(\pi)$ is positive for any form of $U(q)$, $B_{2m+1}$ are
definitely negative at large $m$.  As a result, an isotropic
system with initially repulsive interaction is still unstable
towards pairing, at least with large odd angular momentum of a
Cooper pair.   The harmonics with even $l$ are attractive when
$U(0)/U(\pi) > \sqrt{3}-1$.

The KL scenario for the pairing can be extended in several
directions.  First, one can consider the case of strong regular
screening, when the bare $U(q)$ can be approximated by a constant
(the Hubbard model). In this situation, without $2k_F$
renormalization, the bare interaction is repulsive in $l=0$
($s-$wave) channel, but zero in all other channels.  Once $2k_F$
renormalization is included, ${\bar \Gamma}^{0}$  acquires angular
dependence, and  both odd and even partial components become
attractive because for $U(0)=U(\pi)=U$, $A_{2m} = - U^2 S_{2m}$,
$B_{2m+1} = U^2 S_{2m+1}$, and $S_{2m} >0, S_{2m+1} <0$.   The
issue is: at which $l$ the coupling is most attractive? The
analysis of this last issue requires some caution because at $l =
O(1)$, all transferred momenta $q$, not only those near $2k_F$,
contribute to partial components of ${\bar \Gamma}^{0} (\theta)$.
One has to be careful here because some of regular contributions
from $q$ away from $2k_F$ may be already included into the
renormalization of the  Coulomb interaction into short-range,
Hubbard $U$.  If we just neglect this potential double counting,
i.e., assume that the screening from Coulomb interaction into a
Hubbard $U$ is produced by the processes different from the KL
ones, we can extend the KL analysis for a constant $U$ to
arbitrary $l$.  It then turns out that attraction survives down to
$l=1$, and the $l=1$ component is the strongest~\cite{Fay,kagan}.
In explicit form, \beq B_1 = -\frac{m k_F U^2}{2\pi^2 \hbar^3}~
\frac{(2\log{2}-1)}{5}. \label{04_30} \eeq Substituting this $B_1$
into the pairing channel, we obtain the pole in the triple
component of the full $\Gamma^{full} (\theta)$ at $\Omega = i
\Omega^{l=1}_p$, where \beq \Omega^{l=1}_p \propto e^{-B
\left(\frac{\hbar}{ak_F}\right)^2}, \eeq $B =
5\pi^2/(4(2\log{2}-1))$, and, as before, $a = mU/4\pi \hbar^2$ is
$s-$wave scattering length in Born approximation. The calculation
of the prefactor requires quite serious efforts as one needs to
include terms up to fourth order in the interaction (see
Ref.~\cite{kagan_prefactors}).

The $p-$wave pairing can be rationalized even when the bare
interaction is angle-dependent.  Because the momentum dependence
is via $q^2 = 2k^2_F (1-\cos\theta)$, higher angular harmonics of
$U(q)$ contain higher powers of $k_F$. In particular, $U_1 \sim U
(r_0 k_F/\hbar)^2$, where $r_0$ is the radius of the interaction.
Suppose it is repulsive. The p-wave component of the effective
irreducible interaction, which we obtained, is of order $U
(ak_F/\hbar)$ (see Eq. (\ref{04_30}).
 The ratio $a/r_0$ is the Born parameter.  When it is of order one,
$a$ and $r_0$ are of the same order, i.e., $ak_F \sim r_0 k_F$ are
small.  The induced attraction then wins because it contains a
smaller power of the small parameter~\cite{Fay}.  This reasoning,
however, works only when $a \sim r_0$. If we treat interaction as
small and $k_F$ is arbitrary, bare repulsion  is generally larger
than induced attraction, unless a bare $U(q)$ is a constant.

Before we move forward, let us make a quick remark about KL effect
in 2D systems. The  eigenfunctions of the angular momenta in 2D
are $P^{d=2}_l = \cos (l \theta)$ for $l \neq 0$ and
$P^{d=2}_0=1$. The  expansion of the irreducible interaction in
these eigenfunctions yields \beq\label{eq:spherical harmonis_1}
U(\theta)=U_0 + 2\sum_{l>0} U_l \cos{l \theta} \eeq The situation
in 2D is more tricky than in 3D because in 2D $\Pi_{ph}$ for free
fermions remains flat all the way up to $q = 2k_F$, i.e., for a
constant $U$, $U^2 \Pi_{pp} (q)$ does not depend on the angle
between incoming and outgoing fermions. Then harmonics with
non-zero $l$ do not appear, i.e., there is no KL effect. There is
a non-analyticity in $\Pi_{ph} (q)$ at $2k_F$ also in 2D, but it
is one-sided: at $q \geq 2k_F$, $\Pi_{ph} (q)$ behaves as
$\Pi_{ph} (q) =\Pi_{ph} (2k_F) -a \sqrt{q^2-4k^2_F}$ ($a>0$),
while at $q < 2k_F$, $\Pi_{ph} (q) = \Pi_{ph} (2k_F)$. However,
the non-analyticity at $q >2k_F$ is irrelevant for the pairing
problem because we need the interaction between fermions right on
the FS, and for them the largest momentum transfer is $2k_F$. The
situation changes when we move to next order in $U$ and include
vertex corrections to particle-hole bubble. These corrections make
$\Pi_{ph} (q)$ momentum-dependent also for $q <2k_F$, and, most
important, $2k_F$ non-analyticity becomes two-sided.   At large
$l$, partial harmonics of ${\bar \Gamma}^{0} (\theta)$ scale as
$1/l^2$ and, like in 3D, are attractive for both even and odd $l$,
if we set the bare pairing interaction, fully renormalized by
vacuum corrections,  to be a constant.  The largest interaction is
again in $l=1$ ($p-$wave) channel, and the pole in the
spin-triplet part of the full $\Gamma^{full} (\theta)$ is located
at $\Omega = i \Omega^{l=1}_{2D}$, where $\Omega^{l=1}_{2D}
\propto e^{-0.24/(ak_F/\hbar)^3}$.

Details and other discussion on the KL mechanism and its
application to $p-$wave superconductivity in systems with strong
ferromagnetic fluctuations can be found in
Refs.\cite{Fay,kagan,chubukov93,kagan_prefactors,KL3,Raghu,Raghu_1,KL4,Finn,Gr}.
The rest of these lecture notes will be devoted to discussion of
superconductivity in lattice models, where $k_F$ is generally not
small and rotational symmetry is broken.

\section{Superconductivity in lattice materials: Application to pnictides, cuprates and doped graphene}

\begin{figure*}[htp]
$\begin{array}{ccc}
\includegraphics[width = 2.2in]{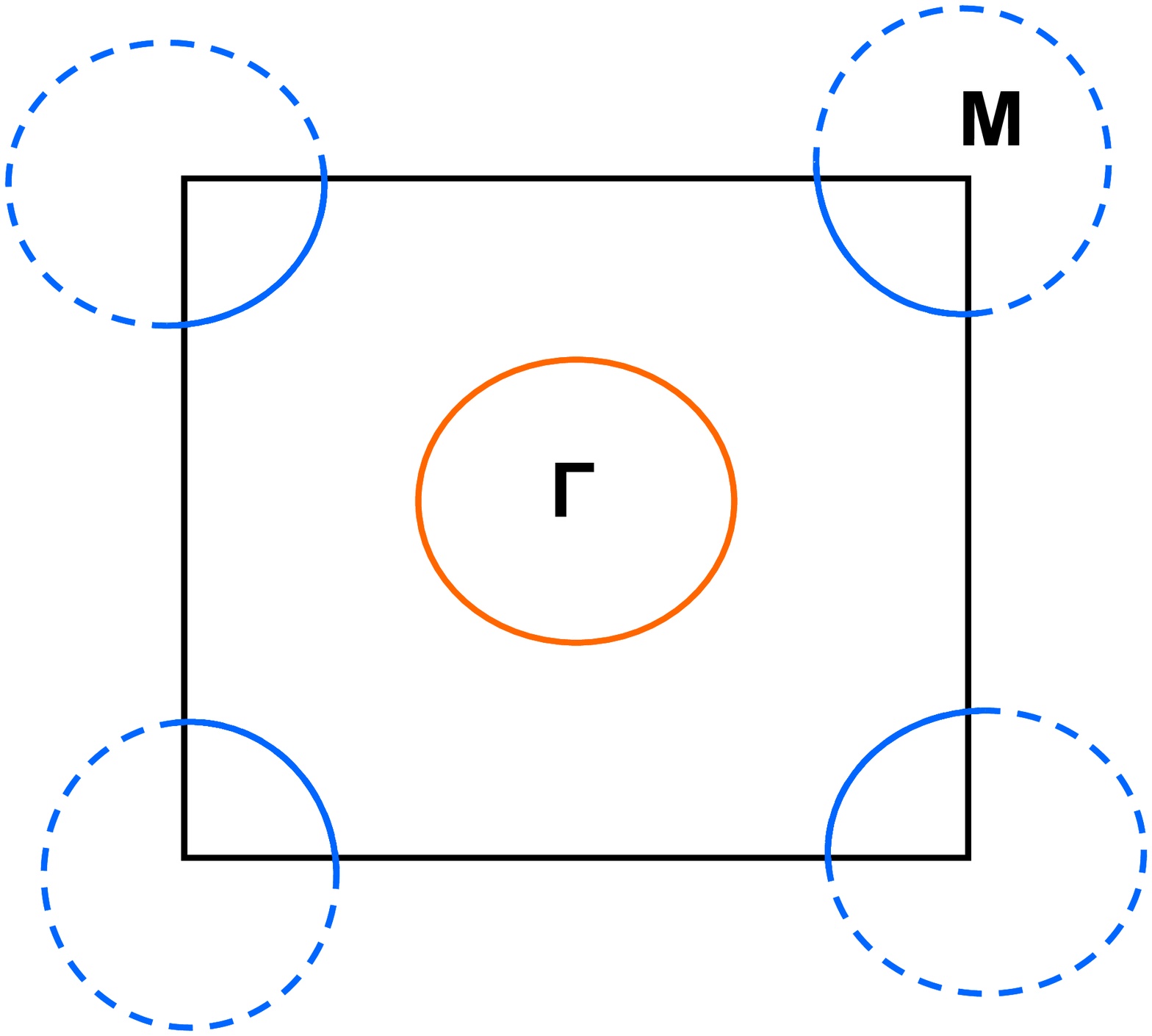}&
\includegraphics[width = 2.2in]{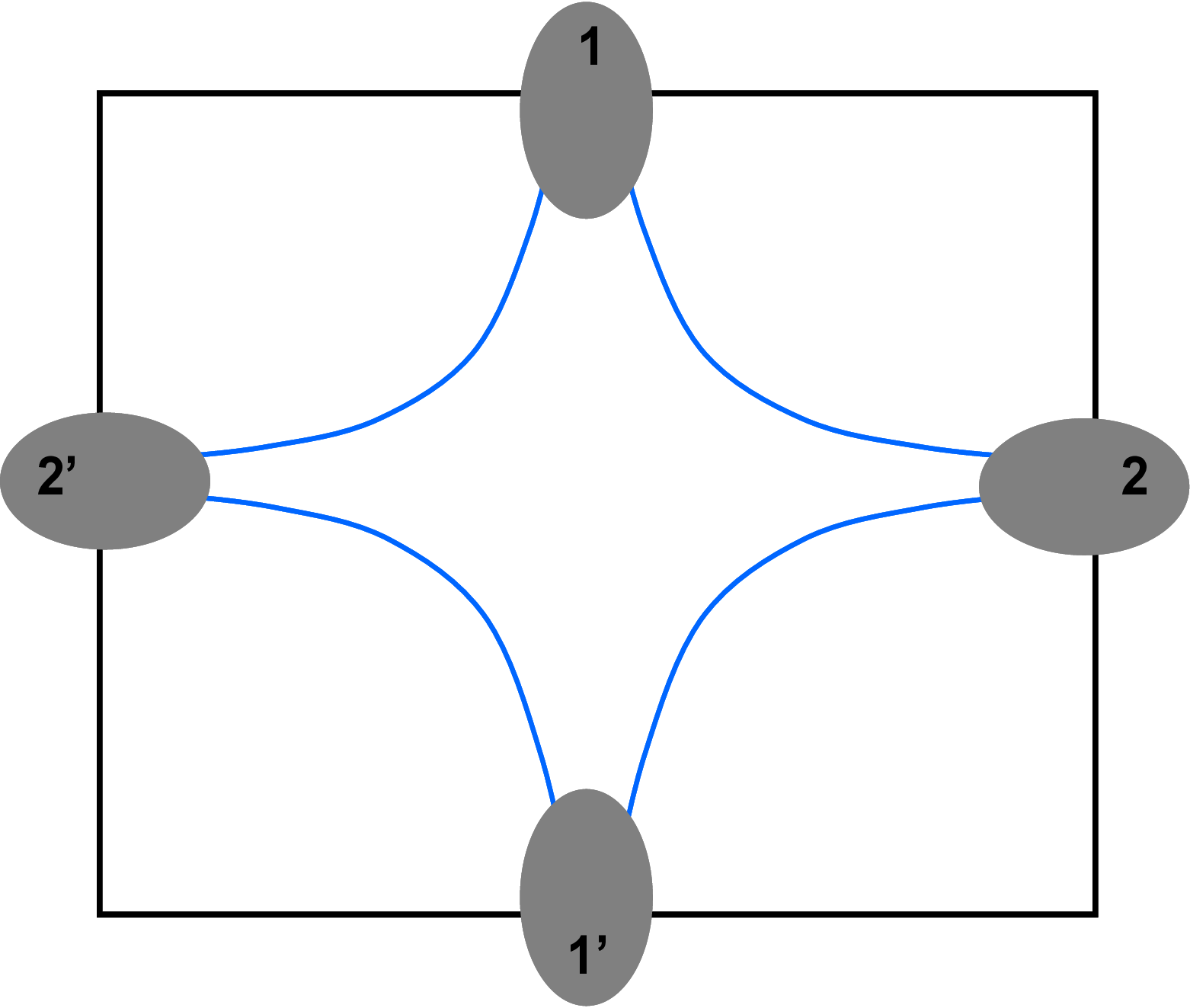}&
\includegraphics[width = 2.2in]{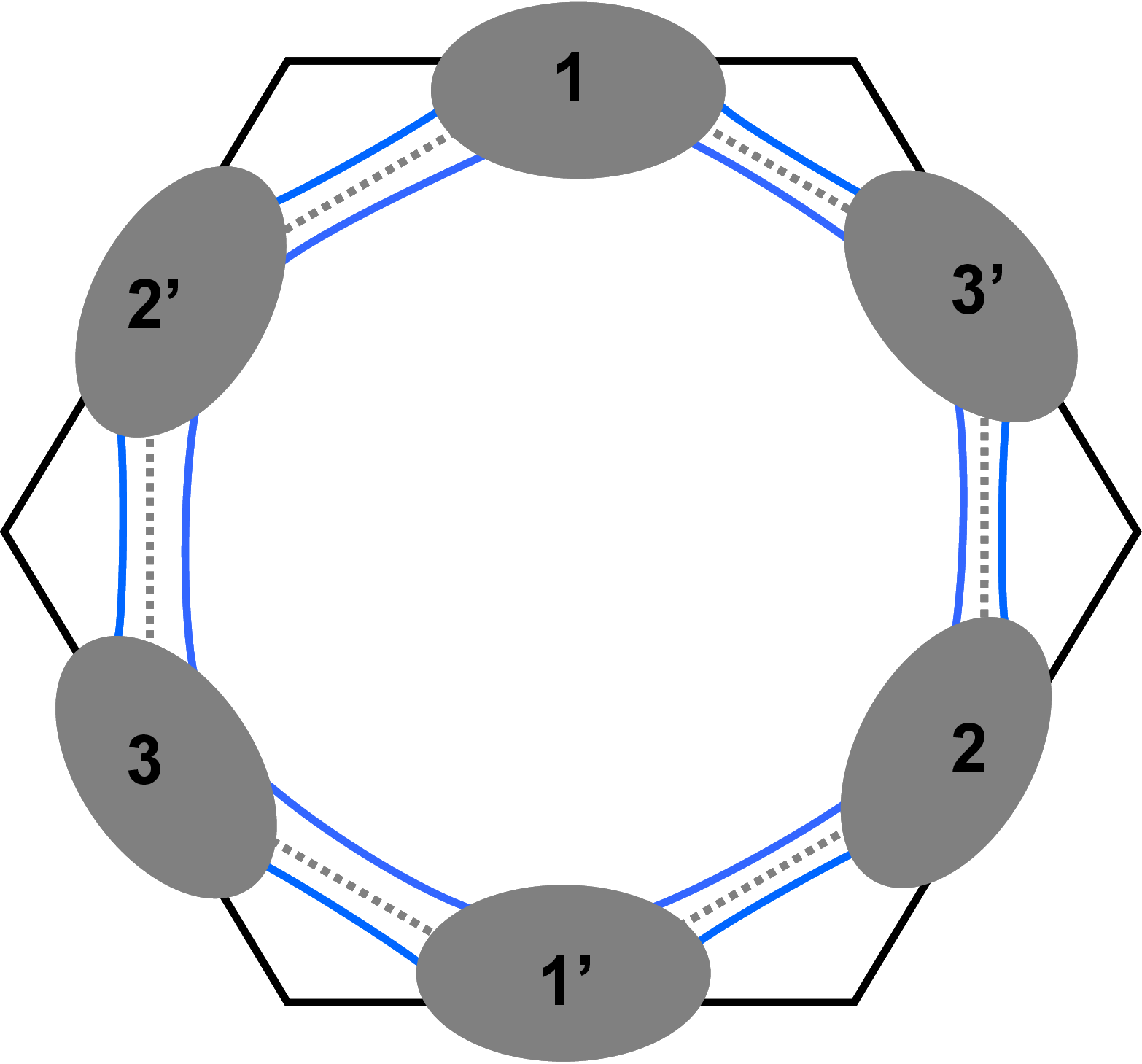}
\end{array}$
\caption{\label{fig:FS}FS topologies for a ``simplified''
pnictide(left), for hole-overdoped cuprate(center) and doped
graphene(right). In pnictides there are two kinds of FS's
electrons and holes(blue and orange circles, respectively). For
cuprates and graphene one can have disconnected pieces or a
singly-connected FS, depending on the doping. The doping at which
an open FS changes its character to a closed one is called the
Van-Hove doping. At this doping the density of states is
logarithmically singular near the saddle points. This points with
enhanced density of states are marked by grey patches.}
\end{figure*}

In studying superconductivity in solid-state systems one has to
deal with fermions moving on a lattice rather than in isotropic
media. Lattice systems have only discrete symmetries, and in
general FS does not have an isotropic form (spherical  in 3D or
circular in 2D) and may even be an open electron FS, meaning that
its does not form a closed object centered at $k=0$ and instead
ends at the boundaries of the Brillouin zone. (The locus of points
where energy is larger than $E_F$  is a closed object in this
situation,  and such a FS is often called a  closed hole FS).
Also, in many cases electronic structure is such that there are
several different FS's which can be either closed or open. We show
examples in Fig. \ref{fig:FS}.

We will consider systems with inversion symmetry and $SU(2)$ spin
symmetry. For such systems, the pairing instability is still
towards a creation of a bound pair of two fermions with momenta
${\bf k}$ and $-{\bf k}$ in either  spin-singlet or spin-triplet
channel.  However, if one attempts to expand the interaction into
eigenfunctions of momenta for isotropic systems (Legendre
polynomials $P_l (\theta)$ in 3D and $\cos (l\theta)$ in 2D), one
finds that different angular components no longer decouple.

One can still do a partial decoupling, however, due to discrete
symmetries of lattice systems. As an example, consider 2D fermions
on a square lattice.
 Fermionic dispersion and interactions are
invariant with respect to rotations by multiples of $\pi/2$ ($x
\to y, y \to -x$ and $x \to -x, y \to -y$),  with respect to
reflections across $x$ or $y$ axis ($x \to -x$ or $y \to -y$),
 and with respect to reflections across diagonals ($x \to y$ or $x \to
-y$). The corresponding symmetry group
 C$_{4v}$ contains  8 elements and has
  four one-dimensional  representations called
$A_{1}$, $A_{2}$, $B_{1}$, $B_{2}$ and one two-dimensional
representation $E$.  Eigenfunctions from $A_{1}$ remain invariant
under rotations and reflections, eigenfunctions from $B_{1}$
change sign under rotation by $\pi/2$ 2and under reflections
across diagonals, but invariant under rotation by $\pi$ and
reflection across $x$ or $y$, eigenfunctions from $B_{2}$ change
sign under rotations by $\pi/2$, and under reflections across $x$,
$y$, and one of diagonals, but remain invariant under rotation by
$\pi$, and under reflection across another diagonal, and so on. In
real 3D systems, interactions are also invariant with respect
 to $z \to -z$ inversion, and the symmetry group extends to
 $D_{4h}$, which includes 16 elements - 8 are even under
  to $z \to -z$ and 8 are odd ($g$ and $u$ subgroups).
  We will restrict our consideration to  spin-singlet
   superconductivity with pair wave-functions symmetric with respect to $z \to -z$.
   Accordingly, we stick with four one-dimensional representations
 $A_{1g}$, $A_{2g}$, $B_{1g}$ and $B_{2g}$. Each
 of these
 representations contains infinite number of eigenfunctions: $1$,
$\cos k_x + \cos k_y$, $\cos 2k_x + \cos 2k_y$, etc for $A_{1g}$,
$ \cos k_x - \cos k_y$,  $\cos 2 k_x - \cos 2 k_y$, etc for
$B_{1g}$, $\sin k_x \sin k_y$, $\sin 2 k_x \sin 2 k_y$, etc for
$B_{2g}$, and so on. (For convenience, for Brillouin Zone
variables we  measure $k$ in units of $\hbar/a$-where $a$ is the
lattice constant. This makes $k$ dimensionless). The basic
functions in each representation are summarized in Table
\ref{tab:rep}.

\begin{table}
\begin{tabular}{lrr}
\hline
A$_{1g}$&$s-$wave& $\cos k_x$+$\cos k_y$, $\cos k_x\cos k_y$\\
A$_{2g}$&$g-$wave&  $\sin k_x \sin k_y (\cos k_x$-$\cos k_y$)\\
B$_{1g}$&$d-$wave& $\cos k_x$-$\cos k_y$\\
B$_{2g}$&$d-$wave& $\sin k_x \sin k_y$\\
\hline
\end{tabular}
\caption{Basic functions in 1D  representations of the
square-lattice group $D_{4h}$} \label{tab:rep}
\end{table}

If we try to group eigenfunctions of momenta from the isotropic
case into these representations, we find that eigenfunctions with
$l = 4n$ belong to $A_{1g}$, eigenfunctions with $l = 4n+1$ belong
to $E$, eigenfunctions with  $l = 4n+2$ belong to $B_{1g}$ or
$B_{2g}$,  and eigenfunctions with $l = 4n+3$ belong to $A_{2g}$.
Because of this decomposition, $A_{1g}$ representation is often
called $s-$wave, $E$ is called p-wave, $B_{1g}$ and $B_{2g}$ are
called $d$-wave ($d_{x^2-y^2}$  and $d_{xy}$, respectively), and
$A_{2g}$ is called g-wave. We will use these notations below.

Now, if we now expand the interactions in eigenfunctions of
$D_{4h}$ group and consider the pairing problem in the same way as
we did before, we find that functions belonging to different
representations decouple, but infinite set of functions within a
given representation remain coupled. In this situation, the KL
result for the isotropic case that the system will eventually be
unstable against pairing with some angular momentum, is no longer
valid because large $l$ components from any given representation
mix with smaller $l$ components from the same representation, and
the latter can be repulsive and larger in magnitude.  Indeed, we
will see below that for lattice systems, there is no guarantee
that the pairing will occur,  i.e., a non-superconducting state
well may survive down to $T=0$.  We refer a reader to several
papers in which superconductivity has either been ruled out at
large $U$~\cite{new} or  found to be present using the
controllable approximation at any $U$ (Ref. \cite{new_1}).

At the same time, we will see that another part of KL-type
analysis can be straightforwardly extended from isotropic to
lattice systems. Namely, if we approximate the bare interaction
$U(q)$ by a constant $U >0$, we get a repulsive interaction in
$s-$wave channel, but nothing in $p-$wave, $d-$wave, and $g$-wave
channel.  Once we include KL contribution to order $U^2$, we do
get interaction in these channels. We recall that in the isotropic
case, the induced interaction in all non-s-wave channels is
attractive. We show that the same happens in lattice systems, at
least in the examples we consider below.

And there is more:  even within $s-$wave channel, the full
$\Gamma^{full} ({\bf k}, {-\bf k}; {\bf p}, -{\bf p})$ is the
solution of the coupled set of equations for infinite number of
$A_{1g}$ eigenfunctions.  Diagonalizing the set  one obtains
infinite number of coupling constants (effective interactions).
For a constant $U$, some eigenfunction are positive (repulsive),
but some are  zero.  This can be easily understood by looking at
the first two wavefunctions: $1$  and $\cos k_x + \cos k_y$. The
first is invariant under the shift ${\bf k} \to {\bf k} +
(\pi,\pi)$, while the second changes sign under this
transforation. A constant $U$ cannot create a pairing wavefunction
which changes sign under ${\bf k} \to {\bf k} + (\pi,\pi)$, hence
the bare coupling for such a state is zero. KL terms produce
momentum dependence of the irreducible interaction in the pairing
channel and shift the eigenvalue for the sign-changing
wavefunction. If this eigenvalue is attractive, then KL physics
gives rise to an s-wave attractive interaction, which may be even
stronger than KL-induced attraction in other channels.

We will discuss KL pairing in three representative families of
materials: Fe-pnictides, cuprates, and doped graphene.
Superconductivity in cuprates and Fe-pnictides (and
Fe-chalcogenides) has been detected in numerous experiments.
Superconductivity in doped graphene has been predicted
theoretically but so far not detected experimentally. Although
historically, cuprates were discovered first in $1986$, for
pedagogical reasons it is convenient to start with Fe-pnictides,
where we show that superconductivity is due to KL-induced
attraction in $A_{1g}$ channel.  We then discuss cuprates
and show that KL renormalization of the pairing interaction gives
rise to attraction in $B_{1g}$
channel. Finally, we consider graphene doped to van-Hove density
(or, equivalently, fermions on a triangular lattice at van-Hove
doping) and show that KL mechanism gives rise to a
doubly-degenerate pairing state, whose components  can be viewed
as $B{1g}$ and $B{2g}$ using square-lattice representations (or
$E_{2g}$ using representations for a hexagonal lattice).

There is extensive literature on all three classes of systems, and
superconductivity is one of many interesting and still puzzling
properties of these materials. Some researchers believe that in
either all or some of these systems superconductivity is
ultimately related to Mott physics\cite{Lee}, and some believe
that superconductivity may be mediated by
phonons\cite{Kontani,Dev,Shen,efetov2,Sokolik}. We will not dwell
into these issues and simply discuss the conditions and
consequences of the electronic mechanism of superconductivity in
these materials for the portions of the phase diagrams where
electronic correlations are not strong enough to localize the
electrons.  The goal of these lectures is to discuss how much
information about pairing one can extract from the analysis of the
KL scenario. Our key conclusion is that the pairing in all three
classes of materials can be traced to the same KL physics, which,
however, predicts different pairing symmetries in each class of
materials.

\subsection{Superconductivity in Fe-Pnictides}

Fe-pnictides are binary compounds of pnictogens, which are the
elements from the 5th group: N, P, As, Sb, Bi.  Superconductivity
in these materials has been discovered in  2008 by Hosono and his
collaborators~\cite{bib:Hosono}. Later, superconductivity has been
found also in Fe-chalcogenides -- Fe-based compounds with elements
from the 16th group: S, Se, Te \cite{hsu,FeChal1,FeChal2,FeChal3}.

The family of $Fe$-based superconductors (FeSCs) is already quite
large and keeps growing. It includes $1111$ systems RFeAsO
($R=$rare earth element) \cite{bib:Hosono,bib:X Chen,bib:G
Chen,bib:ZA Ren}, $122$ systems XFe$_2$As$_2$(B=Ba,Na, K)
\cite{bib:Rotter,bib:Sasmal,bib:Ni,KFeAs_ARPES_QO,KFeAs_exp_nodal}
and AFe$_2$Se$_2$ (A = K, Rb,
Cs)~~\cite{exp:AFESE,exp:AFESE_ARPES},  111 systems like LiFeAs
\cite{bib:Wang}, and 11 systems, like FeTe$_{1-x}$Se$_x$
~~\cite{bib:Mizuguchi}.

Parent compounds of most of FeSCs are metallic
antiferromagnets~\cite{cruz}. Because  electrons, which carry
magnetic moments, can travel relatively freely from site to site,
antiferromagnetic order is often termed as a ``spin-density-wave''
(SDW), by analogy with e.g., magnetism in $Cr$, rather than
``Heisenberg antiferromagnetism" -- the latter term is reserved
for systems in which electrons are ``nailed down" to particular
lattice sites by very strong Coulomb repulsion.

\begin{figure}[tbp]
\includegraphics[angle=0,width=\linewidth]{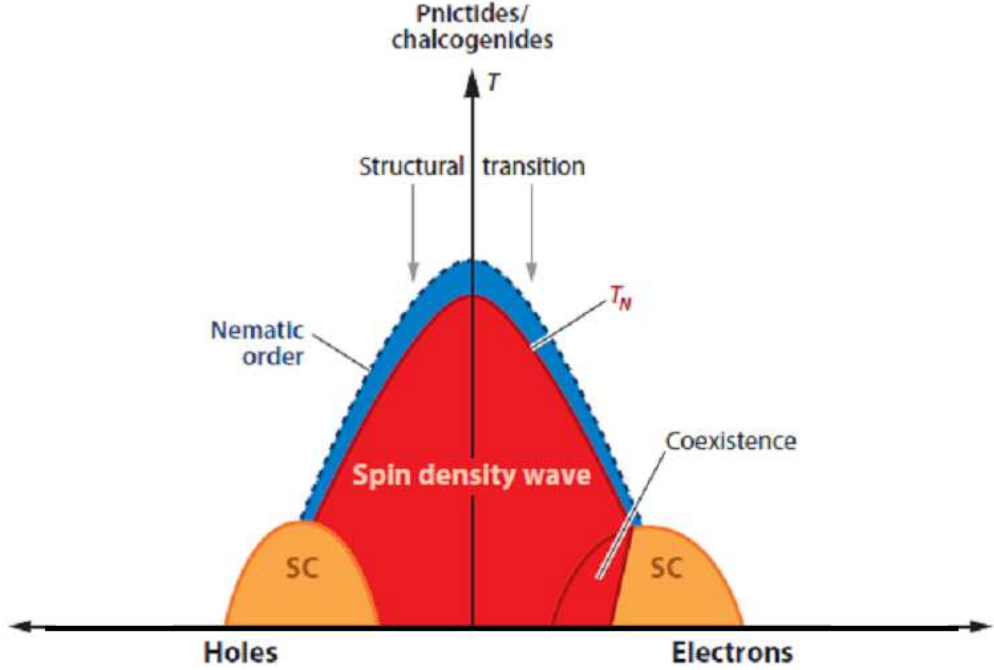}
\caption{Schematic phase diagram of Fe-based pnictides upon hole
or electron doping.  In the shaded region, superconductivity and
antiferromagnetism co-exist. Not all details/phases are shown.
Superconductivity can be initiated not only by doping but also by
pressure and/or isovalent replacement of one pnictide element by
another~~\cite{nakai}. Nematic phase at $T
> T_N$ is another interesting subject but we don't discuss this in the text.
Taken from Ref. \cite{review_we}.} \label{fig:fig1}
\end{figure}

Superconductivity in Fe-pnictides emerges upon either hole or
electron doping  (see Fig. \ref{fig:fig1}), but can also be
induced  by pressure or by isovalent replacement of one pnictide
element by another, e.g., As by P (Ref.~\cite{nakai}). In some
systems, like LiFeAs~~\cite{bib:Wang} and
LaFePO~~\cite{bib:Kamihara}, superconductivity emerges already at
zero doping, instead of a magnetic order.

magnetism, the electronic structure, the normal state properties
of  FeSCs, and the interplay between FeSCs and cuprate
superconductors have been reviewed in several recent
publications~~
\cite{review,review_2,review_3,review_4,Graser,peter,Kuroki_2,rev_physica,mazin_schmalian,review_5,review_we,NormanPhysNews,Xu2008,rev_annual}.
Below we shall not dwell into the intricacies of the phase diagram
but only focus on the superconductivity.

The electronic structure of FeSCs is fairly complex with multiple
FS's surfaces extracted from ARPES and quantum oscillations
measurements. In most systems, there are two or three
near-cylindrical hole FS's centered at $k_x=k_y=0$ and two
electron FS's centered at $(\pi,\pi)$.   For electron pockets,
states inside the pockets are occupied, for hole pockets, states
inside the pockets are empty.

This electronic structure agrees with the one obtained
theoretically from the ten-orbital model, which includes five Fe
d-orbitals and takes into account the fact that an elementary unit
cell contains two Fe-atoms because As atoms are located above and
below an Fe plane. All d-orbitals hybridize, and to convert to
band description one has to diagonalize the Hamiltonian in the
orbital basis. The diagonalized quadratic Hamiltonian $H_2 =
\sum_{i=1}^{10} \epsilon_{i,k} a^\dagger_{i,k} a_{i,k}$ describes
ten fermionic bands, some of which cross chemical potential and
give rise to hole and electron pockets.  The interactions between
these band fermions are the original interactions in the orbital
basis, dressed up by the "coherence factors" associated with the
transformation from orbital to band fermions (the coherence
factors are the coefficients in the linear transformation from
original fermions describing d-orbitals to new fermions which
diagonalize the quadratic Hamiltonian). Interactions in the
orbital basis are local, to a reasonably good accuracy, but the
coherence factors know about fermion hopping from site to site and
depend on momenta. As the consequence, the interactions between
band fermions acquire momentum dependence, which leads to several
new and interesting phenomena like the appearance of accidental
zeros in the two-particle bound state wave function
\cite{AccNodes}.

For  proof-of-concept we consider a simpler problem: a 2D
two-pocket model with one hole and one electron FS, both circular
and of equal sizes (see Fig.\ref{fig:FS}), and approximately
momentum-independent.

The free-fermion Hamiltonian is the sum of kinetic energies of
holes and electrons: \beq H_2 = \sum_{k,\sigma} \e_c
c^{\dag}_{k,\sigma}c_{k,\sigma} +\e_f
f^{\dag}_{k,\sigma}f_{k,\sigma} \eeq where $c$ stands for holes,
$f$ stands for electrons, and $\e_{c,f}$ stand for their
respective dispersions with the property the $\e_c(k)=-\e_f(k+Q)$,
where $\textbf{Q}$ is the momentum vector which connects the
centers of the two fermi surfaces. The  density of states $N_0$ is
the same on both pockets, and the electron pocket `nests'
perfectly within the hole pocket when shifted by $\textbf{Q}$.

\begin{figure*}[htp]
$\begin{array}{ccc}
\includegraphics[width = 2.2in]{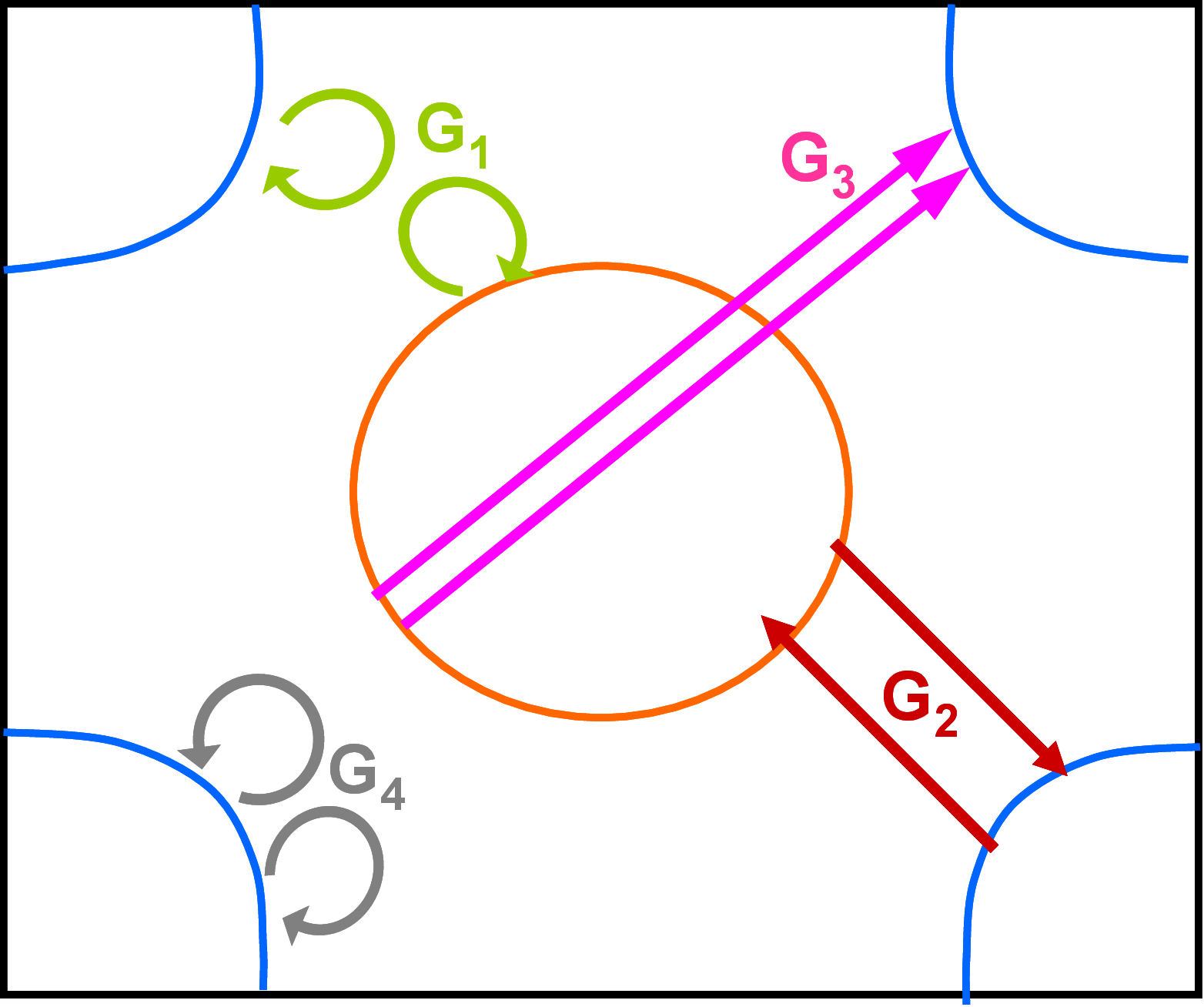}&
\includegraphics[width = 2.2in]{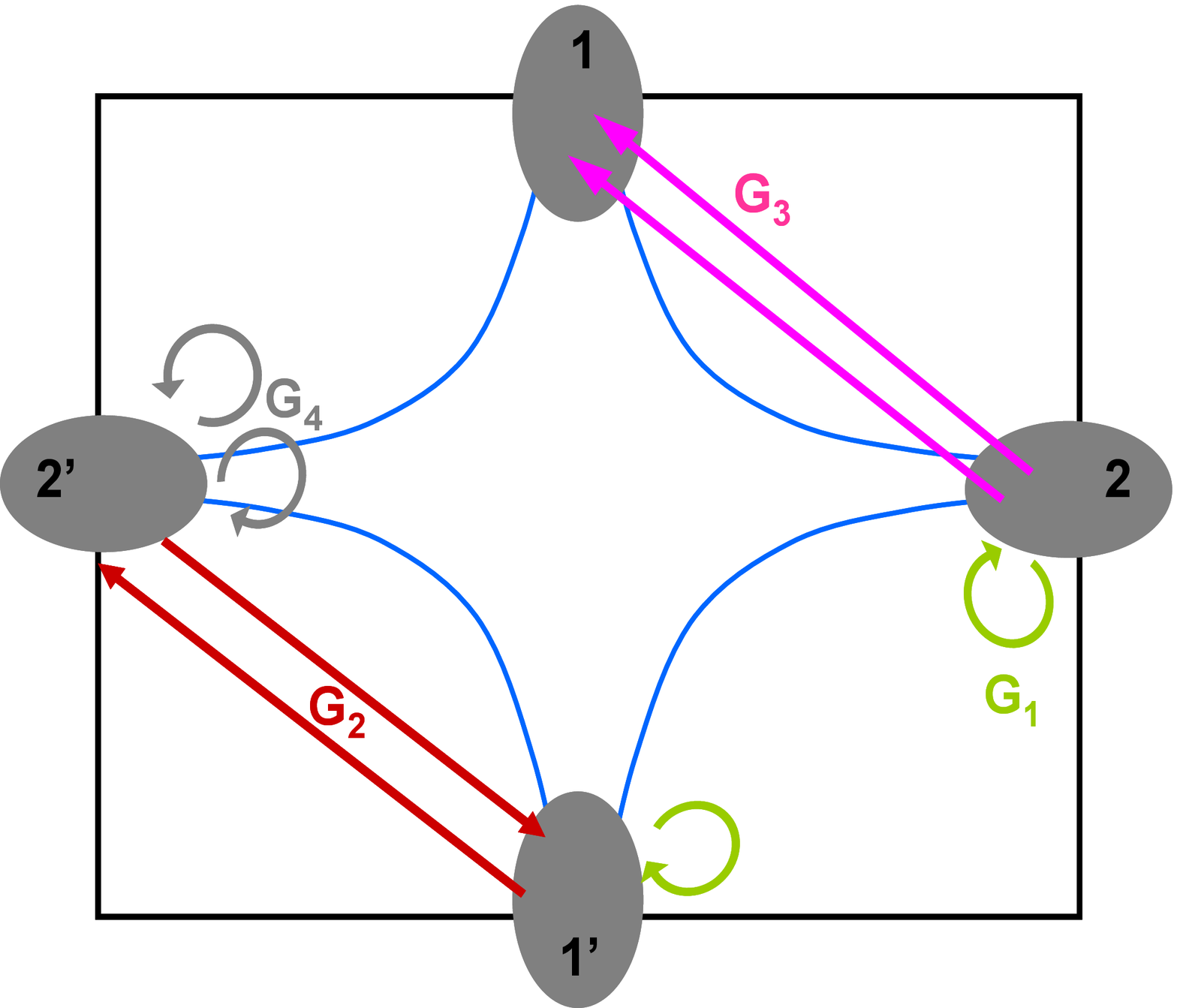}&
\includegraphics[width = 2.2in]{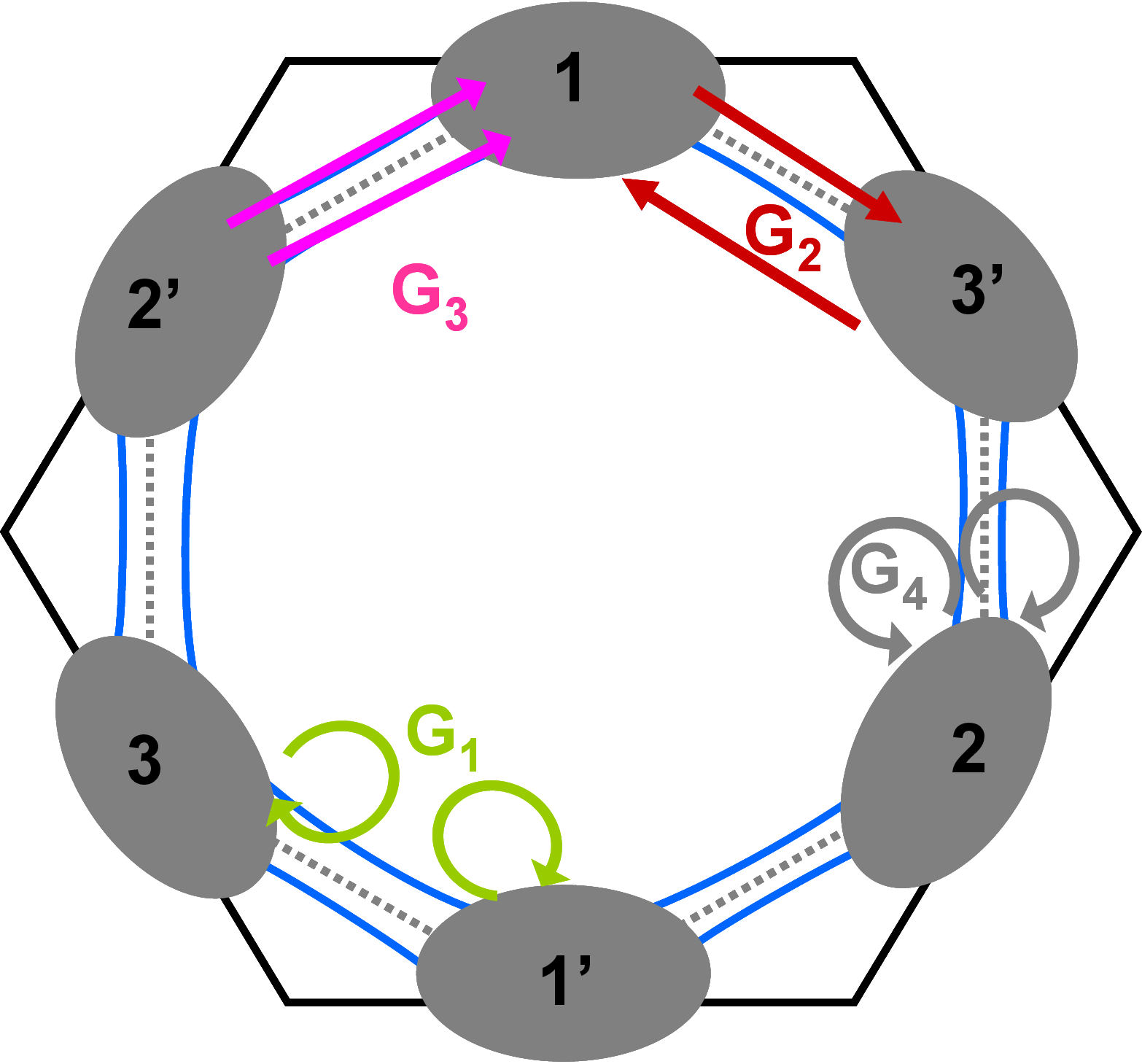}
\end{array}$
\caption{\label{fig:int}The interactions between patches/pockets
in the pnictides(left), cuprates (center) and graphene (right).
$G_1$ is a density-density interaction between fermions from
different pockets/patches. $G_2$ is an exchange between the
pockets/patches, $G_3$ is a pair hopping process between the
pockets/patches, and $G_4$ is a density-density interaction within
the same pocket/patch. All interactions are repulsive (positive).}
\end{figure*}

There are five different types of interactions between low-energy
fermions: two intra-pocket density-density interactions, which we
treat as equal, interaction between densities in different
pockets, exchange interaction between pockets, and pair hopping
term, in which two fermions from one pocket transform into two
fermions from the other pocket. We show these interactions
graphically in Fig \ref{fig:int}.

In explicit form \bea\label{eq:2-poc Ham} H_{\text{int}}&=& G_1
\sum_{[k,\sigma]}  c^{\dagger}_{k_1 \sigma} f^{\dagger}_{k_2
\sigma'}  f_{k_3 \sigma'} c_{k_4 \sigma}
\nonumber\\
&&+G_2 \sum_{[k,\sigma]} f^{\dagger}_{k_1 \sigma} c^{\dagger}_{k_2
\sigma'} f_{k_3 \sigma'} c_{k_4 \sigma}
\nonumber\\
&&+ \sum_{[k,\sigma]}\frac{G_3}{2}
\left(c^{\dag}_{k_1,\sigma_1}c^{\dag}_{k_2,\sigma_2}f_{k_3,\sigma_2}f_{k_4,\sigma_1}
+
\text{h.c}\right)\nonumber\\
&&+\sum_{[k,\sigma]}\left(\frac{G_4}{2}
c^{\dag}_{k_1,\sigma_1}c^{\dag}_{k_2,\sigma_2}c_{k_3,\sigma_2}c_{k_4,\sigma_1}
+ c\leftrightarrow f\right) \nonumber\\ \eea where
$\sum{_{[k,\sigma]}}$ is short for the sum over the spins and the
sum over all the momenta constrained to $k_1+k_2=k_3+k_4$ modulo a
reciprocal lattice vector.

\begin{figure}[tbp]
\includegraphics[width=2.8in]{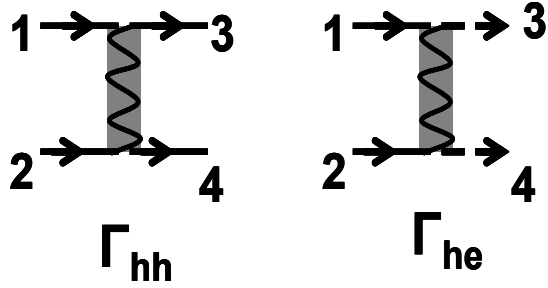}
\caption{Vertices $\Gamma_{hh}=\Gamma_{ee}$ and $\Gamma_{he}$
introduced in the 2 pocket model. } \label{fig:new_vertices}
\end{figure}

As we did for isotropic systems, consider the vertex function for
fermions on the FS, for zero total incoming momentum. Because
there are two pockets, there are three relevant vertices:
$\Gamma_{hh} ({\bf k}_F,-{\bf k}_F, {\bf p}_F,-{\bf p}_F);
\Gamma_{ee} ({\bf k}_F,-{\bf k}_F, {\bf p}_F,-{\bf p}_F)$, where
${\bf k}_F$ and ${\bf p}_F$ belong to the same pocket, and
$\Gamma_{he} ({\bf k}_F,-{\bf k}_F, {\bf p}_F,-{\bf p}_F)$, where
${\bf k}_F$ and ${\bf p}_F$ belong to different pockets (see Fig.
\ref{fig:new_vertices}). To first order in $G_i$, we have \bea
&&\Gamma^{0}_{hh}  ({\bf k}_F,-{\bf k}_F, {\bf p}_F,-{\bf p}_F) = -G_4 \nonumber \\
&&\Gamma^{0}_{ee}  ({\bf k}_F,-{\bf k}_F, {\bf p}_F,-{\bf p}_F) = -G_4 \nonumber \\
&&\Gamma^{0}_{he}  ({\bf k}_F,-{\bf k}_F, {\bf p}_F,-{\bf p}_F) =
-G_3 \eea where the spin dependence for both terms is
$\delta_{\alpha\gamma}\delta_{\beta\delta}  -
\delta_{\alpha\delta}\delta_{\beta\gamma}$. Let's now solve for
the full $G$, restricting with the renormalizations in the pairing
channel (i.e., with only Cooper logarithms). A simple analysis
shows that the full $\Gamma$ is given by \bea \Gamma^{full}_{hh}
&=& - \frac12\left(\frac{G_4 + G_3}{1 + (G_4 + G_3)\Pi_{pp}} +
\frac{G_4 - G_3}{1 + (G_4 - G_3)
\Pi_{pp}}\right)\nonumber\\
\Gamma^{full}_{ee}&=&\Gamma^{full}_{hh}\nonumber\\
\Gamma^{full}_{he} &=& - \frac12\left(\frac{G_4 + G_3}{1 + (G_4 +
G_3)\Pi_{pp}} - \frac{G_4 - G_3}{1 + (G_4 - G_3)
\Pi_{pp}}\right)\nonumber\\ \label{4_2_1} \eea

and $\Pi_{pp} = \Pi_{pp} (q,\Omega)$  has the same logarithmical
form as before. For $q=0$, $\Pi_{pp} (0,\Omega) = N_0
(\log{|\omega_c/\Omega|} +i\pi/2)$, where $N_0$ is the density of
states  at the FS (and is the same on both electron and hole
pockets)

We see that the presence or absence of a pole in $\Gamma^{full}$,
depends on the signs of $G_3 + G_4$ or $G_4-G_3$. If both are
positive, there are no poles, i.e., non-superconducting state is
stable. In this situation, at small $\Omega$, $\Gamma^{full}_{hh}
\approx -1/\Pi_{pp}$, $\Gamma^{full}_{he} \approx
-(G_3/(G^2_4-G^2_3))\Pi^2_{pp}$, i.e., both vertex functions
decrease (inter-pocket vertex decreases faster). If one (or both)
combinations are negative, there are poles in the upper frequency
half-plane and fermionic system is unstable against pairing.  The
condition for the instability is $|G_3| > G_4$. $G_4$ is
inter-pocket interaction, and there are little doubts that it is
repulsive, even if to get it one has to transform from orbital to
band basis.  $G_3$ is interaction at large momentum transfer, and,
in principle, it can be either positive or negative depending on
the interplay between intra- and inter-orbital interactions.   In
most microscopic multi-orbital calculations, $G_3$ turns out to be
positive, and we set $G_3>0$ in our analysis (for the case $G_3<0$
see Ref. \cite{Kontani}).

For positive $G_3$,  the condition for the pairing instability is
$G_3 > G_4$. What kind of a pairing state we get?  First, both
$\Gamma^{full}_{hh}$ and $\Gamma^{full}_{he}$ do not depend on the
direction along each of the two pockets, hence the pairing state
is necessary $s-$wave. On the other hand, the pole is in
$\Gamma_2$, which appears with opposite sign in
$\Gamma^{full}_{hh}$ and $\Gamma^{full}_{he}$. The pole components
of the two vertex functions then also differ in sign, which
implies that the two-fermion pair wave function changes sign
between pockets. Such an $s-$wave state  is often call $s^{+-}$ to
emphasize that the pair wave function changes sign between FSs.
This wave function  much resembles the second wave function from
$A_{1g}$ representation: $\cos k_x + \cos k_y$. It is still
$s-$wave, but it changes sign under ${\bf k} \to {\bf k} +
(\pi,\pi)$, which is precisely what is needed as hole and electron
FSs are separated by $(\pi,\pi)$.  We caution, however, that the
analogy should not be taken too far because the pairing wave
function is defined {\it only} on the two FSs, and any function
from $A_{1g}$ representation which changes sign under ${\bf k} \to
{\bf k} + (\pi,\pi)$ would work equally well.

Having established the pairing symmetry, we now turn to the
central issue: how to get an attraction.  Like we did in the
isotropic case, let's start with the model with a
momentum-independent  (Hubbard) interaction in band basis.  For
such interaction, all $G_i$ are equal, and, in particular, $G_3 =
G_4$. Then $\Gamma_2$ just vanishes, i.e., at the first glance,
there is no pole. However, from KL analysis for the isotropic
case, we know that do decide whether or not there is an attraction
in some channel, we need to analyze the full irreducible vertex
function. To first order in $G_i$, the irreducible vertex function
coincides with the (anti-symmetrized) interaction, but to order
$G^2_i$, there appear additional terms. Let's see how they look
like in the two pocket model.

\begin{figure}[htp]
\includegraphics[width=4in]{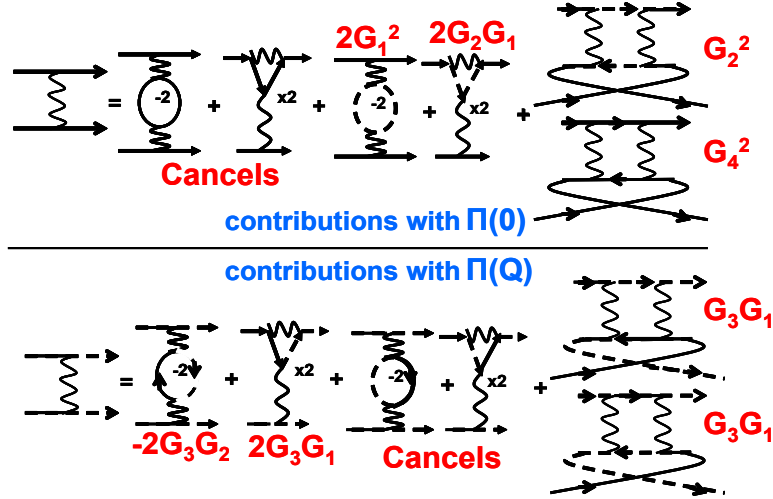}
\caption{\label{fig:Pi0} Contributions to the irreducible vertices
$\bar\Gamma_{hh}^0$(top) and $\bar\Gamma_{he}$(bottom).
$\bar\Gamma_{hh}^0$ only gets contributions form $\Pi(0)$ while
$\bar\Gamma_{he}^0$ gets contribution from $\Pi(Q)$.}
\end{figure}

\begin{figure}[htp]
\includegraphics[width=4in]{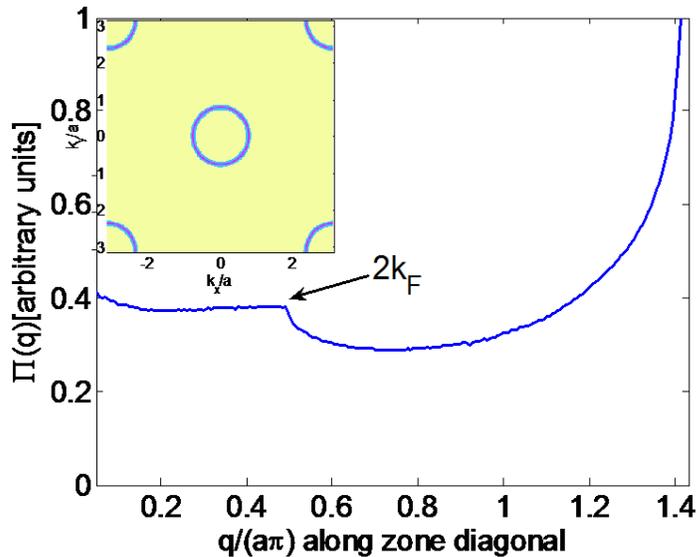}
\caption{\label{fig:PiQ}The plot of $\Pi(q)$ for a 2-pocket model
with $\vec{q}$ along the zone diagonal. When $\vec{q} <2k_F$,
$\Pi(q)$ saturates, as it is expected for a 2D system with a
circular Fermi surface. Note the $2k_F$ cusp-like feature, which
is the one-sided $2k_F$ non-analiticity of $\Pi(q)$ in 2D. At
larger $q$, $\Pi(q)$ gets larger and almost diverges at
$\vec{q}\sim \vec{Q}$ due to near-nesting. The inset shows the FS
topology for which $\Pi(q)$ has been calculated. The arcs at the
corners are parts of the electron pocket and the one in the center
is the hole pocket.}
\end{figure}

The contributions to irreducible ${\bar \Gamma}^{0}_{hh}$ and
${\bar \Gamma}^{0}_{he}$ are shown in Fig \ref{fig:Pi0}. In
analytical form we have (keeping the notations $G_i$ for better
clarity) \bea
{\bar \Gamma}^{0}_{hh} &=& - G_4  - \left(G^2_4 + G^2_2- 2 G_1 (G_1-G_2) \right)\Pi_{ph} (0), \nonumber \\
{\bar \Gamma}^{0}_{he} &=& - G_3 - 2G_3(2 G_1 - G_2) \Pi_{ph} (Q),
\eea where ${\bf Q} = (\pi,\pi)$. For a constant $G$, this reduces
to \bea
{\bar \Gamma}^{0}_{hh} &=& - G\left(1 +  2 G\Pi_{ph} (0)\right), \nonumber \\
{\bar \Gamma}^{0}_{he} &=& - G\left(1 + 2 G \Pi_{ph} (Q) \right),
\eea The relation (\ref{4_2_1}) still holds when we replace  $G_3$
by $- {\bar \Gamma}^{0}_{he}$ and $G_4$ by $- {\bar
\Gamma}^{0}_{hh}$. It can be very easily shown that
$\Gamma^{full}_{ee}=\Gamma^{full}_{hh}$ and thus we will only deal
with $\Gamma^{full}_{hh}$ and $\Gamma^{full}_{he}$ which are given
by \bea
\Gamma^{full}_{hh} &=& \frac{1}{2} \left(\frac{ {\bar \Gamma}^{0}_{he} + {\bar \Gamma}^{0}_{hh}}{1 -({\bar \Gamma}^{0}_{he} + {\bar \Gamma}^{0}_{hh}) \Pi_{pp}} + \frac{{\bar \Gamma}^{0}_{hh} - {\bar \Gamma}^{0}_{he}}{1  - ({\bar \Gamma}^{0}_{hh}-{\bar \Gamma}^{0}_{he}) \Pi_{pp}}\right),\nonumber \\
\Gamma^{full}_{he} &=& \frac{1}{2} \left(\frac{ {\bar \Gamma}^{0}_{he} + {\bar \Gamma}^{0}_{hh}}{1 -({\bar \Gamma}^{0}_{he} + {\bar \Gamma}^{0}_{hh}) \Pi_{pp}} - \frac{{\bar \Gamma}^{0}_{hh} - {\bar \Gamma}^{0}_{he}}{1  - ({\bar \Gamma}^{0}_{hh}-{\bar \Gamma}^{0}_{he}) \Pi_{pp}}\right),\nonumber\\
&& \label{4_2_1_1} \eea and the condition for the pairing
instability becomes $ {\bar \Gamma}^{0}_{hh} >|{\bar
\Gamma}^{0}_{he}|$. Comparing the two irreducible vertex
functions,  we find \bea {\bar \Gamma}^{0}_{hh} - {\bar
\Gamma}^{0}_{he} = 2 G^2 \left(\Pi_{ph} (Q) - \Pi_{ph} (0)\right)
\label{4_3_1} \eea i.e., the condition for the pairing is
satisfied when $\Pi_{ph} (Q) > \Pi_{ph} (0)$.  For a gas of
fermions with one circular FS, $\Pi_{ph} (q)$ either stays
constant or decreases  with $q$, and the condition $\Pi_{ph} (Q) >
\Pi_{ph} (0)$ cannot be satisfied. However, in our case, there are
two FS's separated by ${\bf Q}$, and, moreover, one FS is of hole
type, while the other is of electron type.  One can easily verify
that, in this situation, $\Pi_{ph} (Q)$ is enhanced  comparable to
$\Pi_{ph} (0)$. We present the plot of $\Pi_{ph} (q)$ along
$q_x=q_y$ in Fig \ref{fig:PiQ}. Indeed, $\Pi_{ph} (Q)$ is much
larger than  $\Pi_{ph} (0)$.

We see therefore that for the two-pocket model with circular hole
and electron FSs and a constant repulsive electron-electron
interaction
\begin{itemize}
\item
the KL mechanism -- the renormalization of the bare interaction
into an irreducible pairing vertex,  does give rise to a pairing,
\item
the pair wave function has $A_{1g}$ (s-wave) symmetry, but changes
sign between hole and electron pockets
\end{itemize}

Comparing isotropic and lattice cases, we see two differences.
First, because of the lattice, particle-hole bubble $\Pi_{ph} (q)$
no longer has to be a decreasing function of $q$.  In fact, as we
just found, in the two-pocket model the KL mechanism leads to a
pairing instability precisely because $\Pi_{ph} (Q)$ is larger
than $\Pi_{ph} (0)$. Second, because we deal with fermions with
circular FSs located near particular $k-$points, polarization
operators at small momentum transfer and momentum transfer ${\bf
Q} =(\pi,\pi)$ can be approximated by constants.  Then the
irreducible vertex function has only an $s-$wave ($A_{1g}$)
harmonic, like the bare interaction, i.e. KL renormalization does
not generate interactions in other channels.   Treating pockets as
circular is indeed an approximation, because for square lattice
the only true requirement is that each FS is symmetric with
respect to rotations by multiples of $\pi/2$ ($C_4$ symmetry). For
small pocket sizes, deviations from circular forms are small, but
nevertheless are generally finite.  If we include this effect, we
find that the KL effect does generate interactions in other
channels ($B_{1g}, B_{2g}$, and $A_{2g}$), which may be
attractive, and also leads to more complex structure of the pair
wave function in $s^{+-}$ channel, which now acquires  angular
dependence along hole and electron pockets, consistent with $C_4$
symmetry\cite{AccNodes,Cvetkovic}

The situation changes when we consider the actual bare
interactions $G_i$, extracted from the multi-orbital model.  Then
$G_4-G_3$ is generally non-zero already before KL renormalization.
It is natural to expect that the bare interaction is a decreasing
function of momenta, in which case $G_4$, which is the interaction
at small momentum transfer, is larger than the interaction $G_3$
at momentum transfer near $Q$. Then the KL term has to compete
with the first-order repulsion. As long as $G \Pi_{ph} (Q)$ is
small, KL renormalization cannot overshoot bare repulsion, and the
bound state does not appear. The situation may change when we
include momentum dependence of the interaction and non-circular
nature of the pockets.  In this last case, there appears infinite
number of $A_{1g}$ harmonics, which all couple to each other, and
in some cases one or several eigenfunctions may end up as
attractive \cite{A1g}. Besides, angle dependence generates
$d-$wave and $g-$wave harmonics, and some of eigenfunctions in
these channels may  also become attractive and
compete\cite{B1gA1g}. Still, however, in distinction to the
isotropic case, there is no guarantee that ``some" eigenfunction
from either $A_{1g}$, or $B_{1g}$, or $B_{2g}$, or $A_{2g}$, will
be attractive.  A lattice system may well remain in the normal
state down to $T=0$.

We will discuss how to go beyond second order in $G$ in the next
section. In the remainder of this section we discuss KL physics in
the two other classes of systems -- cuprates and doped graphene.

\subsection{Superconductivity in cuprates}

Cuprates are layered materials with one or more  crystal planes
consisting of Cu and O atoms (two O per Cu), and charge reservoirs
between them. Superconductivity is widely believed to originate
from electron-electron interactions  in these CuO$_2$ planes. The
undoped parent compounds are Mott insulators/Heisenberg
antiferromagnets due to very strong Coulomb repulsion which
prevents electron hopping from Cu to Cu and therefore localizes
electrons near  lattice sites. Doping these insulating CuO$_2$
layers with carriers (by adding/removing electrons from/to charge
reservoir) leads to a (bad) metallic behavior and to the
appearance of high- temperature superconductivity.  A schematic
phase diagram of doped cuprates is shown in Fig. \ref{fig:Cu
phase}. The richness of this phase diagram generated a lot of
efforts, both in experiment and in theory to understand the key
physics of the cuprates (see e.g.
Refs.\cite{acs,Scal,Lee,Hirsch,ZX,Subir,Greene,millis,Norman,sawat,ricereview,Monthoux,lonzarich,AMT1,AMT2}.
There are several features in the phase diagram, like the
pseudogap
in hole-doped cuprates,  which are still not fully understood,
although a substantial progress has been made over the last few
years
 on the issue of the interplay between pseudogap
and superconductivity~\cite{Max,efetov,Physics}

By all accounts, the symmetry of the superconducting state does
not change between small doping, where pseudogap physics is
relevant, and doping above the optimal one.  For these larger
dopings, ARPES and quantum oscillation experiments show a large FS
(see Fig. \ref{fig:Cu_Large_FS}) consistent with Luttinger count
for fermionic states.  In this doping range, it is natural to
expect that the pairing symmetry can be at least qualitatively
understood by performing weak coupling analysis.

\begin{figure}[htp]
\includegraphics[width=4in]{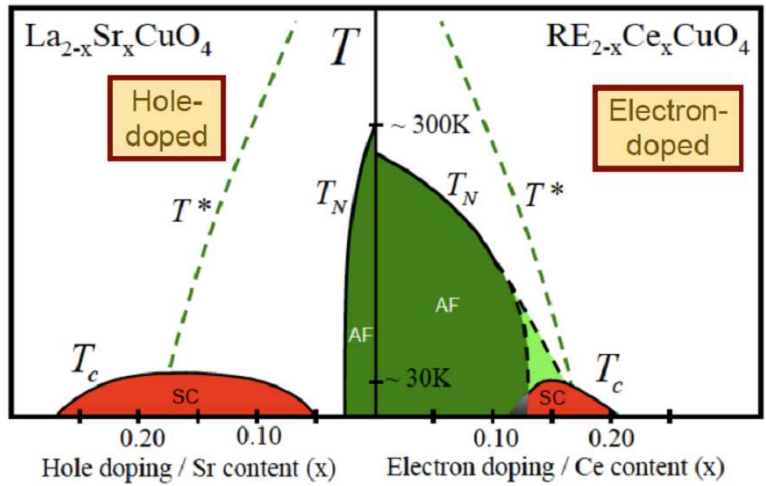}
\caption{\label{fig:Cu phase}Typical phase diagram for the
cuprates for electron and hole doping. The similarity of this
phase diagram with pnictides is the proximity to the
antiferromagnetic phase. Amongst differences, the most important
one is the fact that the antiferromagnetic phase in cuprates stems
out of a Mott insulator in the parent compounds. Others are a
remarkable asymmetry between electron and hole doping and
pseudogap phase indicated by the T$^*$ line. $T_N$ is the
transition into the antiferromagnetic state and $T_c$ is the
transition into the superconducting state. We shall only focus on
the superconducting aspect of this phase diagram. Taken from Ref.
\cite{Greene}. }
\end{figure}

\begin{figure}[htp]
\includegraphics[width=2.5in]{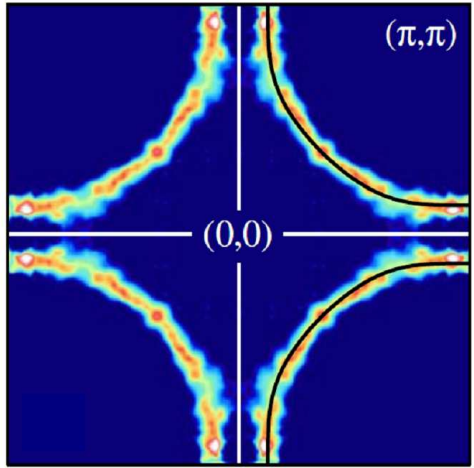}
\caption{\label{fig:Cu_Large_FS}Angle resolved Photoemission data
from Ref.\cite{Cu_large FS}, showing the presence of a large FS
for doped Tl$_2$Ba$_2$CuO$_{6+\delta}$. The FS is extracted from
the position of the peak of the spectral function in the
$k-$space.}
\end{figure}

The FS
for hole-doped cuprates is an is open electron FS shown in Fig.
\ref{fig:FS}(center) and Fig. \ref{fig:Cu_Large_FS}.
Central to our consideration is the fact that the fermionic
density of states is the largest near the points $(0,\pm \pi)$ and
$(\pm \pi,0)$, where two FS lines come close to each other  (one
can show that the density of states is logarithmically enhanced
and actually diverges\cite{VanHove} when the two FS lines merge at
$(0,\pm \pi)$ and $(\pm \pi,0)$). The FS regions with the largest
DOS mostly contribute to superconductivity, and, to first
approximation, one can consider the FS in Fig.
\ref{fig:FS}(center) as consisting of four patches. We focus on
spin-singlet pairing, in which case a pair wave function is an
even function of momentum, and it has the same form in the pairs
of patches which transform into each other under inversion. This
leaves two non-equivalent patches, which for definiteness we
choose to be near $(0,\pi)$ and $(\pi,0)$.

\begin{figure}[htp]
\includegraphics[width=4in]{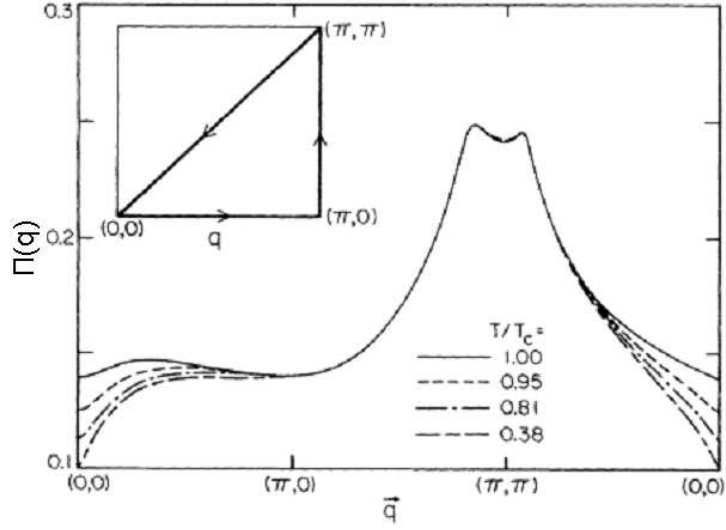}
\caption{\label{fig:Cu_PiQ}The plot of $\Pi(q)$ for a FS topology
shown in Fig. \ref{fig:Cu_Large_FS} with $\vec{q}$ along the
directions in the Brillouin Zone shown in the inset. Different
lines are for different temperatures. Observe that $\Pi(Q)$ is
always larger that $\Pi(0)$. Taken from Ref. \cite{Monthoux}}
\end{figure}

The resulting two-patch model is in many respects similar to the
two-pocket model for Fe-pnictides, only instead of hole-hole,
electron-electron, and hole-electron interaction we now have
intra-patch and inter-patch interactions for two patches, which we
label as 1 and 2. The interaction Hamiltonian  contains four
terms, like in Eq. \ref{eq:2-poc Ham},  and the full pairing
vertices $\Gamma^{full}_{11} = \Gamma^{full}_{22}$ and
$\Gamma^{full}_{12}$ are \bea
\Gamma^{full}_{11} &=& \frac{1}{2} \left(\frac{ {\bar \Gamma}^{0}_{12} + {\bar \Gamma}^{0}_{11}}{1 -({\bar \Gamma}^{0}_{12} + {\bar \Gamma}^{0}_{11}) \Pi_{pp}} + \frac{{\bar \Gamma}^{0}_{11} - {\bar \Gamma}^{0}_{12}}{1  - ({\bar \Gamma}^{0}_{11}-{\bar \Gamma}^{0}_{12}) \Pi_{pp}}\right),\nonumber \\
\Gamma^{full}_{12} &=& \frac{1}{2} \left(\frac{ {\bar \Gamma}^{0}_{12} + {\bar \Gamma}^{0}_{11}}{1 -({\bar \Gamma}^{0}_{12} + {\bar \Gamma}^{0}_{11}) \Pi_{pp}} - \frac{{\bar \Gamma}^{0}_{11} - {\bar \Gamma}^{0}_{12}}{1  - ({\bar \Gamma}^{0}_{11}-{\bar \Gamma}^{0}_{12}) \Pi_{pp}}\right),\nonumber \\
&& \label{4_3_3} \eea or \bea
\Gamma^{full}_{11}  + \Gamma^{full}_{12} &=&  \frac{ {\bar \Gamma}^{0}_{12} + {\bar \Gamma}^{0}_{11}}{1 -({\bar \Gamma}^{0}_{12} + {\bar \Gamma}^{0}_{11}) \Pi_{pp}} \nonumber \\
\Gamma^{full}_{11}  - \Gamma^{full}_{12} &=&   \frac{{\bar
\Gamma}^{0}_{11} - {\bar \Gamma}^{0}_{12}}{1  - ({\bar
\Gamma}^{0}_{11}-{\bar \Gamma}^{0}_{12}) \Pi_{pp}} \label{4_3_4}
\eea where, as before, ${\bar \Gamma}^0$ are irreducible pairing
vertices and $\Pi_{pp} = \Pi_{pp} (q,\Omega)$ contains the Cooper
logarithm. To first order in the interaction ${\bar \Gamma}^0_{11}
={\Gamma}^0_{11} = -G_4$, and ${\bar \Gamma}^0_{12}
={\Gamma}^0_{12} = -G_3$, such that ${\bar \Gamma}^{0}_{12} +
{\bar \Gamma}^{0}_{11} = - (G_4+G_3), {\bar \Gamma}^{0}_{11} -
{\bar \Gamma}^{0}_{12} = -(G_4-G_3)$. Superconductivity requires
${\bar \Gamma}^{0}_{11} + {\bar \Gamma}^{0}_{12}$ or ${\bar
\Gamma}^{0}_{11} - {\bar \Gamma}^{0}_{12}$ to be positive.  For
Hubbard interaction $G_i=G$, the bare ${\Gamma}^{0}_{11} +
{\Gamma}^{0}_{12} = -(G_3+G_4) = -2G$ is negative, hence there is
no pairing instability which would lead to a state with
sign-preserving  wave-function.  At the same time,
${\Gamma}^{0}_{11} - {\Gamma}^{0}_{12} = G_3-G_4 = 0$, hence the
coupling vanishes for a potential instability towards a pairing
with a  wave function which changes sign between patches. To
obtain the information about the sign of the irreducible
${\Gamma}^{0}_{11} - {\Gamma}^{0}_{12}$ one then needs to include
KL renormalization.  The result is, predictably, the same as in
two-pocket model, namely \bea {\bar \Gamma}^{0}_{11} - {\bar
\Gamma}^{0}_{12} = 2 G^2 \left(\Pi_{ph} (Q) - \Pi_{ph} (0)\right)
\label{4_3_5} \eea where $Q = (\pi,\pi)$ is now the distance
between patches. The two particle-hole polarization bubbles can be
straightforwardly calculated for $t-t'$ model of fermionic
dispersion with hopping between nearest and next nearest
neighbors. The result is that $\Pi_{ph} (Q)
> \Pi_{ph} (0)$ (see Fig. \ref{fig:Cu_PiQ}). Then ${\bar \Gamma}^{0}_{11} - {\bar
\Gamma}^{0}_{12} >0$, and the combination of full vertices
$\Gamma^{full}_{11} - \Gamma^{full}_{12}$ has a pole in the upper
frequency half-plane, at $\Omega = i \Omega_p$, which is the
solution of $2 G^2 \left(\Pi_{ph} (Q) - \Pi_{ph} (0)\right)
\Pi_{pp} (i\Omega_p) =1$.

So far, everything is the same as in the two-pocket model. But
there is qualitative difference between the two cases. In the
two-pocket model, the sign-changing pair wave-function changes
sign between different FS pockets, but preserves the same sign
along a given pocket.  Such a wave function belongs to $A_{1g}$
representation. In two-patch model, the sign-changing wave
function changes sign between the two ends of the same "arc" of
the FS.  In other words, it changes sign under $\pi/2$ rotation
from $x$ to $y$ axis. According to classification scheme, such a
wave function belongs to $B_{1g}$ representation, i.e., has a
d-wave symmetry. Further,  if we move along the FS arc away from
patches and assume that the pairing wave function does not vanish
on the FS, except may be special points, we immediately conclude
that it should change sign right at the center of the arc, i.e.,
at the direction along zone diagonal. By symmetry, this should
happen along each diagonal. The prototype wave function for such a
state is $\cos k_x - \cos k_y$. We caution, however, that in the
patch model we only know the wave function near $(0,\pi)$ and
$(\pi,0)$ and its evolution between the patches is generally
described by the whole subset of wave functions from $B_{1g}$
representation with the form $\cos((2m+1) k_x) - \cos((2m+1)
k_y)$.

We see therefore that for two-patch model with a constant
repulsive electron-electron interaction
\begin{itemize}
\item
the KL mechanism  again gives rise to  pairing,
\item
the pair wave function has $B_{1g}$ (d-wave) symmetry, and changes
sign twice  along the open electron FS
\end{itemize}
The KL consideration can also be applied to electron-doped
cuprates~\cite{G}, but the analysis in this case is somewhat
different as hot spots are located close to Brillouin zone
diagonals~\cite{Greene}.

\subsection{Superconductivity in doped graphene}

Graphene is a two-dimensional array of carbon atoms on a honeycomb
lattice. The energy dispersion of graphene has two bands due two
non-equivalent positions of atoms on a honeycomb lattice. The two
bands touch each other at six  points in the Brillouin zone, and
the dispersion near these points is $\pm |{\vec k}|$ what brought
them the name Dirac points. At zero doping, the Fermi level passes
right through Dirac points, what gives rise to highly interesting
low-energy physics\cite{CastroNeto}. Upon doping by either
electron or holes, six separate pockets of carriers emerge. Upon
further doping, these pockets grow, merge at some doping $x_c$,
and at even larger dopings form a large hexagonal FS (see Fig.
\ref{fig:FS} right).
Such a high doping of a single layer graphene has been achieved in
Ref. \cite{roten}
 by placing $Ca$ and $K$ dopants
 above and below a graphene layer.
At $x=x_c$ the system passes through a Van-Hove singularity which
results in an enhanced density of states at the six saddle points
where nearest pockets merge. The fermionic dispersion at $x=x_c$
is very similar to that in the cuprates at the Van-Hove doping,
but the tendency to the nesting (the existence of parallel pieces
of the Fermi surface) is more pronounced here because in graphene
the tight-binding band structure is not sensitive to the second
neighbor hopping\cite{CastroNeto,Gonzalez}.

\begin{figure}[htp]
\includegraphics[width = 4in]{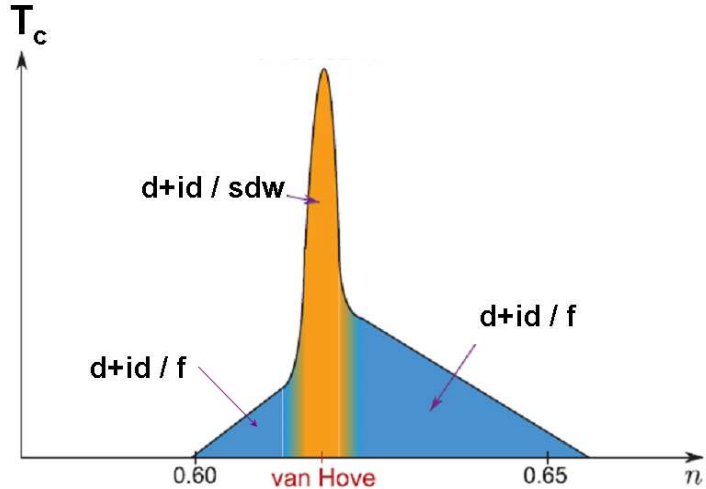}
\caption{\label{fig:Gr_phase} Schematic phase diagram of doped
graphene. $T_c$ is the instability temperature towards spin
singlet $d+id$ or spin triplet $f-$wave SC states, or SDW state.
$T_c$ is plotted against doping ($n$). Doped Graphene is expected
to be mostly superconducting with competition with the SDW phase
near the Van-Hove region. Taken from Ref. \cite{ronny_gr}}
\end{figure}

 The increase of the density of states near Van Hove doping
increases the relative strength of the interaction effects, and
brings in a possibility that already at a weak coupling the Fermi
liquid state will become unstable towards some kind of order. A
number of candidate ordered states has been considered, including
superconductivity, SDW order, nematic order and so on.
See~\cite{andreyd+id,CastroNeto,Gonzalez,Vozmediano,Batista,TaoLi,
MoraisSmith,AndreySDW_gr,ronny_gr}. Because the density of state
diverges at the saddle points at Van Hove doping, each state can
be can be self-consistently obtained at weak coupling. A phase
diagram of doped graphene is shown in Fig. \ref{fig:Gr_phase}.
Superconductivity has also been observed and analyzed in graphite
intercalated compounds like $C_6Ca$ and $C_6Yb$ \cite{Graphite2}.
This superconductivity may be due to electron-electron
interaction~\cite{Graphite1}, but most likely the pairing
interaction in these materials is mediated by intercalant phonons
and/or acoustic phonons.

The presence of saddle (Van-Hove) points along the FS is the
feature that draws our attention and invites us to perform an
analysis similar to that in the cuprates, but with  three rather
than two non-equivalent patches (overall there are six Van Hove
points along the FS, but only three are unique, the three others
are related by inversion symmetry (see Fig\ref{fig:FS} right). The
treatment of doped graphene parallels the description in the above
two subsections, but we will see that there are interesting
details here, not present in the earlier models.

We treat the low-energy physics of doped graphene within the
effective three patch model, just like we did for the cuprates. We
introduce intra-patch and inter-patch vertices $\Gamma_{ij}$,$i,j
\in (1,2,3)$ with $\Gamma_{ij}=\Gamma_{ji}$.  Because the three
patches  are fully symmetric, the
 total number of independent vertices is just two:
\bea\label{eq:gamma constraints}
&&\Gamma^0_{11}=\Gamma^0_{22}=\Gamma^0_{33}=-G_4=\Gamma^0_u\nonumber\\
&&\Gamma^0_{12}=\Gamma^0_{13}=\Gamma^0_{23}=-G_3=\Gamma^0_v \eea

We follow the same line of reasoning as before. The full pairing
vertices $\Gamma^{full}_{11}$ and $\Gamma^{full}_{12}$  are
expressed in terms of irreducible vertices ${\bar \Gamma}^0_u$ and
${\bar \Gamma}^{0}_{v}$ as

\bea\label{eq:gr-vertexfull1} \Gamma^{full}_{11}&=&\bar\Gamma^0_u
+ \bar\Gamma^0_u \Gamma^{full}_{11} \Pi_{pp} + \bar\Gamma^0_v
\left(\Gamma^{full}_{21}+\Gamma^{full}_{31}\right)\Pi_{pp}\nonumber\\
\Gamma^{full}_{12}&=&\bar\Gamma^0_v + \bar\Gamma^0_u
\Gamma^{full}_{12} \Pi_{pp} + \bar\Gamma^0_v
\left(\Gamma^{full}_{22}+\Gamma^{full}_{32}\right)\Pi_{pp}\nonumber\\
\eea

The solutions of this set are

\bea\label{eq:gr-vertexfull3}
\Gamma^{full}_{11}&=& \frac{1}{3}\left( \frac{\bar\Gamma^0_{u}+2\bar\Gamma^0_{v}}{1-(\bar\Gamma^0_{u}+2\bar\Gamma^0_{v})\Pi_{pp}}+2\frac{\bar\Gamma^0_{u}-\bar\Gamma^0_{v}}{1-(\bar\Gamma^0_{u}-\bar\Gamma^0_{v})\Pi_{pp}}\right)\nonumber\\
\Gamma^{full}_{12}&=& \frac{1}{3}\left( \frac{\bar\Gamma^0_{u}+2\bar\Gamma^0_{v}}{1-(\bar\Gamma^0_{u}+2\bar\Gamma^0_{v})\Pi_{pp}}-\frac{\bar\Gamma^0_{u}-\bar\Gamma^0_{v}}{1-(\bar\Gamma^0_{u}-\bar\Gamma^0_{v})\Pi_{pp}}\right)\nonumber\\
\eea

or \bea\label{eq:graphene_vertex_full}
&&\Gamma^{full}_{11}+2\Gamma^{full}_{12}=\frac{\bar\Gamma^0_{u}+2\bar\Gamma^0_{v}}{1-(\bar\Gamma^0_{u}+2\bar\Gamma^0_{v})\Pi_{pp}}\nonumber\\
&&\Gamma^{full}_{11}-\Gamma^{full}_{12}=\frac{\bar\Gamma^0_{u}-\bar\Gamma^0_{v}}{1-(\bar\Gamma^0_{u}-\bar\Gamma^0_{v})\Pi_{pp}}\nonumber\\
\eea we see that to get the pairing we need either
$\bar\Gamma^0_{u}+2\bar\Gamma^0_{v}$ or
$\bar\Gamma^0_{u}-\bar\Gamma^0_{v}$ to be positive. To first order
in the interaction we have  $\bar\Gamma^0_{u}= \Gamma^0_{u} =
-G_4$ and $\bar\Gamma^0_{v}= \Gamma^0_{v} = -G_3$, hence the
conditions for the pairing are $G_4+2G_3 <0$ or $G_3 > G_4$. The
first condition is analogous to $G_3 + G_4 <0$ for the two-patch
model and is never satisfied for a repulsive interaction, when
$G_4$ and $G_3$ are both positive.  The second condition is
exactly the same as in two-patch model and requires inter-patch
interaction to be larger than intra-patch interaction. If the bare
interaction is momentum-independent, $G_3 = G_4 = G$, and one of
the two pairing channels is neither repulsive nor attractive.

Continue with the Hubbard interaction. To second order in $U$, we
have from KL renormalization
 \bea\label{eq:gr-2nd order} \bar\Gamma^0_{u}&=& -
G_4-\Pi_{ph}(0)\left[ G_4^2 + 2G_2^2 - 4G_1(G_1-G_2)
\right]\nonumber\\
\bar\Gamma^0_{v}&=&-G_3-\Pi_{ph}(Q)\left[ 2G_3(2G_1-G_2)
\right]\eea and $\bar\Gamma^0_{u} - \bar\Gamma^0_{v}$ becomes
\beq\label{eq:gr condition} \bar{\Gamma}^0_{u}-\bar{\Gamma}^0_{v}=
G^2(2\Pi_{ph}(Q)-3\Pi_{ph}(0))\eeq Like in the previous two
examples, if $\Pi_{ph}(Q)$ is larger than $\Pi_{ph} (0)$
(specifically, if $\Pi_{ph}(Q) > (3/2)\Pi_{ph}(0)$) the
irreducible pairing interaction is attractive. The particle-hole
bubble can be straightforwardly computed and the result is,
predictably, that near Van-Hove doping, $\Pi_{ph}(Q) >
(3/2)\Pi_{ph}(0)$.
This result was fist obtained by Gonzales\cite{Gonzalez} and
reproduced in more recent work~\cite{andreyd+id}.

\begin{figure*}[t]
$\begin{array}{cc}
\includegraphics[width = 2.2in]{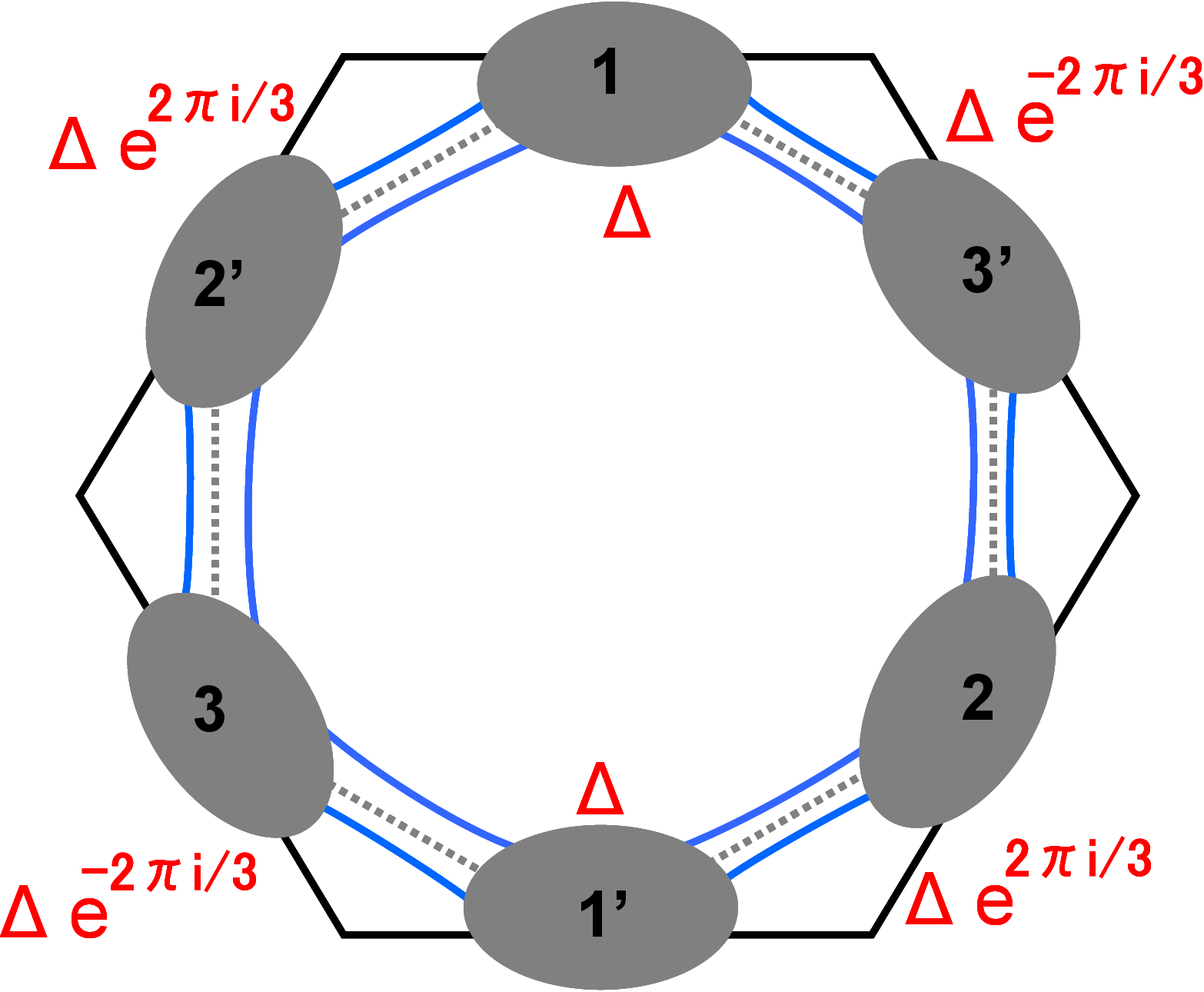}&
\includegraphics[width = 2.2in]{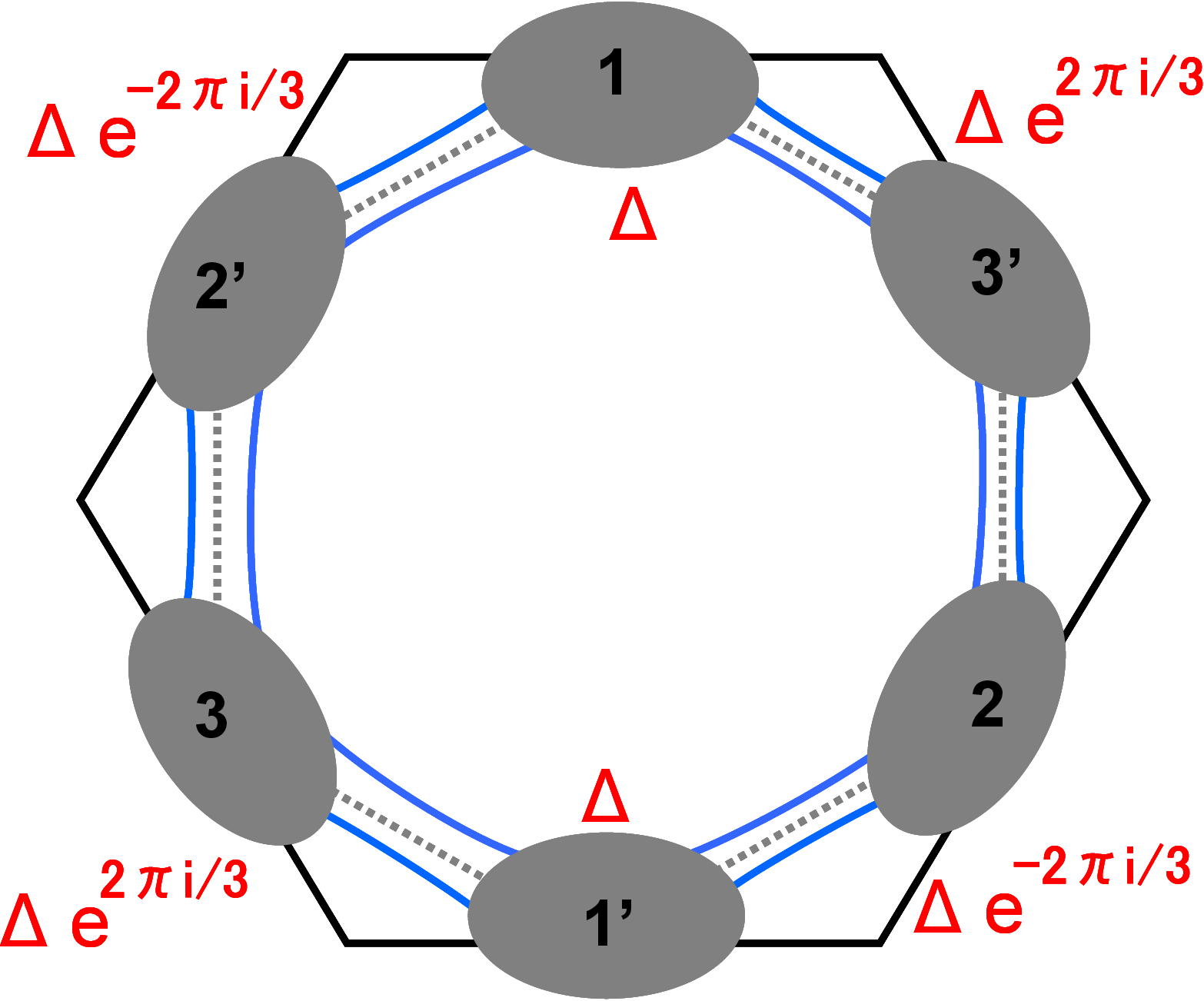}
\end{array}$
\caption{\label{fig:Gr_gap}The phases of the pair-wave functions
at the patches (regions with enhanced density of states). The left
and right represent the two Z$_2$ breaking $d-$wave solutions
$d+id$ or $d-id$.}
\end{figure*}

So far, the results are virtually undistinguishable from the
previous two cases.  The new physics in the three-patch model
reveals itself when we note that the presence of the pole for the
combination $\Gamma^{full}_{11}-\Gamma^{full}_{12}$ and its
absence for $\Gamma^{full}_{11}+2\Gamma^{full}_{12}$ in Eq.
(\ref{eq:graphene_vertex_full}) implies that near the instability
the fully renormalized intra-patch and inter-patch pairing
vertices must satisfy \beq \label{eq:1}
\Gamma^{full}_{11}=-2\Gamma^{full}_{12} \eeq together with the
symmetry-imposed conditions $\Gamma^{full}_{11}=
\Gamma^{full}_{22} = \Gamma^{full}_{33}$ and $\Gamma^{full}_{12} =
\Gamma^{full}_{13} = \Gamma^{full}_{23}$. In other words if
intra-patch $\Gamma^{full}_{ii}=D$, then inter-patch
$\Gamma^{full}_{ij}= -\frac D 2 $ for $i\neq j$. Now, if we view
each $\Gamma^{full}_{ii}$ as the modulus square of the
superconducting order parameter $|\Delta_i|^2$
 and $\Gamma^{full}_{ij}$ as $Re[\Delta_i \Delta^*_j]$, we immediately find that Eq. (\ref{eq:1}) implies that
the relative phase of the superconducting order parameter must
change by $\pm\frac{2\pi}{3}$ between each pair of patches
($\cos\frac{2\pi}{3}=-\frac12$). In other words, if the order
parameter in patch 1 is $\Delta_1$, then $\Delta_2 = \Delta_1
e^{\pm 2\pi i/3}$ and $\Delta_3 = \Delta_1 e^{\pm 4\pi i/3}$. The
two resulting $\Delta$ structures are shown in Fig.
\ref{fig:Gr_gap}. We used the fact that this is spin-singlet
pairing, hence $\Delta (-k) = \Delta (k)$. This is a $d-$wave gap
because if we extend the gap structure to all FS, we find that the
gap changes sign twice along the FS. However, we also need to pick
one sign of the phase change or  the other, and this choice breaks
the Z$_2$ symmetry, which in our case is time-reversal symmetry
because it changes the order parameter to its complex conjugate.
Putting it differently, $Z_2$ symmetry corresponds to the freedom
of choice of counter clockwise or clockwise phase winding by
$4\pi$ along the full FS. Such a state is called $d+id$ or $d-id$.
It has a rich phenomenology and is highly desirable for
applications\cite{Volovik,Sirgist,Voj,Moore,Ivanov,kane,Qi,cheng}.


Although intuitively it seems obvious that $Z_2$ symmetry is
broken in a $d+id$ state, one actually needs to do full
Ginzburg-Landau (GL)  analysis and make sure that the
superconducting condensation energy is the largest when only
$d+id$ or only $d-id$ solution develops, but not both of them.
This, however, requires one to go beyond the instability point,
while our goal is to get as much information as possible from the
normal state analysis. We just refer to Ref \cite{andreyd+id}
where GL functional has been derived and analyzed. The result of
that study is that $Z_2$ symmetry is indeed broken below $T_c$.

Another way to see that the two d-wave states are degenerate by
symmetry is to look at the representations of the symmetry group
D$_{6h}$. The two d-wave wave-functions $\cos k_x-\cos k_y$ and
$\sin k_x \sin k_y$ belong to a two-dimensional $E_{2g}$
representation of D$_{6h}$ and must indeed be degenerate by
symmetry.

We see therefore that for a three-patch model with a constant
repulsive electron-electron interaction
\begin{itemize}
\item
the KL mechanism  again gives rise to  pairing,
\item
the pair wave function has $d+id$ or $d-id$ symmetry. In each of
the two states  phase of the wave-function winds by $4\pi$ along
the FS either clockwise or anticlockwise and time reversal
symmetry is broken.
\end{itemize}

We have seen therefore that in  all the three systems which we
analyzed  the  condition for the pairing instability (the
emergence of the pole in the vertex function in the upper
frequency half plane) is that irreducible inter-pocket/inter-patch
pairing vertex should be larger by absolute value than the
absolute value of the irreducible intra-pocket/intra-patch vertex.
For a momentum-independent bare interaction, this  reduces to the
condition of having $\Pi_{ph}(Q)>a \Pi_{ph}(0)$, where $a$ is some
numerical factor of $O(1)$ that depends on the model. Now we will
discuss how one still get an attraction if the bare interaction is
momentum-dependent.

A final remark about doped graphene. The KL mechanism has been
also applied to  somewhat smaller dopings, when the FS still
contains six disconnected pieces. In this doping range, KL-based
analysis yields a novel spin-triplet f-wave
superconductivity~\cite{roten,Gonzalez,Raghu,ronny_gr}.

\section{What to do if the bare irreducible vertex is repulsive}

We recall that setting $G_3=G_4$ is indeed a crude approximation.
In reality, $G_4$ is the interaction at small momentum transfer,
while $G_3$ is the interaction at momentum transfer comparable to
inverse lattice spacing. By conventional wisdom, one should expect
$G_4$ to be larger $G_3$, and microscopic calculations generally
confirm this, although in multi-orbital systems the interplay
between $G_3$ and $G_4$  is more involved as both appear (in the
band basis) even if we only consider on-site interaction in the
orbital basis. In this latter case, $G_4 >G_3$ if Hubbard
interaction for fermions belonging to the same orbital is larger
than Hubbard interaction between fermions belonging to different
orbitals.

If $G_4 >G_3$, the attractive KL contribution of order $G^2
\Pi_{ph}$ has to compete with the repulsive first-order term.  At
weak coupling and for a non-nested FS,  second-order term is
expected to be smaller than first-order term, i.e.,
superconductivity does not occur. The situation may change when we
include momentum dependence of the interaction and non-circular
nature of the pockets.  In this case, there appears an infinite
number of  harmonics in each of the channels $A_{1g}$, $B_{1g}$,
$B_{2g}$, or $A_{2g}$, which all couple to each other, and in some
cases one or several eigenfunctions may end up as attractive.
Still, however, in distinction to the isotropic case, there is no
guarantee that ``some'' eigenfunction from  will be attractive.
In other words, a lattice system well may remain in the normal
state down to $T=0$.

How can we still get superconductivity in this situation?  A first
step is to realize that there is a similarity between the
renormalization of the irreducible particle-particle interaction
to order $G^2$ and the renormalization to the same order in the
particle-hole channel at small momentum transfer. Recall that for
the latter case, the renormalization can be extended to higher
orders because the momentum transfer $q$ and frequency transfer
$\Omega$, $\Pi_{ph} (q, \Omega)$ can be large and one can sum up
series in $G\Pi(q,\Omega)$ by restricting in perturbation series
only with the particle-hole bubbles at small $q$ and $\Omega$. In
this situation, charge and spin components of the full
$\Gamma^{full}_{\alpha\beta,\gamma\delta} (k_1,k_2;k_3,k_4) =
\Gamma_c \delta_{\alpha\gamma}\delta_{\beta\delta}+ \Gamma_s
\vec{\sigma}_{\alpha,\gamma}\cdot\vec{\sigma}_{\beta\delta}$ can
be viewed as effective interactions mediated by collective
excitations in charge and spin channel, respectively,

\beq\label{eq:rpa_1} \Gamma_c=-\frac{G}{2} \frac{1}{1+G\Pi_{ph}(q,
\Omega)}~~~\Gamma_s = \frac{G}{2} \frac{1}{1-G\Pi_{ph}(q,\Omega)}
\eeq

We have ${\bar\Gamma}^{0}_{\alpha\beta,\gamma\delta} (k,-k;p,-p) =
\Gamma_c (k-p) \delta_{\alpha\gamma}\delta_{\beta\delta}+ \Gamma_s
(k-p)
\vec{\sigma}_{\alpha,\gamma}\cdot\vec{\sigma}_{\beta\delta}$.
Restricting with the diagrams containing $\Pi_{ph} (k-p)$ and
neglecting terms with $\Pi_{ph} (k+p)$ (this is often called the
RPA), we obtain for $k$ and $p$ at the same pocket/patch, when
$k-p$ is small
 \beq
\Gamma_c (0) = -\frac{G_4}{2}
\frac{1}{1+G_4\Pi_{ph}(0)}~~~\Gamma_s (0) = \frac{G_4}{2}
\frac{1}{1-G_4\Pi_{ph}(0)} \eeq
 and  for $k$ and $p$ at different pockets/patches, when $k-p \approx Q$
  \beq
\Gamma_c (Q) = -\frac{G_3}{2}
\frac{1}{1+G_3\Pi_{ph}(Q)}~~~\Gamma_s (Q) = \frac{G_3}{2}
\frac{1}{1-G_3\Pi_{ph}(Q)} \label{16_5} \eeq Re-expressing
${\bar\Gamma}^{0}_{\alpha\beta,\gamma\delta} (k,-k;p,-p)$ in terms
of singlet and triplet components as \bea
&&{\bar\Gamma}^{0}_{\alpha\beta,\gamma\delta} (k,-k;p,-p) = \nonumber \\
&&\Gamma_{\text{s=0}} (k-p) \left(\delta_{\alpha\gamma}\delta_{\beta\delta}- \delta_{\alpha\delta}\delta_{\beta\gamma}\right) + \nonumber \\
&&
\Gamma_{\text{s=1}} (k-p) \left(\delta_{\alpha\gamma}\delta_{\beta\delta}+ \delta_{\alpha\delta}\delta_{\beta\gamma}\right) \nonumber \\
\eea We obtain

\bea\label{eq:corr1}
\Gamma_{\text{s=0}}  &=& \frac{1}{2} \left(\Gamma_c-3\Gamma_s\right) \nonumber \\
\Gamma_{\text{s=1}}  &=& \frac{1}{2}
\left(\Gamma_c+\Gamma_s\right) \eea

i.e.

\bea
\Gamma_{\text{s=0}} (0) &=& -\frac{G_4}{4} \left(\frac{1}{1+ G_4 \Pi_{ph} (0)} + \frac{3}{1- G_4 \Pi_{ph} (0)} \right) \nonumber \\
\Gamma_{\text{s=1}} (0) &=& \frac{G_4}{4} \left(\frac{1}{1- G_4 \Pi_{ph} (0)} - \frac{1}{1+ G_4 \Pi_{ph} (0)}\right) \nonumber \\
\Gamma_{\text{s=0}} (Q) &=& -\frac{G_3}{4} \left(\frac{1}{1+ G_3 \Pi_{ph} (Q)} + \frac{3}{1- G_3 \Pi_{ph} (Q)} \right) \nonumber \\
\Gamma_{\text{s=1}} (Q) &=& \frac{G_3}{4} \left(\frac{1}{1- G_3
\Pi_{ph} (Q)} - \frac{1}{1+ G_3 \Pi_{ph} (Q)}\right) \label{16_1}
\eea Let's compare this result with what we obtained in the KL
formalism. Focus on the singlet channel and expand in (\ref{16_1})
to second order in $G_{3,4}$.
 We have
 \bea
&&\Gamma_{\text{s=0}} (0) \approx - \frac{G_4}{2} \left(1 + \frac{1}{1- G_4 \Pi_{ph} (0)}\right)\nonumber \\
&&\approx - G_4 \left(1+ 0.5 G_4\Pi_{ph} (0) \right) \nonumber \\
&&\Gamma_{\text{s=0}} (Q)\approx - \frac{G_3}{2} \left(1 + \frac{1}{1- G_3 \Pi_{ph} (Q)}\right) \nonumber \\
&&\approx - G_3 \left(1 + 0.5 G_3 \Pi_{ph} (Q) \right)
\label{16_2} \eea Apart from the factor of $1/2$ (which is the
consequence of an approximate RPA scheme) $\Gamma_{\text{s=0}}
(0)$ is the same as irreducible vertex ${\bar \Gamma}^0_{11}$,
which we obtained in KL calculation in the previous section, and
$\Gamma_{\text{s=0}} (Q)$ the same as ${\bar \Gamma}^0_{12}$  By
itself, this is not surprising, as in
 $\Gamma_{\text{s=0}}$ we included
the same particle-hole renormalization of the bare pairing
interaction as in the KL formalism.

The outcome of this formula is the observation that KL term is the
first term in the series for the irreducible pairing vertex. In
the RPA scheme, the full series gives,

\bea
&&\Gamma_{\text{s=0}}(0) = -\frac{1}{4} \left(\frac{G_4}{1+G_4 \Pi_{ph} (0)} + \frac{3G_4}{1-G_4 \Pi_{ph} (0)}\right) \nonumber \\
&&\Gamma_{\text{s=0}}(0) = -\frac{1}{4} \left(\frac{G_3}{1+G_3
\Pi_{ph} (Q)} + \frac{3G_3}{1-G_4 \Pi_{ph} (Q)}\right), \eea For
repulsive interaction, the charge contribution only gets smaller
when we add higher terms in $G$, but spin contribution gets
larger. A conventional recipe is to neglect all renormalizations
in the charge channel and approximate ${\Gamma_{\text{s=0}}}$ with
the sum of a constant and the interaction in the spin channel. The
irreducible interaction in the $s+-$ channel in the pnictides or
in the $d-$wave channel in the cuprates and in doped graphene is
then
\bea &&\Gamma_{\text{s=0}}(0) - \Gamma_{\text{s=0}}(Q)=\nonumber\\
&&-G_4+G_3 - \frac{3}{4} \left(\frac{G_4}{1-G_4 \Pi_{ph} (0)} -
\frac{G_3}{1-G_3 \Pi_{ph} (Q)}\right) \label{16_3} \nonumber\\
\eea

Like we said before if $G_4\Pi_{ph}(0)$ and $G_3 \Pi_{ph} (Q)$ are
both small, $G_4-G_3$ term is the largest, and the pairing
interaction is repulsive for $G_4 > G_3$. However, we see that
there is a way to overcome the initial repulsion: if $G_3\Pi_{ph}
(Q)
> G_4\Pi_{ph} (0)$, one can imagine a situation when $G_3\Pi_{ph}
(Q) \approx 1$, and the correction term in (\ref{16_3}) becomes
large and positive and can overcome the negative first-order term.

What does it mean from physics perspective?  The condition
$G_3\Pi_{ph} (Q) =1$ implies that the spin component of the vertex
function, viewed as a function of transferred momentum, diverges.
This obviously implies an instability of a metal towards SDW order
with momentum Q. We don't need the order to develop, but we need
SDW fluctuations to be strong and to mediate pairing interaction
between fermions. Once spin-mediated interaction  exceeds bare
repulsion, the irreducible pairing interaction in the
corresponding channel becomes attractive. Notice in this regard
that we need magnetic fluctuations to be peaked at large momentum
transfer $Q$. If they are peaked at small momenta, $\Pi_{ph} (0)$
exceeds $\Pi_{ph} (Q)$, and the interaction in the singlet channel
remains repulsive.

The importance of spin fluctuations for spin-singlet pairing was
pointed out by many authors, starting from mid-80s. With respect
to the cuprates, d-wave pairing in the Hubbard model near
half-filling was first analyzed by Scalapino, Loh, and
Hirsch~\cite{Hirsch} for the Hubbard model near half-filling. They
used RPA to obtain irreducible pairing vertices in spin-singlet
and spin-triplet channels and found spin-singlet d-wave pairing to
be the dominant instability in the situation when   $\Pi_{ph}
({\bf q}) $ is peaked at $\vec{q}$ near $(\pi,\pi)$. This work and
subsequent works ~\cite{Hlubina,Ruvalds} also analyzed the role of
FS nesting for d-wave superconductivity.

In general, spin-mediate pairing interaction
can be obtained either within RPA\cite{peter,Kuroki_2,Hirsch} or,
using one of several advanced numerical methods developed over the
last decade, or just introduced semi-phenomenologically.  The
semi-phenomenological model is often called the spin-fermion
model\cite{acs}.  Quite often, interaction mediated by spin
fluctuations also critically affects single-fermion propagator
(the Green's function), and this renormalization has to be
included into the pairing problem. As another complication, the
interaction mediated by soft spin fluctuations has a strong
dynamical part due to Landau damping -- the decay of a spin
fluctuation into a particle-hole pair.  This dynamics also has to
be included into consideration, which makes the solution of the
pairing problem near a magnetic instability quite involved
theoretical problem.

There are two crucial aspects of the spin-fluctuation approach.
First, magnetic fluctuations have to develop at energies much
larger than the ones relevant for the pairing, typically at
energies comparable to the bandwidth $W$. It is crucial for
spin-fluctuation approach that SDW magnetism is the {\it only}
instability which develops at such high energies. There may be
other instabilities (e.g., charge order), but the assumption is
that they develop at small enough energies  and can be captured
within the low-energy  model with spin fluctuations already
present\cite{Efetov3,Senthil}. Second, spin-fluctuation approach
is fundamentally not a weak coupling approach. In the absence of
nesting, $\Pi_{ph} (Q)$ and $\Pi_{pp} (0)$ are generally of order
$1/W$, and $\Pi_{ph} (Q)$ is only larger numerically. Then the
interaction $G_{3}$, required to get a strong
magnetically-mediated component of the pairing interaction, must
be of order $W$.

Generally one way to proceed in this situation is to introduce the
spin-fermion model with static magnetic fluctuations built into
it, and then assume that within this model the interaction between
{\it low-energy} fermions ${\bar g}$ is smaller than $W$ and do
controlled low-energy analysis treating $\bar{g}/W$ as a small
parameter\cite{acs,Efetov3,Senthil}. There are several ways to
make the assumptions ${\bar g} \ll W$ and $G \sim W$ consistent
with each other, e.g., if microscopic interaction has length
$\Gamma_0$ and $\Gamma_0k_F/\hbar \gg 1$, then ${\bar g}$ is small
in $1/(\frac{\Gamma_0k_F}{\hbar})$ compared to $G$
(Refs.\cite{cm,Dzero}). At the same time, the properties of the
spin-fermion model do not seem to crucially depend on ${\bar g}/W$
ratio, so the hope is that, even if the actual ${\bar g}$ is of
order $G \sim W$, the analysis based on expansion in ${\bar g}/W$
captures the essential physics of the pairing system behavior near
a SDW instability in a metal.

We will not discuss in detail spin-fluctuation approach, which
requires a separate review. Instead, we ask the question can we
possibly get an attraction in at least one pairing channel already
at {\it weak} coupling, despite that $G_4 > G_3$, i.e., the bare
pairing interaction is repulsive in all channels.  The answer is
actually yes, it is possible, but under a special condition that
$\Pi_{ph} (Q)$ is singular and diverges logarithmically at zero
frequency or zero temperature, in the same way as the
particle-particle bubble $\Pi_{pp} (0)$.  This condition is
satisfied exactly when there is a perfect nesting between
fermionic excitations separated by $Q$. A situation with a perfect
nesting can be found for all three examples for which we analyzed
KL mechanism (another example is quasi-1D organic
conductor~\cite{rg2}).  For Fe-pnictides, it implies that hole and
electron FSs perfectly match each other when one is shifted by
$Q$, for cuprates and doped graphene nesting implies the existence
of parallel pieces of the FS.

We show below that $\Pi_{ph} (Q)$ and $\Pi_{pp} (0)$ do have
exactly the same logarithmic singularity at perfect nesting. At
the moment, let's take it for granted and compare the relevant
scales. First, no fluctuations develop  at energies/temperatures
of order $W$ because at such high scales the logarithmical
behavior of $\Pi_{pp}$ and $\Pi_{ph}$ is not yet developed and
both bubbles scale as $1/W$.  At weak coupling $G/W <<1$, hence
corrections to bare vertices are small. Second, we know that the
pairing vertex evolves at $(G_3-G_4)\Pi_{pp} (0) \sim O(1)$, and
that corrections to the bare irreducible pairing vertex become of
order one when $G_3\Pi_{ph} (Q) \sim O(1)$.  But we also know
from,  e.g., (\ref{16_5}) that at the same scale the SDW vertex
begins to evolve. Moreover other inter-pocket interactions, which
we didn't include so far: density-density and exchange
interactions (which here and below we label  as $G_1$ and $G_2$,
respectively) also start evolving because their renormalization
involves terms $G_1 \Pi_{ph} (Q)$ and $G_2\Pi_{ph} (Q)$ which also
become of $O(1)$ when all bare interactions are of the same order.
Once $G_{1,2} \Pi_{ph} (Q)$ becomes of order one, the
renormalization of $G_3$ by $G_1$ and $G_2$ interactions also
becomes relevant. The bottom line here is that renormalization of
all interactions become relevant at the same scale. At this scale
we can expect superconductivity, if the corrections to $G_4-G_3$
overcome the sign of the pairing interaction, and at the same time
we can expect an instability towards SDW and, possibly, some other
order. The issue then is whether it is possible to construct a
rigorous description of the system behavior in the situation when
all couplings are small compared to $W$, but $G_i \Pi_{ph} (Q)$
and $G_i\Pi_{pp} (0)$ are of order one. The answer is yes, and the
corresponding procedure is called a parquet renormalization group
(pRG).

To re-iterate:  the pRG approach is a controlled weak coupling
approach.  It assumes that no correlations develop at energies
comparable to the bandwidth, but that there are several competing
orders whose fluctuations develop simultaneously at a smaller
scale. Superconductivity is one of them, others include SDW and
potential charge-density-wave (CDW), nematic and other orders. The
pRG approach treats superconductivity, SDW, CDW and other
potential instabilities on equal  footings. Correlations in each
channel grow up with similar speed, and fluctuations in one
channel affect the fluctuations in the other channel and vise
versa. For superconductivity, once the corrections to the pairing
vertex become of order one, and there is a potential to convert
initial repulsion into an attraction. We know that second-order
contribution to the pairing vertex from SDW channel works in the
right direction, and one may expect that higher-order corrections
continue pushing the pairing interaction towards an attraction.
However even if attraction develops, there is no guarantee that
the system will actually undergo a SC transition because it is
entire possible that SDW instability comes before SC instability.

The pRG approach addresses both of these issues. It can be also
applied to a more realistic case of non-perfect nesting if
deviations from nesting are small in the sense that there exist a
wide range of energies where $\Pi_{ph} (Q)$ and $\Pi_{pp} (0)$ are
approximately equal.  Below some energy scale, $\omega_0$, the
logarithmical singularity in $\Pi_{ph} (Q)$ is cut.  If this scale
is smaller than the one at which the leading instability occurs, a
deviation from a perfect nesting is an irrelevant perturbation. If
it is larger, then pRG runs up to $\omega_0$, and at smaller
energies only SC channel continues to evolve in BCS fashion.

There also exists a well-developed numerical computational
procedure called functional RG (fRG)\cite{frg1,frg2}. Its
advantage is that it is not restricted to a small number of
patches and captures the evolution of the interactions in various
channels  even if the interactions depend on the angles along the
FS. The ``price" one has to pay is the reduction in the control
over calculations -- fRG includes both leading and subleading
logarithmical terms. If only logarithmical terms are left, the
angle dependence of the interaction does not change in the RG
flow, only the overall magnitude changes\cite{RG_SM} So far, the
results of fRG and pRG analysis for various systems fully agree.
Below we focus on the pRG approach. For a thorough tutorial on the
RG technique, we direct the reader to Ref. \cite{shankar}.

\subsection{Parquet Renormalization Group}

We follow the same order of presentation as before --  first
consider Fe-pnictides and then discuss patch models for cuprates
and doped graphene. We recall that in Fe-pnictides a bubble with
momentum transfer $Q$ contains one hole (c) and one electron (f)
propagator, and at perfect nesting the dispersions of holes and
electrons are just opposite. $\varepsilon_c (k) =-\varepsilon_f
(k+Q)$.  The particle-hole and particle-particle bubbles are

\bea \label{eq:bubbles} \Pi_{pp}(0)  &=& -i \int \frac{d^2
k\;d\omega}{(2\pi\hbar)^3}
G^c(k,\omega)G^c(-k,-\omega) \nonumber\\
\Pi_{ph}(Q) &=&i \int \frac{d^2 k\;d\omega}{(2\pi\hbar)^3}
G^c(k,\omega)G^f(Q+k,\omega) \eea where

$G^{c,f} = \frac{1}{\omega - \varepsilon^{c,f}_k/\hbar+i\delta
sgn(\omega)}$. Substituting into Eq. \ref{eq:bubbles} and using
$\varepsilon_c (k) =-\varepsilon_f (k+Q)$ one can easily make sure
that the two expressions in Eq. \ref{eq:bubbles} are identical.
Evaluating the integrals we obtain \beq \Pi_{pp}(0) = \Pi_{ph}(Q)
= N_0 L +... \eeq where $N_0 = m/2\pi\hbar^2$ is the 2D density of
states, \beq L = \frac{1}{2} \log \left( \frac{W}{E} \right), \eeq

$E$ is typical energy of external fermions, and the dots stand for
non-logarithmic terms.  The factor $1/2$ is specific to the pocket
model and accounts for the fact that for small pocket sizes, the
logarithm comes from integration  over positive energies $W > E >
E_F$. At non-perfect nesting, the particle-particle channel is
still logarithmic, but the particle-hole channel gets cut by the
energy difference ($\delta E$) associated with the nesting
mismatch, such that

\beq \Pi_{ph}(Q) = N_0 \log \frac{W}{\sqrt{E^2 + \delta E^2}} \eeq

The main idea of pRG (as of any RG procedure) is to consider $E$
as a running variable, assume that initial $E$ is comparable to
$W$ and $G_i \log \left( \frac{W}{E} \right) = G_i L$ is small,
calculate the renormalizations of all couplings by fermions  with
energies larger than $E$, and find how the couplings evolve as $E$
approaches the region where $G_i L = O(1)$.

This procedure can be carried out already in BCS theory, because
Cooper renormalizations are logarithmical. For an isotropic
system, the evolution of the interaction $U_l$ in a channel with
angular momentum $l$ due to Cooper renormalization can be
expressed in RG treatment as \beq \frac{d U^{full}_l}{dL} = - N_0
\left(U^{full}_L\right)^2. \label{17_1} \eeq
  The solution of (\ref{17_1}) is
\beq U^{full}_l (L)  = \frac{U_l}{1 + U_l N_0 L} \eeq

which is the same as Eq. (\ref{sa_1}).  Similar formulas can be
obtained in lattice systems when there are no competing
instabilities, i.e., only renormalizations in the pairing channel
are relevant. For example, in the two-pocket model for the
pnictides, the equations for the full vertices $\Gamma^{full}_{hh}
= -G^{full}_4$ and $\Gamma^{full}_{he} = - G^{full}_3$, Eqs.
(\ref{4_2_1}),  can be reproduced by solving the two coupled RG
equations \bea
&&\frac{d G^{full}_3}{dL} = -2 N_0 G^{full}_3 G^{full}_4 \nonumber \\
&&\frac{d G^{full}_4}{dL} = -N_0 \left(\left(G^{full}_3\right)^2 +
\left(G^{full}_4\right)^2\right) \label{17_2} \eea with boundary
conditions $G^{full}_4 (L=0) = G_4$, $G^{full}_3 (L=0) = G_3$. The
set can be factorized by introducing $G^{full}_A = G^{full}_3 +
G^{full}_4$ and $G^{full}_B = G^{full}_4 - G^{full}_3$
 to
\beq \frac{d G^{full}_A}{dL} = -N_0 \left(G^{full}_A\right)^2,
~~\frac{d G^{full}_B}{dL} = -N_0 \left(G^{full}_B\right)^2 \eeq
The solution of the set yields \bea
&&G^{full}_A = G^{full}_4 + G^{full}_3 = \frac{G_3 + G_4}{1 + N_0 L (G_3 + G_4)} \nonumber \\
&&G^{full}_B = G^{full}_4 - G^{full}_3 = \frac{G_4-G_3}{1 + N_0 L
(G_4 - G_3)} \eea Solving this set and using $\Gamma^{full}_{hh} =
-G^{full}_4$, $\Gamma^{full}_{he} = -G^{full}_3$, we reproduce
(\ref{4_2_1}). This returns us to the same issue as we had before,
namely if $G_4 > G_3$, the fully renormalized pairing interaction
does not diverge at any $L$ and in fact decays as $L$ increases:
$G^{full}_4$ decays as $1/L$ and $G^{full}_3$ decays even faster,
as $1/L^2$.

We now consider how things change when $\Pi_{ph} (Q)$ is also
logarithmical and the renormalizations in the particle-hole
channel have to be included on equal footings with
renormalizations in the particle-particle channel.

\subsubsection{pRG in a 2-pocket model}

\begin{figure}[t]
\includegraphics[width = 2.7in]{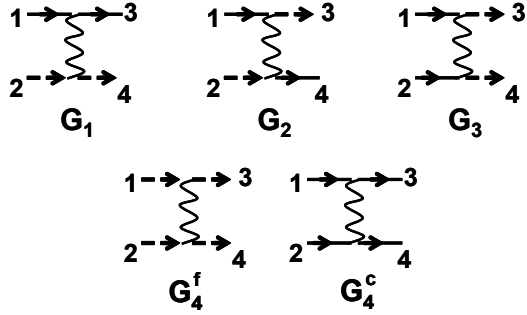}
\caption{\label{fig:couplings}The couplings $G_1$ (inter-pocket
density-density interaction), $G_2$ (fermion exchange), $G_3$
(pair hopping), $G_4^c$ and $G_4^f$ (intra-pocket density-density
interaction). For equivalent hole pockets $G_4^c=G_4^f=G_4$. The
solid lines correspond to hole Green's functions and the dashed
lines to electron Green's functions.}
\end{figure}

Because two types of renormalizations are relevant, we need to
include into consideration all vertices with either small total
momentum or with momentum transfer near $Q$ i.e., use the full
low-energy Hamiltonian of Eq. (\ref{eq:2-poc Ham}). There are
couplings $G_3$ and $G_4$ which are directly relevant for
superconductivity, and also the couplings $G_1$ and $G_2$ for
density-density and exchange interaction between hole and electron
pockets, respectively. These are shown in Fig \ref{fig:couplings}.

The strategy to obtain one-loop pRG equations, suitable to our
case, is the following: One has to start with perturbation theory
and obtain the variation of each full vertex $\delta G_i$ to order
$G_iG_j L$. Then one has to replace $\delta G_i/L$ by $d
G^{full}_i/dL$ and  also replace $G_i G_j$ in the r.h.s. by
$G^{full}_i G^{full}_j$. The result is the set of coupled
differential equations  for $d G^{full}_i/dL$ whose right sides
are given by bilinear combinations of  $G^{full}_i G^{full}_j$.
The procedure may look a bit formal, but one can rigorously prove
that it is equivalent to summing up series of corrections to $G_i$
in powers of $G_i L$, neglecting corrections terms with higher
powers of $G_i$ than of $L$. One can go further  and collecting
correction terms of order $G_i G_j G_k  L$. This is called 2-loop
order, and 2-loop terms give contributions of order $(G^{full})^3$
to the right side of the equations for $d G^{full}_i/dL$.  2-loop
calculations are, however, quite involved\cite{brazil} and below
we only consider 1-loop pRG equations.

\begin{figure*}[htp]
$\begin{array}{cc}
\includegraphics[width = 2.5in]{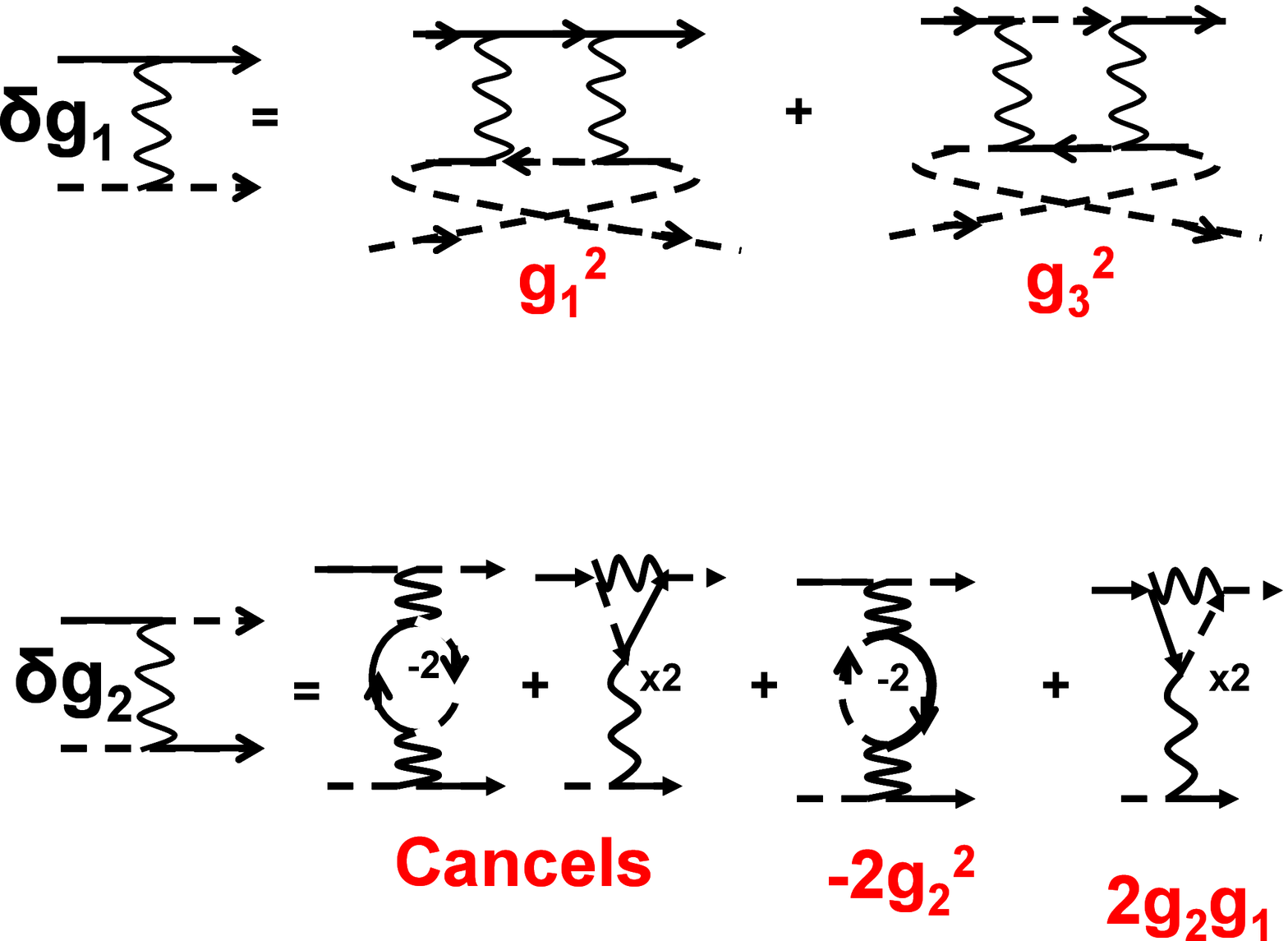}&
~~~~~~\includegraphics[width = 2.5in]{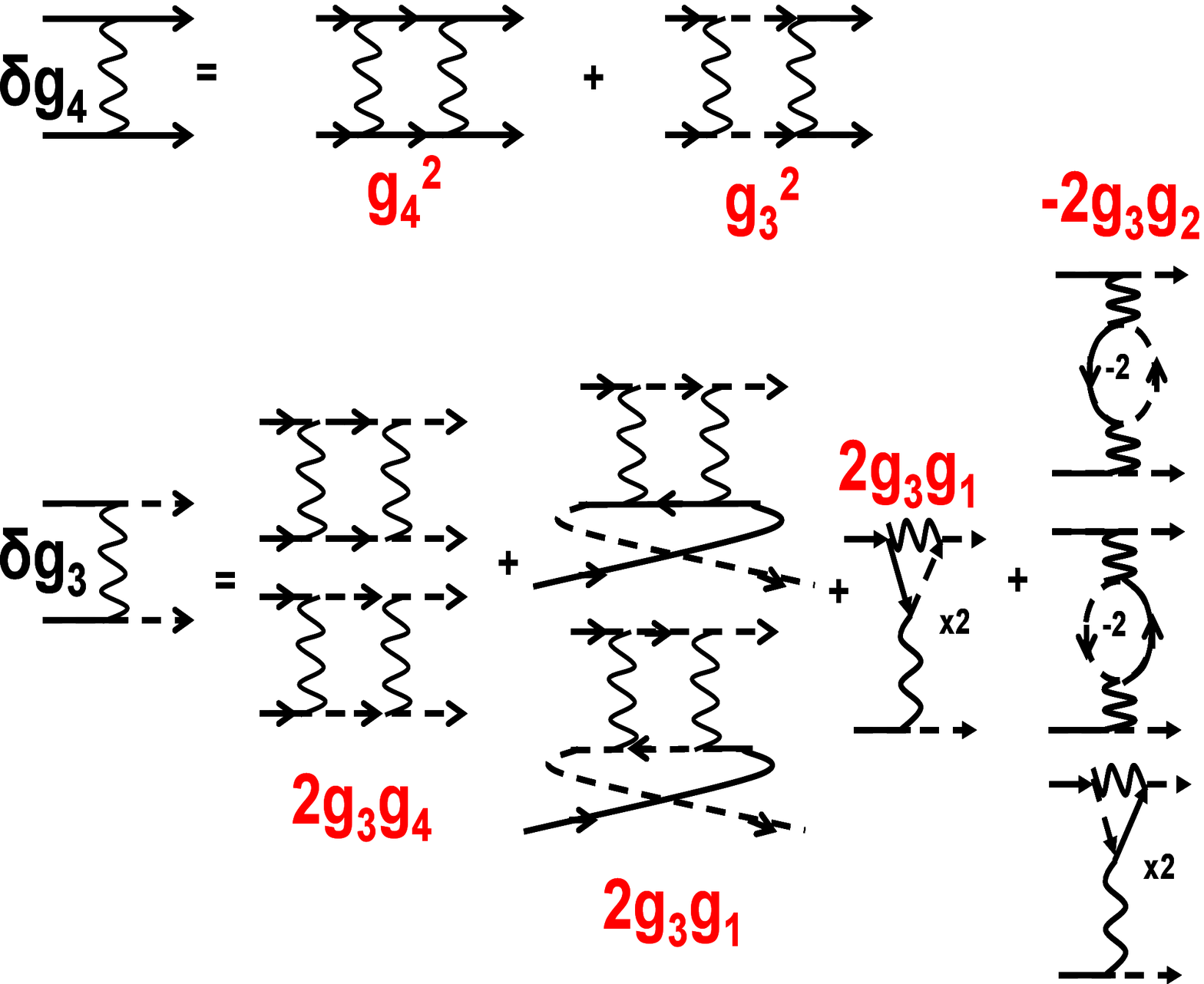}
\end{array}$
\caption{\label{fig:g_3g_4}The diagrams to 1 loop order, which
contribute to the parquet flow of  $g_1$, $g_2$, $g_3$ and $g_4$
vertices.}
\end{figure*}

The $G^2$ corrections to all four couplings are shown in
Fig.\ref{fig:g_3g_4}. Evaluating the integrals and following the
recipe we obtain
\begin{eqnarray}\label{eq:RG_2poc}
&&\dot{g}_1 = g_1^2 + g_3^2\nonumber \\
&&\dot{g}_2 = 2  g_2(g_1 - g_2 ) \nonumber \\
&&\dot{g}_3 =  2 g_3(2 g_1 - g_2 - g_4)\nonumber \\
&&\dot{g}_4 = -g_3^2 - g_4^2 \nonumber \\
\end{eqnarray}
where we introduced $g_i = G^{full}_i N_0$ and $\dot{g}_i =
dg_i/dL$

\begin{figure}[t]
\includegraphics[width = 4in]{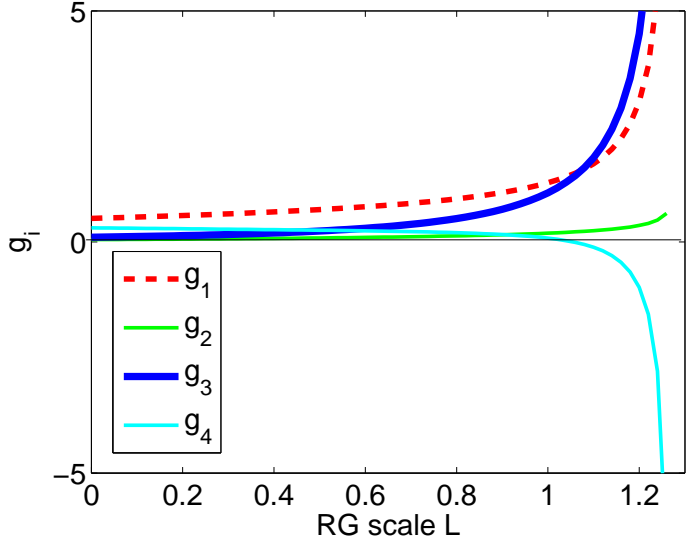}
\caption{\label{fig:Fe_flow} The flow of dimensionless couplings
$g_{1,2,3,4}$. $g_3$ grows and eventually crosses $g_4$, which
becomes negative at a large enough RG scale.}
\end{figure}

We note that the renormalizations of $g_4$ are still only in the
Cooper channel and causes $g_4$ to reduce. But for $g_3$ we now
have a counter-term from $g_1$, which pushes $g_3$ up. And the
$g_1$ term is in turn pushed up by $g_3$. Thus already at this
stage one can qualitatively expect $g_3$ to eventually get larger.
Fig \ref{fig:Fe_flow} shows the solution of (\ref{eq:RG_2poc})--
the flow of the four couplings for this model. We see that, even
if $g_3$ is initially smaller than $g_4$, it flows up with
increasing $L$, while $g_4$ flows to smaller values. At some
$L_0$, $g_3$ crosses $g_4$, and at larger $L$ the pairing
interaction $g_4-g_3$ becomes negative (i.e., attractive). In
other words, in the process of pRG flow, the system self-generates
attractive pairing interaction. We remind that the attraction
appears in the $s^{+-}$ channel. The pairing interaction in
$s^{++}$ channel: $g_3 + g_4$ remains positive (repulsive) despite
that $g_4$ eventually changes sign and becomes negative. It is
essential that for $L \sim L_0$ the renormalized $g_i$ are still
of the same order as bare couplings, i.e., are still small, and
the calculations are fully under control. In other words, the sign
change of the pairing interaction is a solid result, and
higher-loop corrections may only slightly shift the value of $L_0$
when it happens.

At some larger $L= L_c$, the couplings diverge,  signaling the
instability towards an ordered state (which one we discuss later).
One-loop pRG is valid "almost" all the way to the instability, up
to $L_c - L \sim O(1)$, when the renormalized $g_i$ become of
order one.  At smaller distances from $L_c$ higher-loop
corrections become relevant. It is very unlikely, however, that
these corrections will change the physics in any significant way.

The sign change of the pairing interaction can be detected also if
the nesting is not perfect and $\Pi_{ph} (Q)$ does not behave
exactly in the same way as $\Pi_{pp} (0)$.  The full treatment of
this case is quite involved. For illustrative purposes we  follow
the approach first proposed in Ref.\cite{Furukawa} and measure the
non-equivalence between $\Pi_{pp} (0)$ and $\Pi_{ph} (Q)$ by
introducing a phenomenological parameter $d_1 = \Pi_{ph}
(Q)/\Pi_{pp} (0)$ and treat $d_1$ as an $L-$ independent constant
$0<d_1<1$, independent on $L$.  This is indeed an approximation,
but it is at least partly justified by our earlier observation
that the most relevant effect for the pairing is the sign change
of $g_4-g_3$ at some scale $L_0$, and around this scale $d_1$ is
not expected to have strong dependence on $L$.  The case $d_1 =1$
corresponds to perfect nesting, and the case $d_1 =0$ implies that
particle-hole channel is irrelevant,  in which case, we remind,
$g_4-g_3$ remains positive for all $L$.

\begin{figure*}[t]
$\begin{array}{ccc}
\includegraphics[width = 2.2in]{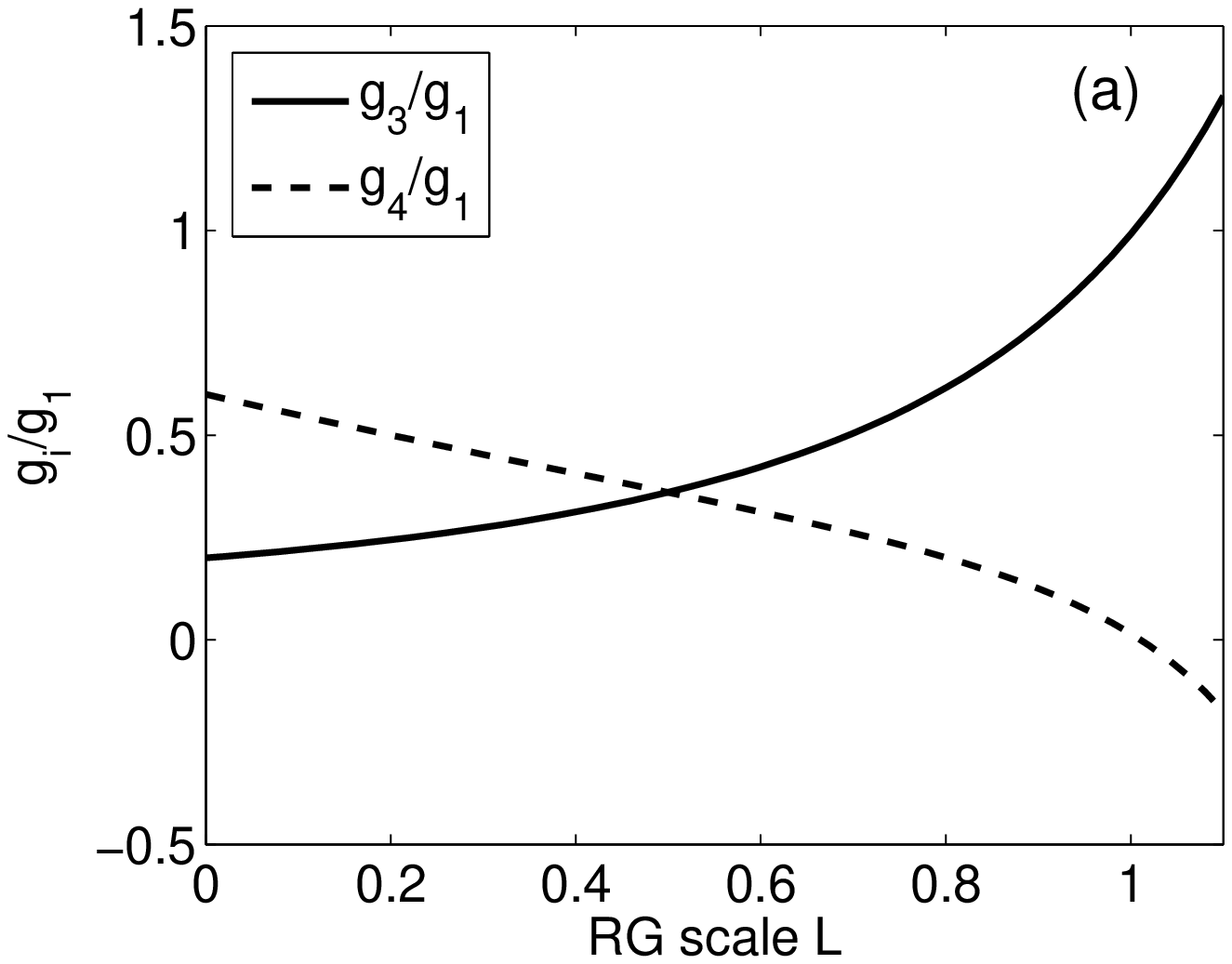}&
\includegraphics[width = 2.2in]{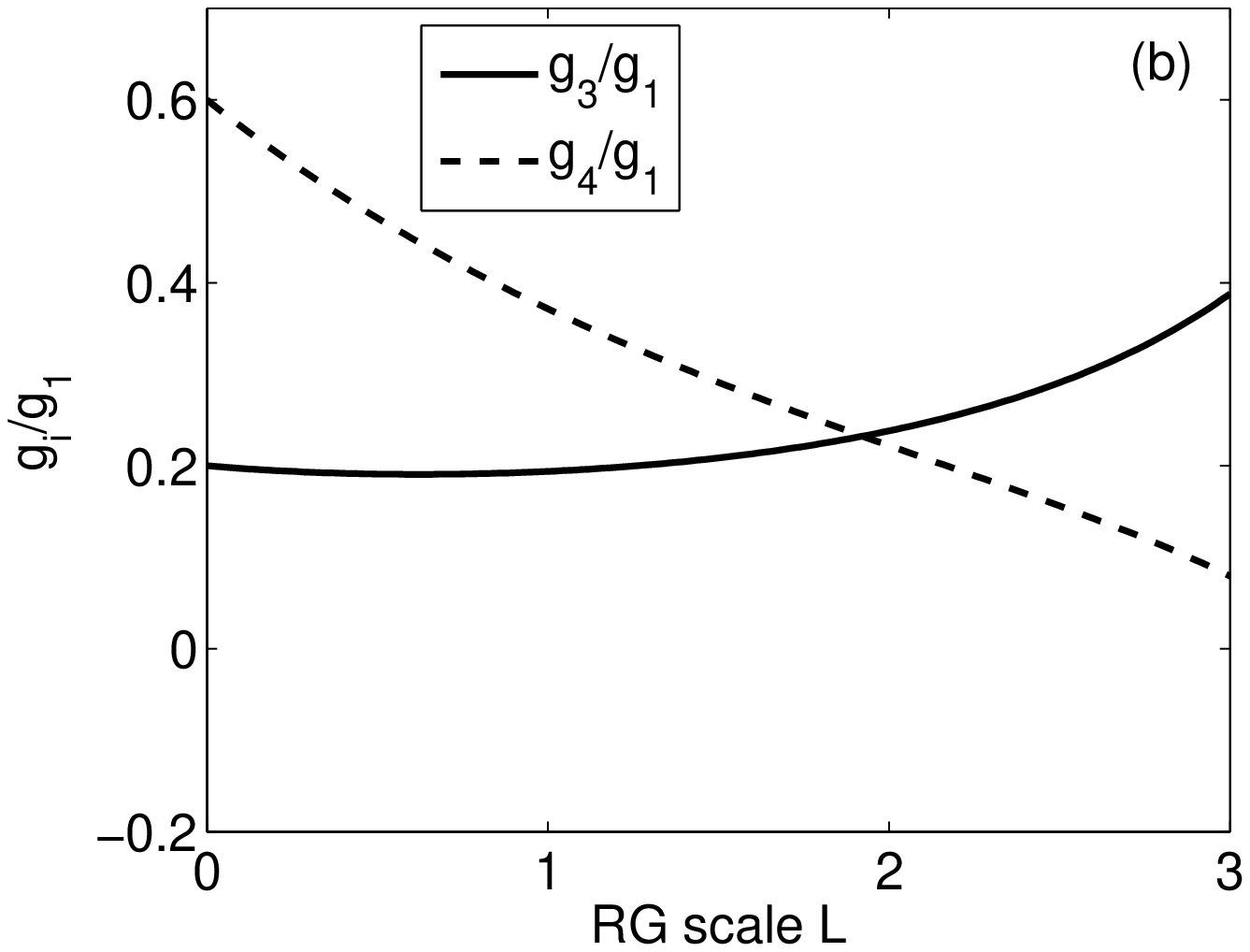}&
\includegraphics[width = 2.2in]{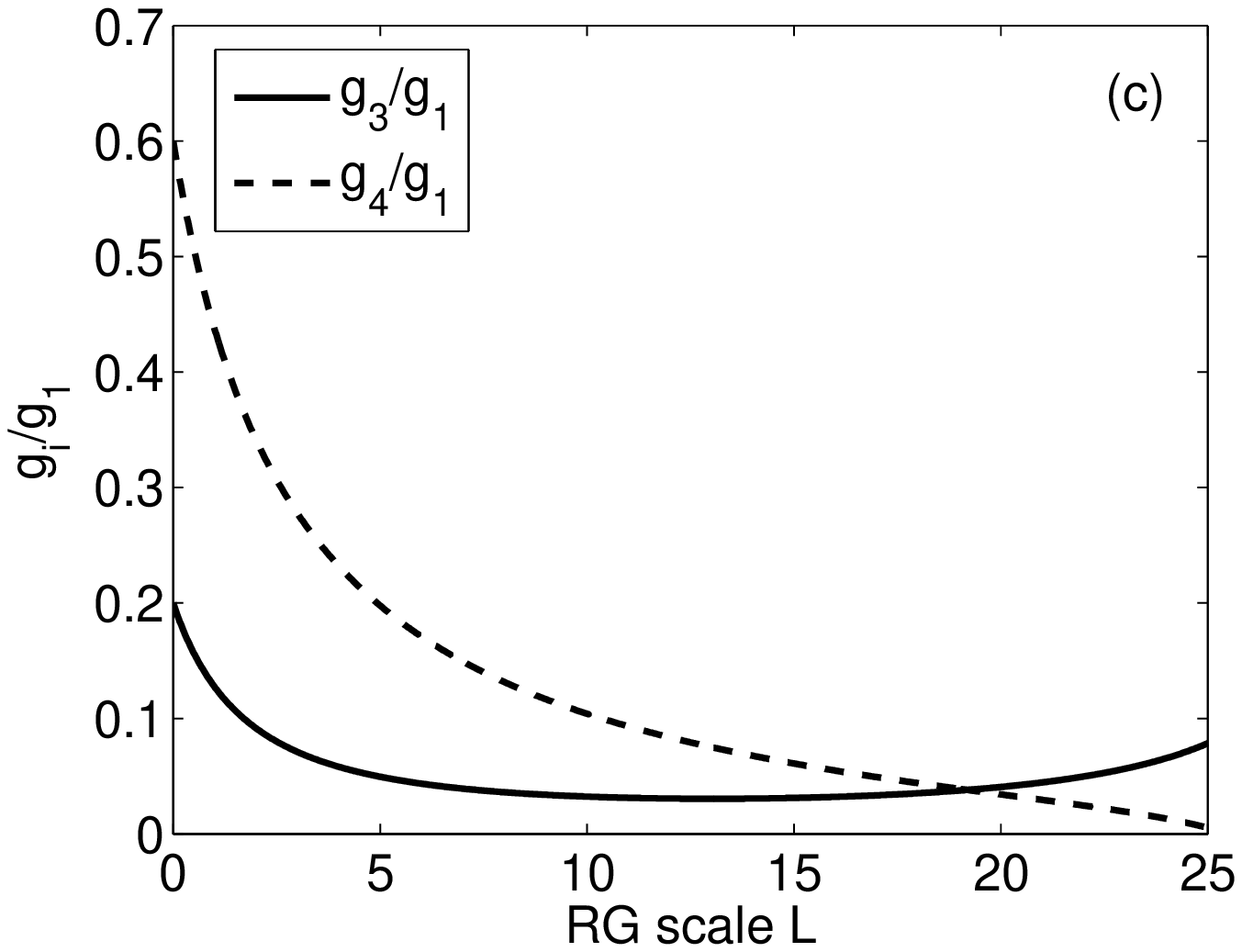}
\end{array}$
\caption{\label{fig:Fe_flows} The flow of ratio of couplings
$g_3/g_1$ and $g_4/g_1$ for different nesting parameters
$d_1=1$(a), $d_1=0.3$(b),$d_1=0.05$(c). All cases are
qualitatively similar in that $g_3/g_1$ eventually crosses
$g_4/g_1$. The smaller is the nesting parameter, the `later' is
this crossing. If $d_1=0$, this crossing will never happen and
$g_4>g_3$ for all $L$.}
\end{figure*}

The pRG equations for arbitrary $d_1$ are straightforwardly
obtained using the same strategy as in the derivation of
(\ref{eq:RG_2poc}), and the result is
\begin{eqnarray}\label{eq:RG_2poc_1}
&&\dot{g}_1 = d_1(g_1^2 + g_3^2)\nonumber \\
&&\dot{g}_2 = 2 d_1 g_2(g_1 - g_2 ) \nonumber \\
&&\dot{g}_3 =  2 d_1g_3(2 g_1 - g_2)- 2g_3 g_4\nonumber \\
&&\dot{g}_4 = -g^2_3 - g^2_4 \nonumber \\
\end{eqnarray}
In Fig \ref{fig:Fe_flows} we show the behavior of the couplings
for representative $0<d_1<1$. Like before, we take bare value of
$g_4$ to be larger than the bare $g_3$, i.e., at high energies the
pairing interaction is repulsive. This figure and analytical
consideration shows that for {\it any} non-zero $d_1$ the behavior
is qualitatively the same as for perfect nesting, i.e., at some
$L_0 < L_c$  the running couplings $g_3$ and $g_4$ cross, and for
larger $L$ (smaller energies) pairing interaction in $s^{+-}$
channel becomes attractive. The only effect of making $d_1$
smaller is the increase in the value of $L_0$. Still, for
sufficiently small bare couplings, the range where the pairing
interaction changes sign is fully under control in 1-loop pRG
theory.

A way to see analytically that $g_3-g_4$ changes sign and becomes
positive is to consider the system behavior near $L = L_c$ and
make sure that in this region $g_3 > g_4$. One can easily make
sure that all couplings diverge at $L_c$, and their ratios tend to
some constant values (see discussion around Eq. (\ref{eq:NL1})
below for more detail). Introducing $g_2 = a g_1, g_3= b g_1$, and
$g_4 = c g_1$, and substituting into (\ref{eq:RG_2poc_1}) we find
an algebraic set of equations for $a, b$, and $c$. Solving the
set, we find that
$b=\frac{\sqrt{\sqrt{16d_1^4-4d_1^2+4}+2-d_1^2}}{d_1}$and $c =
\frac{d_1}{2}(3-b^2)$. The negative sign of $c$ and positive sign
of $b$, combined with the fact that $g_1$ definitely increases
under the flow and surely remains positive,  imply that near
$L_c$, $g_4$ is negative, while $g_3$ is positive (this is also
evident from the Fig \ref{fig:Fe_flows}). Obviously then, $g_3$
and $g_4$ must cross at some $L_0 < L_c$.

\subsubsection{pRG in  patch models}

We now  show that similar behavior holds in patch models. Since
the only difference between patch models for cuprates and for
graphene  is the number of patches (2 vs 3), we consider a generic
model of $n$ - patches with fermion-fermion interaction in the
form \bea \label{eq: action} &&H_{\text{int}} =\frac12
\sum_{\alpha=1}^n G_4 \psi^{\dag}_{\alpha }\psi^{\dag}_{\alpha}
\psi_{\alpha}
\psi_{\alpha} \nonumber\\
&&+\frac{1}{2} \sum_{\alpha \neq \beta} \big[G_1
\psi^{\dag}_{\alpha }\psi^{\dag}_{\beta} \psi_{\beta}
\psi_{\alpha}+G_2 \psi^{\dag}_{\alpha }\psi^{\dag}_{\beta}
\psi_{\alpha} \psi_{\beta} + G_3\psi^{\dag}_{\alpha
}\psi^{\dag}_{\alpha} \psi_{\beta} \psi_{\beta}\big]\nonumber\\
\eea

Keeping again all  diagrams which diverge logarithmically, we end
up with the following set of pRG equations (using the same
notations as before)

\bea\label{eq:n_RG}\dot{g}_1&=&d_1(g_1^2+g_3^2)\nonumber\\
\dot{g}_2&=&2d_1g_2(g_1-g_2)\nonumber\\
\dot{g}_3&=&-(n-2)g_3^2-2g_3g_4+2d_1g_3(2g_1-g_2)\nonumber\\
\dot{g}_4&=&-(n-1)g_3^2-g_4^2\eea

The equations look similar to the ones for the pocket model, up to
the dependence on $n$, but there is one important difference: the
derivative in the l.h.s. is with respect to $\log^2(\Lambda/E)$
rather than a first power of the logarithm. The extra logarithm
comes from the logarithmical enhancement of the density of states
near Van-Hove density. The presence of extra logarithms makes the
theory somewhat less controlled because already at second order
there are terms of order $g^2 \log^2$ and $g^2 \log$.  The set of
equations (\ref{eq:n_RG}) corresponds to keeping $g^2 \log^2$
neglecting $g^2 \log$ terms, and $\dot{g}_i$ in (\ref{eq:n_RG}) is
$\dot{g}_i = dg_i/d \log^2(\Lambda/E)$.
 Strictly speaking, this implies that RG scheme can be applied only
  at one-loop order, while extending  Eq, \ref{eq:n_RG} to two-loop and higher orders will require one to go beyond RG.

Like before, $d_1$, subject to $0< d_1 <1$, accounts for relative
strength of $\Pi_{ph} (Q)$ compared to $\Pi_{pp} (0)$.  In
reality, $d_1 = \Pi_{ph} (Q)/\Pi_{pp} (0)$ depends on the running
scale $L = \log^2(\Lambda/E)$, but we approximate it by a
constants using the same reasoning as for the pocket model.

We show the solution of the set (\ref{eq:n_RG}) in Fig. \ref{fig:
rgintegration} for $n=3$ ($n=2$ result is identical to Fig.
\ref{fig:Fe_flow}).
Combining again the numerical analysis and the analytical
reasoning similar to the one for the pocket model, we find that,
for any $n$ and any $d_1
>0$, there exists a scale $L_0$ at which $g_3$ and $g_4$ cross,
and at larger $L$ (i.e., at smaller energies) the pairing
interaction in the d-wave channel (for which the pairing vertex is
proportional to  $g_4-g_3$) changes sign and becomes attractive.

\begin{figure}[t]
\includegraphics[width = \columnwidth]{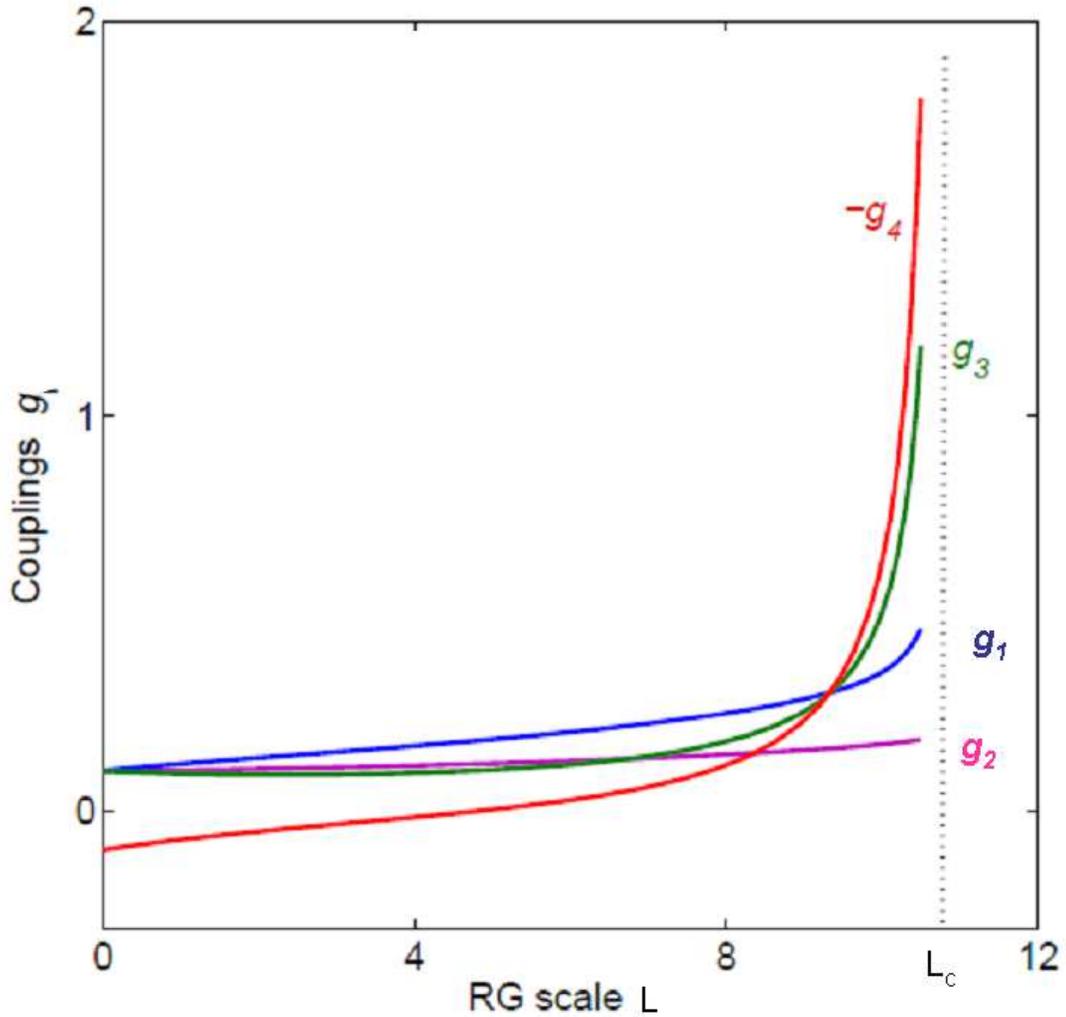}
\caption{ The flow of the couplings vs the pRG scale $L$ in the
3-patch model. We assume that all couplings are repulsive. We see
that $g_3$ increases under the flow, while $g_4$ decreases.
Observe that the coupling $g_4$ eventually gets overscreened and
changes sign. Taken from Ref. \cite{andreyd+id}. \label{fig:
rgintegration}}
\end{figure}

The outcome of these studies is that in all three systems which we
considered, the system self-generates attraction below some
particular energy $E_0$, which is of order $\Lambda e^{-1/(N_0
G)}$ for the pocket model and of order  $\Lambda e^{-1/(N_0
G)^{1/2}}$ for the patch models.

The reason for the sign change of the pairing interaction is clear
from the structure of the pRG equation for $g_3$ the r.h.s. of
which contains the term $4 d_1 g_3 g_4$, which pushes $g_3$ up. We
know from second-order KL calculation that the upward
renormalization of $g_3$ comes from the magnetic channel and can
be roughly viewed as the contribution from spin-mediated part of
effective fermion-fermion interaction.  Not surprisingly, we will
see below that $g_1$ does, indeed, contribute to the SDW vertex.
From this perspective, the physics of the attraction in pRG (or in
fRG, which brings in the same conclusions as pRG) and in
spin-fermion model is the same:  magnetic fluctuations push
inter-pocket/inter-patch interaction up, and below some energy
scale the renormalized inter-pocket/inter-patch interaction
becomes larger than repulsive intra-pocket/intra-patch
interaction.

There is, however, one important difference between the RG
description and the description in terms of spin-fermion model. In
the spin-fermion model, magnetic fluctuations are strong, but the
system is {\it assumed} to be
at
 some distance away from an SDW
instability. In this situation, SC instability definitely comes
ahead of SDW magnetism. There may be other instabilities produced
by strong spin fluctuations, like bond
CDW\cite{Max,efetov,Physics}, which compete with SC and, by
construction, also occur before SDW order sets in
In RG treatment (pRG or fRG), SDW magnetism and SC instability
(and other potential instabilities) compete with each other, and
which one develops first needs to be analyzed.  So far, we only
found that SC vertex changes sign and becomes attractive. But we
do not know whether superconductivity is the leading instability,
or some other instability comes first. This is what we will study
next. The key issue, indeed, is whether superconductivity can come
ahead of SDW magnetism, whose fluctuations helped convert
repulsion in the pairing channel into an attraction.

\section{Competition between density wave orders and superconductivity}

Thus far, we identified an instability in a particular channel
with the appearance of a pole in the upper frequency half-plane in
the corresponding vertex -- the vertex with zero total momentum in
the case of SC instability, and the vertex with the
 total momentum $Q$ in the case of SDW instability.
Since our goal is to address the competition between these states,
it is actually advantageous to use a slightly different approach:
introduce all potentially relevant fluctuating fields, use them to
decouple 4-fermion terms into a set of terms containing two
fermions and a fluctuating field, compute the renormalization of
these ``three-legged" vertices and use these renormalized vertices
to obtain the susceptibilities in various channels and check which
one is the strongest.  We will see that the renormalized vertices
in different channels (most notably, SDW and SC) do diverge near
$L_c$, but with different exponents. The leading instability will
be in the channel for which the exponent is the largest. There is
one caveat in this approach --- for a divergence of the
susceptibility the exponent for the vertex should be larger than
$1/2$ (Ref.\cite{oskar}), but we will see below that this
condition is satisfied for all three cases which we consider, at
least for the leading instability.

\subsection{Two pocket model}

\begin{figure}[t]
\includegraphics[width=4in]{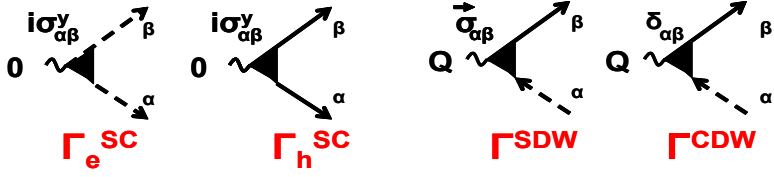}
\caption{\label{fig:new_vertex} Superconducting and density-wave
three legged vertices. Divergence of any of these vertices
indicates that the system is likely to be unstable to the
corresponding order. $\Gamma^{SC}_{h,e}$ are superconducting
vertices, $\Gamma^{SDW}$ is SDW vertex and $\Gamma^{CDW}$ is CDW
vertex.}
\end{figure}

\begin{figure}[t]
\includegraphics[width=3in]{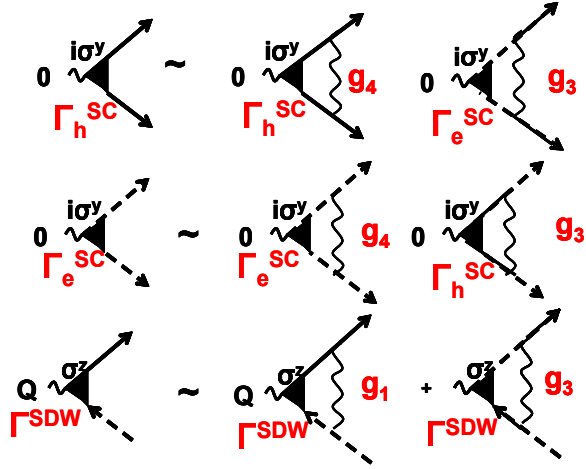}
\caption{\label{fig:new_vertex2} Diagrams to analyze the flow of
the effective vertices: SC vertex  (top two) and SDW vertex
(bottom). The couplings $g_i$'s here are running couplings in RG
sense.}
\end{figure}

Let us see how it works for a two-pocket model. There are two
particle-particle three legged vertices $\Gamma_{h,e}$ as shown in
Fig \ref{fig:new_vertex}.  To obtain the flow  of these vertices,
i.e.,  $\Gamma^{SC}_{h,e} (L)$ we assume that external fermions
and a fluctuating field have energies  comparable to some  E
(i.e.,$L = \log \Lambda/E$) and collect contributions from all
fermions with energies larger than $E$. To do this with
logarithmical accuracy we  write all possible diagrams, choose a
particle-particle cross-section at the smallest internal energy
$E' \geq E$ and sum up all contributions to the left and to the
right of this cross-section, as shown in Fig
\ref{fig:new_vertex2}. The sum of all contributions to the left of
the cross-section gives the three legged vertex at energy $E'$ (or
$L' =  \log \Lambda/E'$), and the sum of all contributions to the
right of the cross-section gives the interaction $g_i$ at energy
$L'$. The integration over the remaining cross-section gives
$\int^L d L'$ (with our normalization of $g_i$), and the equation
for, e.g., $\Gamma_{h} (L)$ becomes \bea \Gamma^{SC}_h (L) =
\int^L d L' \left(\Gamma^{SC}_h (L') g_4 (L') + \Gamma^{SC}_e(L')
g_3(L')\right) \eea Differentiating over the upper limit, we
obtain differential equation for $d \Gamma^{SC}_h (L)/dL$ whose
r.h.s. contains $\Gamma^{SC}_{h,e} (L)$ and $g_{3,4} (L)$ at the
same scale $L$.

 Collecting the contributions for $\Gamma^{SC}_h (L)$ an $\Gamma^{SC}_e (L)$ we obtain
 \bea\label{eq:run1}
\frac{d\Gamma^{SC}_h}{dL}&=&\Gamma^{SC}_h g_4 + \Gamma^{SC}_e g_3\nonumber\\
\frac{d\Gamma^{SC}_e}{dL}&=&\Gamma^{SC}_e g_4 + \Gamma^{SC}_h g_3\nonumber\\
\eea

or

\bea\label{eq:run2}
\frac{d\Gamma_{++}}{dL}&=&(g_4 + g_3) \Gamma_{++}\nonumber\\
\frac{d\Gamma_{+-}}{dL}&=&(g_4 - g_3)\Gamma_{+-}\nonumber\\
\eea

where $\Gamma_{++}\equiv\Gamma^{SC}_h+\Gamma^{SC}_e$ and
$\Gamma_{+-}\equiv \Gamma^{SC}_h-\Gamma^{SC}_e$.  The first one is
for $s^{++}$ pairing, the second is for $s^{+-}$ pairing.
 We have seen in the previous section that
the running couplings $g_{3,4}$ diverge at some critical RG scale
$L_c$. The flow equation near $L_c$ is in
 the form $\dot{g}\sim g^2$, hence
\beq\label{eq:div}g_i=\frac{\alpha_i}{L_c-L}.\eeq Substituting
this into Eq. \ref{eq:run2} and solving the differential equation
for $\Gamma$ we find that the two SC three legged vertices
 behave as
\beq\label{eq:div2}\Gamma_{s^{++}} \propto
\frac{1}{(L_c-L)^{-\alpha_3-\alpha_4}},~~\Gamma_{s^{+-}}=\frac{1}{(L_c-L)^{\alpha_3-\alpha_4}},
\eeq The requirement for the divergence of $\Gamma_{s^{+-}}$ is
$\alpha_3 >\alpha_4$, which is obviously the same as $g_3 > g_4$
(see (\ref{eq:div})).

We follow the same procedure for an SDW vertex ${\vec
\Gamma}^{SDW}$. We introduce a particle-hole  vertex with momentum
transfer $Q$ and spin factor ${\vec \sigma}_{\alpha \beta}$, as
shown in  Fig \ref{fig:new_vertex}, and obtain the equation for $d
{\vec \Gamma}^{SDW} (L)/dL$  in the same way as we did for SC
vertices. We obtain (see Fig. \ref{fig:new_vertex2})
\bea\label{eq:run3}
\frac{d{\vec\Gamma}^{SDW}}{dL}&=&d_1(g_1 + g_3){\vec \Gamma}^{SDW}\nonumber\\
\eea Using Eq. \ref{eq:div} and following the same steps as above
we obtain at $L\approx L_c$

\beq\label{eq:div3}{\vec \Gamma}^{SDW} \propto
\frac{1}{(L_c-L)^{d_1(\alpha_1+\alpha_3)}}\eeq

For CDW vertex (the one with the overall factor
$\delta_{\alpha\beta}$ instead $\sigma_{\alpha\beta}$), the flow
equation is \bea\label{eq:run4}
\frac{d\Gamma^{CDW}}{dL}&=&d_1(g_1 +g_3 -2g_3 -2g_2)\Gamma^{CDW}\nonumber\\
&=& d_1(g_1 -g_3 -2g_2)\Gamma^{CDW} \eea

Using the same procedure as before we obtain
\beq\label{eq:div3_1}\Gamma^{CDW}=\frac{1}{(L_c-L)^{d_1(\alpha_1-\alpha_3-2\alpha_2)}}\eeq

The exponents $\alpha_i$ can be easily found by plugging in the
asymptotic forms in Eq. \ref{eq:div} into the RG equations. This
gives the following set of non linear algebraic equations in
$\alpha_i$ \bea\label{eq:NL1}
\alpha_1 &=& d_1(\alpha_1^2 + \alpha_3^2)\nonumber \\
\alpha_2 &=& 2 d_1 \alpha_2(\alpha_1 - \alpha_2 ) \nonumber \\
\alpha_3 &=&  2 d_1\alpha_3(2 \alpha_1 - \alpha_2)-2\alpha_3\alpha_4\nonumber \\
\alpha_4 &=& -\alpha_3^2 - \alpha_4^2 \nonumber \\
\eea

Consider first the case of perfect nesting, $d_1=1$. The solution
of the set of equations is $\alpha_1=\frac{1}{6}$, $\alpha_2=0$,
$\alpha_3=\frac{\sqrt{5}}{6}$ and $\alpha_4=-\frac{1}{6}$;
Combining $\alpha$'s, we find that the exponents  for
superconducting and spin density wave instabilities and positive
and equal:

\bea\label{eq:expo}
\alpha_{s\pm}&\equiv&\alpha_3-\alpha_4=\frac{1+\sqrt{5}}{6}\approx 0.539\nonumber\\
\alpha_{SDW}&\equiv&\alpha_1+\alpha_3=\frac{1+\sqrt{5}}{6}\approx 0.539\nonumber\\
\eea while the exponent for CDW and $s++$ vertices are negative
\bea \alpha_{CDW} &=&
\alpha_1+\alpha_3=\frac{1-\sqrt{5}}{6}\approx -
0.206 \nonumber\\
\alpha_{s++} &=& -\alpha_3-\alpha_4=\frac{1-\sqrt{5}}{6}\approx -
0.206 \eea

We see that the superconducting ($s+-$) and SDW channels have
equal susceptibilities in this approximation, while CDW channel is
not a competitor.

\begin{figure}[htp]
\includegraphics[width=4in]{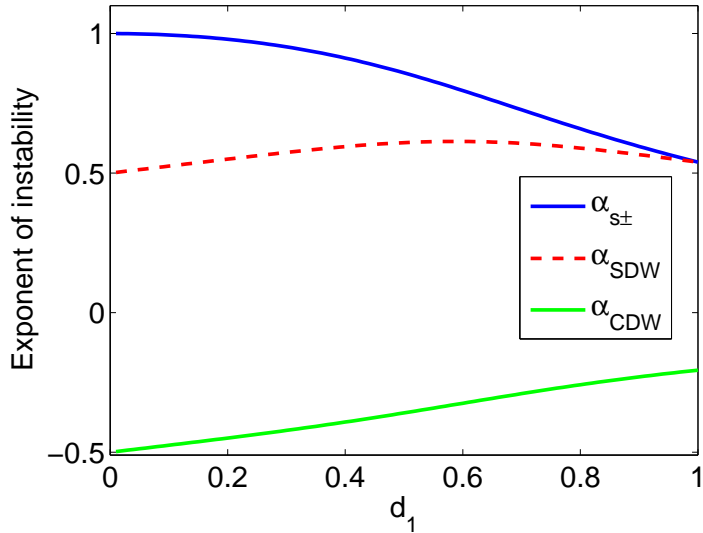}
\caption{\label{fig:alpha2} Exponents ($\alpha_{s\pm}$,
$\alpha_{SDW}$ and $\alpha_{CDW}$) for different values of the
nesting parameter $d_1$ calculated near the critical RG scale,
where the couplings diverge. The state with the largest exponent
wins. SDW and SC are degenerate when $d_1=1$ (perfect nesting) and
superconductivity wins for all other values of $d_1$. CDW is not a
competitor.}
\end{figure}

The analysis can be extended to  $d_1<1$. We  define $\beta\equiv
\alpha_4/\alpha_1$, $\gamma\equiv \alpha_3/\alpha_1$ and obtain
\bea\label{eq:expo_sol}
\gamma^2&=&\frac{\sqrt{16d_1^4-4d_1^2+4}+2-d_1^2}{d_1^2}\nonumber\\
\beta&=&\frac{d_1}{2}\left(3-\gamma^2\right)\nonumber\\
\alpha_1&=&\frac{1}{d_1}\frac{1}{1+\gamma^2} \eea In
Fig\ref{fig:alpha2} we plot $\alpha_{s\pm}=\alpha_3-\alpha_4$,
$\alpha_{SDW}=\alpha_1+\alpha_3$, and
$\alpha_{CDW}=\alpha_1-\alpha_3$, We clearly see that (i) CDW
channel is never a competitor, and (ii) as $d_1$ decreases (the
nesting gets worse), the pairing vertex diverges with a higher
exponent that SDW channel, hence $s^{+-}$ superconductivity
becomes the leading instability, overshooting the channel which
helped SC vertex to change sign in the first place.

In real systems, pRG equations are only valid up to some distance
from the instability at $L_c$. Very near $L_c$ three-dimensional
effects, corrections from higher-loop orders and other
perturbations likely affect the flow of the couplings. Besides, in
pocket models, the pRG equations are only valid for $E$ between
the bandwidth $W$ and the Fermi energy $E_F$. At $E < E_F$,
internal momenta in the diagrams which account for the flow of the
couplings become smaller than external $k_F$, and  the
renormalization of $g_i$ start depending on the interplay between
all four external momenta in the vertices\cite{rev_physica,RG_SM}.
The calculation of the flow in this case is technically more
involved, but the result is physically transparent -- SDW and
$s^{+-}$ SC channels stop talking to each other, and the vertex
evolves according to Eqs. (\ref{eq:div2}) and (\ref{eq:run3}),
with $g_i$ taken at the scale $E_F$ (or $L_F = \log \Lambda/E_F$).
If $L_F > L_c$, the presence of the scale set by the Fermi energy
is irrelevant, but if $L_F < L_c$ (which is the case for the
Fe-pnictides because superconducting $T_c$ and magnetic $T_{SDW}$
are much smaller than $E_F$), then one should stop pRG flow at
$L_{E_F}$. At perfect nesting, the SDW combination $g_1 + g_3$ is
larger than $s^{+-}$ combination $g_3-g_4$ at any $L < L_c$, hence
SDW channel wins, and the leading instability upon cooling down
the system is towards a SDW order. At  non-zero doping, $\Pi_{ph}
(Q)$ is cut by a deviation from nesting, what in our language
implies that $d_1 <1$. If bare $g_3$ and $g_4$ are not to far
apart,  there exists a critical $d_1$ at which $g_3-g_4$ crosses
$d_1 (g_1 + g_3)$ at $L_{F}$, and at larger $d_1$ the crossing
occurs before $L_F$. In this situation, $s^{+-}$ SC becomes the
leading instability upon cooling off the system.

The comparison between different channels can be further extended
by considering current SDW and CDW vertices (imaginary
$\Gamma^{SDW}$ and $\Gamma^{CDW}$) and so on. We will not dwell
into this issue because for all three cases we consider the real
competition is between SDW and SC vertices.

\begin{figure}[htp]
\includegraphics[width=4in]{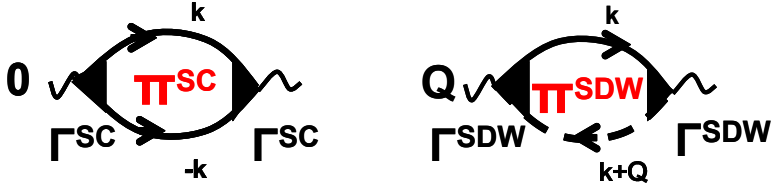}
\caption{\label{fig:fluc} (Left) The fluctuation correction to SC
pairing susceptibility. (Right) The fluctuation correction to SDW
susceptibility. 
}
\end{figure}

Before moving on, we need to clarify one more point. So far we
found that the vertices $\Gamma^{SC}$ and $\Gamma^{SDW}$ diverge
and compared the exponents. However, to actually analyze the
instability in a particular channel one has to compute fluctuation
correction to susceptibility

\beq \label{eq:susdiv}
 \chi^{i}_{fl} (L) \sim \int^L d L' \left(\Gamma^{i} (L') \right)^2
\eeq

where $\Pi_i$ is either $\Pi^{SDW} = \Pi_{ph}$ or $\Pi^{SC} =
\Pi_{pp}$ (see Fig \ref{fig:fluc})

The fully renormalized susceptibility in a given channel is \beq
\chi^{-1} (L) =  r_0 - \chi^{i}_{fl} (L) \eeq where $r_0$ is some
bare value of order one.  The true instability occurs at $L^*$
when $ \chi^{i}_{fl} (L^*) = r_0$. At weak coupling, the critical
$L^*$ is close to $L_c$, and, indeed, the instability occurs first
in the channel with the largest exponent for $\Gamma^{i}$.
However, we need $\chi^{i}_{fl} (L)$ to diverge at $L_c$,
otherwise there will no instability at weak coupling~\cite{oskar}.
This requirement sets the condition that the exponent for the
corresponding $\Gamma$ must be larger than $1/2$. Fortunately,
this  condition is satisfied in the two-pocket model. For $d_1=1$,
this is evident from (\ref{eq:expo}). For $d_1 <1$, the exponent
for the SC channel only increases, while the one in SDW channel
decreases but still remains larger than $1/2$ as it is evidenced
from Fig\ref{fig:alpha2} where we plotted the exponents for SC and
SDW vertices as a function of $d_1$. In the limit $d_1\rightarrow
0$, \beq\alpha_{SDW}\approx \frac{1}{2}+\frac{d_1}{4}\eeq.

The fact that both $\alpha_{SC}$ and $\alpha_{SDW}$ are larger
than $1/2$ implies that in Landau-Ginzburg expansion in powers of
SC and SDW order parameters ($\Delta$ and $M$, respectively), not
only the prefactor for $\Delta^2$ changes sign at $T_c$, but also
the prefactor for $M^2$ term changes sign and becomes negative
below some $T_m < T_c$. This brings in the possibility that at low
T SC and SDW orders co-exist. The issue of the co-existence,
however, requires a careful analysis of the interplay of
prefactors for fourth order terms $M^4$, $ \Delta^4$, and $M^2
\Delta^2$. We do not discuss this specific issue. For details
see~\cite{VVC,FS}

\subsubsection{Multi-pocket models}

\begin{figure}[htp]
\includegraphics[width=4in]{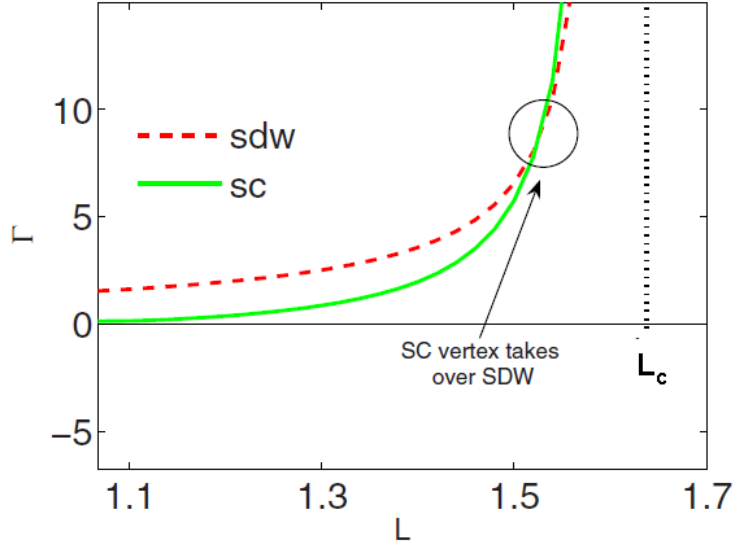}
\caption{\label{fig:RG_3poc} The flow of the SC and SDW vertices
with the RG scale. Both diverge at a critical scale, $L_c$, but
the SC vertex diverges stronger. Taken from Ref. \cite{RG_SM}.}
\end{figure}

The interplay between SDW and SC vertices is more involved in more
realistic multi-pocket models Fe-pnictides, with several electron
and hole pockets. We recall that weakly doped Fe-pnictides have 2
electron pockets and 2-3 hole pockets. In multi-pocket models one
needs to introduce a larger number of  intra-and inter-pocket
interactions  and analyze the flow of all couplings to decide
which instability is the leading one. This does not provide any
new physics compared to what we have discussed, but in several
cases the interplay between SC and SDW instabilities becomes such
that superconductivity wins already at perfect nesting. In
particular, in 3-pocket models (two electron pockets and one hole
pockets) the exponent for the SC vertex gets larger than the
exponent for the SDW vertex already at $d_1=1$. We show the flow
of SC an SDW couplings for 3-pocket model in
Fig.\ref{fig:RG_3poc}. Once $d_1$ becomes smaller than one, SC
channel wins even bigger compared to SDW channel.

Superconductivity right at zero doping has been detected in
several Fe-pnictides, like LaOFeAs and LiFeAs,  and it is quite
possible that this is at least partly due to the specifics of pRG
flow.

\subsection{Patch models}

The analysis of the patch model show a very similar behavior --
SDW and d-wave SC vertices compete, and which one wins depends on
the number of patches and (for $n=2$) on the value of $d_1$.

\begin{figure}[t]
\includegraphics[width = 4in]{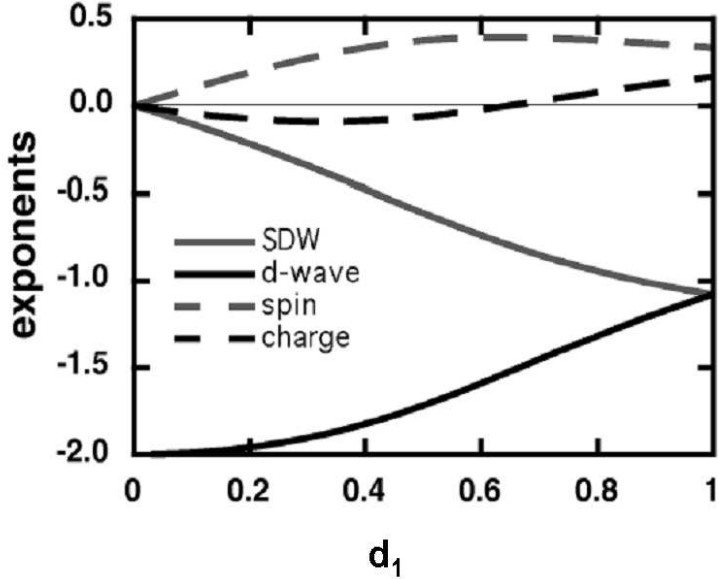}
\caption{\label{fig:Cu_flows2} The exponents for various
instabilities computed for different nesting parameters $d_1$. At
perfect nesting ($d_1=1$) the SDW and SC channels have the same
exponent, for $d_1<1$. The larger exponent is in the
superconducting channel. Compare with Fig. \ref{fig:alpha2}.
(Taken from Ref. \cite{Furukawa}.)}
\end{figure}

For 2-patch model, the equations and the results are the same as
in 2-pocket model: SDW wins at perfect nesting and SC winds at
non-perfect nesting (see Fig \ref{fig:Cu_flows2}).

For 3-patch model we have

\bea\label{eq:gr_vertices}
\frac{d\Gamma^{SC}}{dL}&=& 2(g_3-g_4)\Gamma^{SC}\nonumber\\
\frac{d\Gamma^{SDW}}{dL}&=& 2d_1(g_1+g_3)\Gamma^{SDW}\eea and $g_i
= \alpha_i/(L_c -L)$, so which channel wins depends on the
interplay between $\alpha_{SC} = 2 (\alpha_3-\alpha_4)$ and
$\alpha_{SDW} = 2d_1(\alpha_1+\alpha_3)$

Substituting $g_i = \alpha_i/(L_c -L)$ into the set of pRG
equations (\ref{eq:n_RG}) we obtain \bea\label{eq:gr_expo}
\alpha_1&=&d_1(\alpha_1^2+\alpha_3^2)\nonumber\\
\alpha_2&=&2d_1\alpha_2(\alpha_1-\alpha_2)\nonumber\\
\alpha_3&=&-\alpha_3^2-2\alpha_3\alpha_4+2d_1\alpha_3(2\alpha_1-\alpha_2)\nonumber\\
\alpha_4&=&-\alpha^2_4-2\alpha^2_3 \eea For $d_1 =1$, the solution
is \beq\label{eq:gr_expo_sol} \alpha_1\approx0.14, \alpha_2=0,
\alpha_3 =0.35, \alpha_4 \approx -0.4.  \eeq Hence
\beq\label{eq:gr_expo_sol_1}
\alpha_{SC}=0.74;~~~\alpha_{SDW}\approx 0.48\eeq

We see that already at perfect nesting SC vertex has a larger
exponent, i,e superconductivity is the first instability of a
system upon cooling.
 The same result has been obtained in fRG approach~\cite{ronny_gr}.
Observe that $\alpha_{SC} >1/2$, i.e., the divergence of the SC
three legged vertex does indeed lads to a SC instability (which,
we recall, leads to a $d + id$ or $d-id$ state, each breaks
time-reversal symmetry). However, $\alpha_{SDW}<1/2$ what implies
that in Ginzburg-Landau expansion the prefactor for the
$M^2_{SDW}$ remains positive, at least around superconducting
$T_c$.  This generally makes the possibility that SC and SDW
states co-exist below $T_c$ less likely~\cite{rahul_ch}

\begin{figure}[htp]
\includegraphics[width=4in]{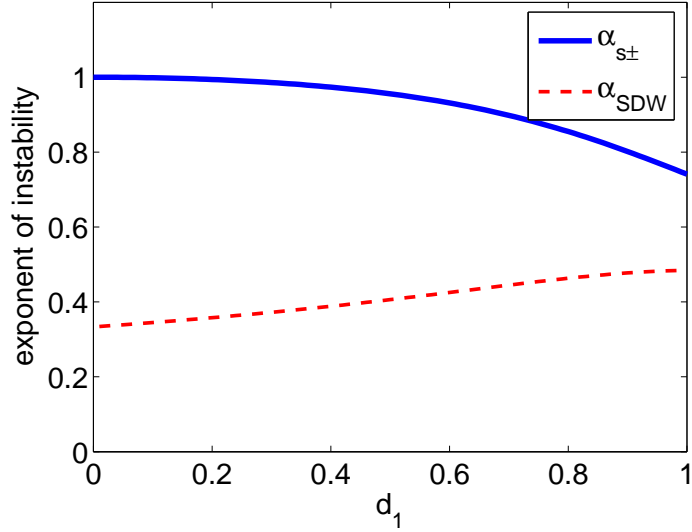}
\caption{\label{fig:Gr_alpha} Plot of $\alpha_{SC}$ and
$\alpha_{SDW}$ in 3-patch model vs $d_1$. Observe that
$\alpha_{SC}$ is larger already at $d_1=1$. In this respect, the
flow in 3-patch model is different from that in two-patch model
(Fig \ref{fig:alpha2}),  where $\alpha_{SDW}$ and $\alpha_{SC}$
were degenerate (to leading) order for $d_1=1$. }
\end{figure}

When $d_1 <1$, $\alpha_{SC}$ gets larger and $\alpha_{SDW}$ gets
smaller, i.e, SC instability becomes even more dominant. We show
the behavior of $\alpha_{SC}$ and $\alpha_{SDW}$ at different
$d_1$ in Fig\ref{fig:Gr_alpha}.

To summarize the results of pRG analysis:

\begin{itemize}
\item The SC vertex starts out as repulsive, but
it  eventually changes sign at some RG scale ($L_0$). This happens
due to the "push" from SDW channel, which rives rise to upward
renormalization of the inter-pocket/inter-patch interaction $g_3$.
\item
Both SDW and SC  vertices diverge at RG scale $L_c$ which is
larger than $L_0$. The leading instability is in the channel whose
vertex diverges with a larger exponent. At perfect nesting, SDW
instability occurs first in 2-pocket and two-patch models, however
in 3-patch model (and in some multi-pocket models) SC vertex has a
larger exponent that the SDW vertex and SC becomes the leading
instability.
\item Deviations from perfect nesting (quantified by $d_1<1$) act against SDW order by
reducing the corresponding exponent.  At sufficiently small $d_1$
SC instability becomes the leading one in all models which we
considered
\item
The necessary condition for the instability is the diverges of the
fluctuating component of the susceptibility. This sets up a
condition $\alpha >1/2$, where $\alpha$ is the exponent for the
corresponding vertex. For the leading instability, we found
$\alpha >1/2$ in all cases. For the subleading instability,
$\alpha$ can be either larger or smaller than $1/2$.  This affects
potential co-existence of the leading and subleading orders at a
lower $T$.
\end{itemize}

\section{Summary}

The goal of these lecture notes was  two-fold. First, to discuss
Kohn-Luttinger mechanism of superconductivity in systems with
nominally repulsive interaction, and, second, to provide a
guideline how to perform calculations to analyze SC instability
and its interplay with other potential instabilities, most notably
SDW instability. These lecture notes  is by no means a
comprehensive review of this wonderful field and we apologize if
we have missed some of the viewpoints and references. We have
presented how a weak coupling perspective can be used (quite
successfully) to describe not only superconductivity in many high
$T_c$ systems but interplay between competing orders.

We started by providing a very brief outline of the Fermi Liquid
theory that is necessary to deal with interacting fermions. We
used the Green's function formalism to define a 4-fermion
scattering vertex
$\Gamma_{\alpha\beta,\gamma\delta}(k_1,k_2;p_1,p_2)$. Quite
generally, the appearance of a pole in this vertex function in the
upper frequency half plane indicates that the Fermi Liquid is
unstable against a particular order.  We  showed the presence of
one such pole in the particle-particle (Cooper) channel at zero
total momentum of the interacting particles,
 for arbitrary small attractive interaction.

In the bulk of these lecture notes we addressed the issue how one
can get SC in systems with repulsive interaction. For isotropic
systems, the first step is the observation that the pairing
problem decouples between pairing channels with different angular
momentum $l$, and to get SC one needs an attraction for just one
value of $l$.  The second step is the observation, made by Kohn
and Luttinger, that Friedel oscillations of the screened repulsive
fermion-fermion interaction give rise to the appearance of
attractive components of the pairing interaction at large odd $l$,
no matter how the screening affects the regular (non-oscillating)
part of the interaction potential. Mathematically, the attraction
is due to non-analyticity of the screened interaction at the
maximum momentum transfer $2k_F$ between particles on the FS. We
applied KL reasoning to weak coupling and showed that in 3D the
attraction persists down to $l=1$, and the partial component with
$l=1$ is the largest by magnitude. The outcome is that an
isotropic 3D system with weak repulsive electron-electron
interaction is unstable towards a $p-$wave pairing.  The $p-$wave
pairing is the leading pairing instability also in 2D case, but to
get it one has to go to third order in the perturbation, while in
3D systems the attraction emerges already at second order.

Such a decomposition into decoupled angular momentum harmonics is,
however,  not possible in lattice systems due to reduced symmetry.
One can only decouple between  partial harmonics of the
interaction belonging to different representations of a discrete
lattice symmetry group.  We showed that KL reasoning can be
applied to lattice cases as well and considered as examples three
2D models: two-pocket model with small electron and hole pockets
separated by ${\bf Q} = (\pi,\pi)$, two-patch model with one large
FS on which there are two distinct regions with large density of
states, and three-patch model, with three such regions.  We argued
that the first model is applicable to Fe-pnictides, the second one
to optimally doped and overdoped cuprates, and the third one to
graphene doped to a vicinity of a topological transition from
multiple small FSs sheets to a single large FS. For each model we
found that superconductivity is possible if the interaction at
large momentum transfer $Q$ exceeds the interaction at a small
momentum transfer ($g_3 > g_4$ in our notations).  The emerging
pairing state has $s^{+-}$ symmetry for the Fe-pnictides,
$d_{x^2-y^2}$ symmetry for the cuprates, and $d+id$ symmetry for
doped graphene. In the latter case, superconductivity breaks
time-reversal symmetry.

We found that KL renormalization, taken to order $g^2$, does
produce an attractive component of the interaction.  If bare $g_3$
and $g_4$ are identical (the case of the momentum-independent
Hubbard-like interaction),  KL mechanism is sufficient to explain
the emergence of the attractive pairing interaction.  However, in
a more realistic case, $g_4$ (the interaction at small momentum
transfer), is larger than $g_3$ (the interaction with large
momentum transfer). In this situation, KL attraction has to
overcome bare repulsion, and this is generally not possible,
particularly at weak coupling. As a result, a lattice system can
remain in the normal (non-SC) state down to $T=0$.

We speculated how one can get SC by going beyond weak coupling and
briefly discussed spin-fermion model in this context. We argued
that the KL renormalization can be viewed as the  first term in
the series which gives rise to effective interaction mediated by
collective spin fluctuations. We then explored a peculiar
situation when the renormalization in particle-hole channel is
almost as strong as the renormalization in the particle-particle
(Cooper) channel. This is the case when the FS is nested, and we
argued that nesting is present in all three cases which we
considered. We argued that the nesting case can be studied beyond
second-order by applying a parquet renormalization group
technique.  This is a fully controlled weak coupling theory which
neglects higher terms in the dimensional couplings $g_i$ but keeps
corrections in $g_i \Pi_{pp} (0)$ and $g_i \Pi_{ph} (Q)$ to all
orders.

We found that in all three examples which we considered, RG flow
of the couplings is such that the system self-generates an
attraction below some energy scale. Specifically, we demonstrated
that at some RG scale the initially repulsive pairing interaction
changes sign and beyond this scale (at smaller energies) becomes
attractive. We argued that this conversion of repulsion into an
attraction is a universal phenomenon which does not depend on the
details of the underlying model, as long as particle-hole bubble
is comparable to particle-particle bubble and RG analysis is
applicable.

Finally, we analyzed the interplay between superconductivity and
other orders. The competition with SDW order is a particularly
relevant issue because SDW fluctuations are responsible for the
appearance of an attraction in the SC channel. We argued that in
some cases of near-perfect nesting  SDW order occurs first, but at
deviations from  nesting SC instability eventually occurs prior to
a magnetic instability. In other cases, SC instability comes first
even at perfect nesting, overshooting the interaction which made
attraction in the pairing channel possible.

We hope to have fairly addressed the phenomenon of
superconductivity in systems with repulsive interactions, but we
fully understand that we left a near-infinite amount  of
interesting physics that comes along with it. Our main hope is
that the readers, particularly graduate students, will find this
subject interesting and worth studying in more detail and depth.

\begin{theacknowledgments}
We are thankful to Profs. A. Avella and F. Manchini for organizing
the school in which one of us (A.C.) was a lecturer. It was
time-consuming but quite interesting task to write these notes in
the combination "professor-graduate student", and we have had
numerous discussions between the two of us about various aspects
of the electronic pairing mechanism. One of us (A.C.) worked on KL
mechanism about 25 years ago with M. Kagan, and it was quite
interesting to return to this issue after so many years and look
at high $T_c$ superconductivity though the prism of KL physics.
The work by Raghu et. al. (Ref. \cite{Raghu}) was very stimulating
in this regard.

The two of us  discussed various aspects of electronic pairing
with a large number of our colleagues, and we want to thank all of
them for fruitful discussions and various comments. This work has
been supported by the DOE grant DE-FG02-ER46900. SM acknowledges
support from ICAM-DMR-084415.
\end{theacknowledgments}

\bibliographystyle{aipproc}

\begin{thebibliography}{9}
\bibitem{oskar}
The authors are thankful to Oskar Vafek for clarifying this point
to them.
\end{thebibliography}

\end{document}